\documentclass[a4paper,11pt]{article}
\pdfoutput=1
\usepackage{jheppub} 
\usepackage[utf8x]{inputenc}      
\usepackage[T1]{fontenc}          
\usepackage{amsmath,amssymb}
\usepackage{booktabs,tabularx}
\usepackage{graphicx}
\usepackage{xspace}
\usepackage{tikz}
\usepackage[super]{nth}
\usepackage[font=small,labelfont=bf,format=plain,margin=0.05\textwidth]{caption}
\usepackage{listings}

\tikzstyle{arrow} = [draw, -latex, thick]
\tikzstyle{line} = [draw, -latex]
\usetikzlibrary{decorations.pathmorphing}	
\usetikzlibrary{decorations.markings}
\usetikzlibrary{arrows.meta}
\tikzset{
	fermion/.style={draw=black, postaction={decorate},
		decoration={markings,mark=at position .58 with {\arrow[draw=black,rotate=8]{Latex}}}},
	scalarnoarrow/.style={dashed,draw=black},
}

\title{\Large Improved MSSM Higgs mass calculation using the 3-loop
  FlexibleEFTHiggs approach including $x_t$-resummation }
\author[a]{Thomas Kwasnitza,}
\author[a]{Dominik St\"ockinger,}
\author[b]{Alexander Voigt}

\affiliation[a]{Institut f\"ur Kern- und Teilchenphysik,
  TU Dresden,\\ Zellescher Weg 19, 01069 Dresden, Germany}

\affiliation[b]{Institut für mathematische, naturwissenschaftliche und technische Bildung,
   Europa-Universität Flensburg, Auf dem Campus 1, 24943 Flensburg, Germany}

\emailAdd{thomas.kwasnitza@mailbox.tu-dresden.de}
\emailAdd{dominik.stoeckinger@tu-dresden.de}
\emailAdd{alexander.voigt@uni-flensburg.de}

\abstract{
    We present an improved calculation of the light CP-even Higgs
    boson pole mass in the MSSM based on the \feft\ hybrid
    method.  The calculation resums large logarithms to all orders and
    includes power-suppressed terms at fixed order.  It uses 
    state-of-the-art 2- and 3-loop matching of the quartic Higgs
    coupling and renormalization group running up to 4-loop,
    resulting in a resummation of large logarithmic corrections up to
    \NCLL\ level.  A conceptually novel ingredient is the expansion of
    the matching conditions in terms of high-scale MSSM parameters
    instead of SM parameters.  In this way leading QCD-enhanced 
    terms in the stop-mixing parameter are effectively resummed,
    leading to an improved numerical convergence of the perturbative
    expansion.  Furthermore, the avoidance of double counting of loop
    corrections is more transparent than in other approaches and more
    independent of the high-scale model.  We present numerical results
    and a detailed discussion of theoretical uncertainties for
    standard benchmark scenarios.
}
  

\arxivnumber{}

\hypersetup{
  pdftitle    = {Improved MSSM Higgs mass calculation using the 3-loop FlexibleEFTHiggs approach including xt-resummation},
  pdfauthor   = {Thomas Kwasnitza, Dominik Stöckinger, Alexander Voigt},
  pdfkeywords = {MSSM, Higgs, Mass}
}

\allowdisplaybreaks

\newcommand{\sarah}{\texttt{SARAH}\@\xspace}
\newcommand{\spheno}{\texttt{SPheno}\@\xspace}
\newcommand{\fs}{\texttt{FlexibleSUSY}\@\xspace}

\newcommand{\HSSUSY}{\texttt{HSSUSY}\xspace}
\newcommand{\susyhd}{\texttt{SusyHD}\@\xspace}
\newcommand{\MhEFT}{\texttt{MhEFT}\@\xspace}
\newcommand{\FeynHiggs}{\texttt{FeynHiggs}\@\xspace}
\newcommand{\Himalaya}{\texttt{Himalaya}\@\xspace}

\newcommand{\feft}{Flex\-ib\-le\-EFT\-Higgs}
\newcommand{\fefts}{FEFT\@\xspace} 
\newcommand{\prog}{\texttt{MSSMEFTHiggs3L}}

\newcommand{\ol}[1]{\overline{#1}}
\newcommand{\MSbar}{\ensuremath{\ol{\text{MS}}}\xspace}

\newcommand{\DRbarPrime}{\ensuremath{\ol{\text{DR}}'}\xspace}

\newcommand{\pole}{\text{pole}}
\newcommand{\EP}{\text{EP}}
\newcommand{\unit}[1]{\,\text{#1}}
\newcommand{\GeV}{\unit{GeV}}
\newcommand{\MeV}{\unit{MeV}}
\newcommand{\TeV}{\unit{TeV}}
\newcommand{\aemhatfive}{\ensuremath{\aemhat^{\SM(5)}}}
\newcommand{\ashatfive}{\ensuremath{\ashat^{\SM(5)}}}
\newcommand{\aemMZ}{\ensuremath{\aemhatfive(M_Z)}}
\newcommand{\asMZ}{\ensuremath{\ashatfive(M_Z)}}
\newcommand{\aemQ}{\ensuremath{\aemhatfive(Q)}}
\newcommand{\asQ}{\ensuremath{\ashatfive(Q)}}

\newcommand{\ord}{\mathcal{O}}
\newcommand{\order}[1]{\ord(#1)}
\newcommand{\full}{\ensuremath{\text{full}}\xspace}
\newcommand{\eft}{\ensuremath{\text{eft}}\xspace}
\newcommand{\EFT}{\ensuremath{\text{EFT}}\xspace}
\newcommand{\SM}{\ensuremath{\text{SM}}\xspace}
\newcommand{\BSM}{\ensuremath{\text{BSM}}\xspace}
\newcommand{\MSSM}{\ensuremath{\text{MSSM}}\xspace}

\newcommand{\QCD}{\ensuremath{\text{QCD}}}

\newcommand{\MS}{\ensuremath{M_S}\xspace}

\newcommand{\Qmatch}{\ensuremath{Q_\text{match}}}

\newcommand{\barlog}{\overline{\log}}
\newcommand{\tL}{{\ensuremath{0\ell}\xspace}}
\newcommand{\oneL}{{\ensuremath{1\ell}\xspace}}
\newcommand{\twoL}{{\ensuremath{2\ell}\xspace}}
\newcommand{\thrL}{{\ensuremath{3\ell}\xspace}}
\newcommand{\fourL}{{\ensuremath{4\ell}\xspace}}
\newcommand{\nL}{{\ensuremath{n\ell}\xspace}}

\newcommand{\NCLO}{\ensuremath{\text{N}^3\text{LO}}\xspace}
\newcommand{\NCLL}{\ensuremath{\text{N}^3\text{LL}}\xspace}
\newcommand{\NKLL}[1]{\ensuremath{\text{N}^{#1}\text{LL}}\xspace}
\newcommand{\figref}[1]{figure~\ref{#1}}
\newcommand{\Figref}[1]{Figure~\ref{#1}}
\newcommand{\secref}[1]{section~\ref{#1}}
\newcommand{\subsecref}[1]{subsection~\ref{#1}}
\newcommand{\appref}[1]{appendix~\ref{#1}}
\newcommand{\tabref}[1]{table~\ref{#1}}
\newcommand{\Qpole}{\ensuremath{Q_\pole}}
\newcommand{\Qlow}{\ensuremath{Q_{\text{low}}}}
\newcommand{\rd}{\ensuremath{\text{d}}}
\newcommand{\mbfive}{\ensuremath{\hat{m}_b^{\SM(5)}}}
\newcommand{\mbmb}{\ensuremath{\mbfive(\mbfive)}}

\newcommand{\mbQ}{\ensuremath{\mbfive(Q)}}
\newcommand{\msf}[1]{\ensuremath{m_{\tilde{#1}_3}}}
\newcommand{\msfi}[2]{\ensuremath{m_{\tilde{#1}_{#2}}}}
\newcommand{\msq}{\msf{q}}
\newcommand{\msu}{\msf{u}}
\newcommand{\msd}{\msf{d}}
\newcommand{\msl}{\msf{l}}
\newcommand{\mse}{\msf{e}}

\newcommand{\DMhQmatch}{\ensuremath{\Delta M_h^{\Qmatch}}}
\newcommand{\DMhQpole}{\ensuremath{\Delta M_h^{\Qpole}}}
\newcommand{\DMhtwoYt}{\ensuremath{\Delta M_h^{y_t,\twoL}}}
\newcommand{\DMhthrYt}{\ensuremath{\Delta M_h^{y_t,\thrL}}}
\newcommand{\DMhnYt}{\ensuremath{\Delta M_h^{y_t,\nL}}}
\newcommand{\DMhEW}{\ensuremath{\Delta M_h^{\text{imp},g_{1,2},\twoL}}}
\newcommand{\DMhtwoQCD}{\ensuremath{\Delta M_h^{\text{imp},g_3y_t,\twoL}}}
\newcommand{\DMhthrQCD}{\ensuremath{\Delta M_h^{\text{imp},g_3y_t,\thrL}}}
\newcommand{\DMhnQCD}{\ensuremath{\Delta M_h^{\text{imp},g_3y_t,\nL}}}
\newcommand{\DMhlam}{\ensuremath{\Delta M_h^{\lambda,\nL}}}
\newcommand{\DMhrepQCD}{\ensuremath{\Delta M_h^{\text{rep},g_3,\thrL}}}
\newcommand{\DMhrepHiggs}{\ensuremath{\Delta M_h^{\text{rep},g_3y_t,\thrL}}}
\newcommand{\DMhHS}{\ensuremath{\Delta M_h^{\text{HS}}}}
\newcommand{\DMhLS}{\ensuremath{\Delta M_h^{\text{LS}}}}
\newcommand{\DMh}{\ensuremath{\Delta M_h}}
\newcommand{\DlQmatch}{\ensuremath{\Delta \lambda_h^{\Qmatch}}}
\newcommand{\Dltwo}{\ensuremath{\Delta\lambda^{\twoL}}}
\newcommand{\Dlthr}{\ensuremath{\Delta\lambda^{\thrL}}}
\newcommand{\DltwoEW}{\ensuremath{\Dltwo_{g_1,g_2}}}
\newcommand{\DlthrGL}{\ensuremath{\Dlthr_{g_3,y_t}}}
\newcommand{\Dlrep}{\ensuremath{\Delta\lambda^{\text{rep}}}}
\newcommand{\mstop}[1]{\ensuremath{m_{\tilde{t}_{#1}}}}

\DeclareMathOperator{\re}{Re}

\def\elm{\text{em}}
\def\at{\alpha_t}
\def\athat{\hat{\alpha}_t}

\def\as{\alpha_s}
\def\ashat{\hat{\alpha}_s}
\def\atau{\alpha_{\tau}}
\def\aem{\alpha_{\elm}}
\def\aemhat{\hat{\alpha}_{\elm}}

\def\twoloopfm{g_3^2(y_t^4+y_b^4) + (y_t^2+y_b^2)^3 + (y_t^2+y_\tau^2)^3}
\def\twoloopeft{\hat g_3^2(\hat y_t^4+\hat y_b^4) + (\hat y_t^2+\hat y_b^2)^3 + (\hat y_t^2+\hat y_\tau^2)^3}
\def\thrloopfm{g_3^4y_t^4}
\def\thrloopeft{\hat g_3^4\hat y_t^4}

\begin{document}
\maketitle
\newpage

\section{Introduction}

Since the discovery of the Higgs boson
\cite{Aad:2012tfa,Chatrchyan:2012xdj}, the Higgs boson mass
$M_h = (125.10 \pm 0.14)\GeV$ \cite{Aad:2015zhl,Tanabashi:2018oca} has
become a high-precision observable \cite{Khachatryan:2016vau}, which
represents another useful tool to search for physics beyond the
Standard Model (SM) and constrain the large zoo of proposed SM
extensions, such as supersymmetric (SUSY) models.  The latter are
particularly interesting, as they require the existence of scalar
fields and predict the quartic Higgs coupling and thus the Higgs
boson mass.
The precise prediction of the SM-like Higgs boson mass in the Minimal
Supersymmetric Standard Model (MSSM), however, is a long-standing
challenge, because in viable MSSM scenarios large radiative loop
corrections of the order $\Delta m_h^2 \sim (100\GeV)^2$ are required,
resulting in a large truncation error of the perturbation series.

There are two main mechanisms which can generate such large loop
corrections: $(i)$ Large SUSY masses $\MS$ (in particular stop masses)
lead to large logarithmic corrections of the form
$\log\left(\MS/v\right)$, where $v$ represents the electroweak
scale. $(ii)$ A large mixing in the stop sector, governed by the
parameter $X_t$, leads to power corrections of the order
$(X_t/\MS)^n$.  Effective field theory (EFT) techniques are a
well-known tool to perform a resummation of the large logarithmic
corrections, thus effectively avoiding a large truncation error of the
contributions from mechanism $(i)$.  Concerning the $(X_t/\MS)^n$ power
corrections, however, no similar resummation technique has been used
so far.  In the present work we present a technique to effectively
resum leading terms in $X_t$ in the prediction of the light Higgs
boson mass.

There are different approaches to calculate the Higgs boson mass in
supersymmetric models, which can be roughly classified into
fixed-order
\cite{Hempfling:1993qq,Heinemeyer:1998kz, Heinemeyer:1998jw,
  Heinemeyer:1998np, Degrassi:2001yf,
  Brignole:2001jy,Martin:2001vx,Martin:2002iu,Martin:2002wn,Dedes:2002dy,
  Brignole:2002bz, Dedes:2003km,Martin:2003it,
  Allanach:2004rh,Martin:2004kr, Heinemeyer:2004xw, Martin:2005eg,
  Martin:2007pg, Heinemeyer:2007aq, Harlander:2008ju,
  Kant:2010tf,Hollik:2014wea,
  Hollik:2014bua,Degrassi:2014pfa,Borowka:2014wla,Borowka:2015ura,
  Goodsell:2016udb, Harlander:2017kuc,
  Passehr:2017ufr,Stockinger:2018oxe, Borowka:2018anu, R.:2019ply,
  Goodsell:2019zfs}%
\footnote{Here we focus on multi-loop calculations. For further 
references see the review \cite{Draper:2016pys} and the references therein.},
EFT
\cite{Draper:2013oza,Bagnaschi:2014rsa,Vega:2015fna,Lee:2015uza,
  Bagnaschi:2017xid,Braathen:2018htl,Gabelmann:2018axh,Allanach:2018fif,
  Harlander:2018yhj,Bagnaschi:2019esc,Kramer:2019fwz,Bahl:2019wzx},
and hybrid
\cite{Hahn:2013ria,Bahl:2016brp,Athron:2016fuq, Staub:2017jnp,Athron:2017fvs,
  Bahl:2017aev,Bahl:2018jom,R.:2019irs,Harlander:2019dge,
  Bahl:2019hmm}
approaches, which combine the virtues of the former two.  Fixed-order
approaches truncate the perturbation series at a certain order in
loops and couplings, neglecting in particular large logarithmic
corrections arising at higher orders.  Thus, when $\MS \gg v$, the
fixed-order approaches usually suffer from a large uncertainty due to
missing large higher-order corrections.
EFT approaches, on the other hand, resum the large logarithmic
corrections to all orders, but usually neglect terms of the order
$v^2/\MS^2$.  As a consequence, EFT approaches become imprecise when
$\MS \sim v$.

Hybrid approaches combine the virtues of fixed-order and EFT
calculations: They resum large logarithmic corrections to all orders
and include terms suppressed by $v^2/\MS^2$ at fixed order.  A first
variant of such a hybrid approach was presented in
ref.~\cite{Hahn:2013ria} and implemented into \FeynHiggs.  This
approach uses a ``subtraction method'', where the large logarithmic
corrections are subtracted from a fixed-order calculation and are
replaced by resummed logarithms, avoiding  double counting.  This method was refined in
refs.~\cite{Bahl:2016brp,Bahl:2017aev}
and applied in the context of the
\DRbarPrime\ scheme in ref.~\cite{Harlander:2019dge}.

An alternative way to realize a hybrid approach was presented in
refs.~\cite{Athron:2016fuq,Athron:2017fvs}. This so-called \feft\
approach is an EFT calculation in which the matching condition is
suitably modified such that terms suppressed by powers of $v^2/\MS^2$
are included in the quartic Higgs coupling.  One advantage of this
method is the structural simplicity of the matching condition.  As a
result, the method is well suited for automation and has thus been
implemented into the generic spectrum generators \fs\
\cite{Athron:2014yba,Athron:2017fvs}
and \sarah/\spheno\ \cite{Staub:2017jnp}.
A difficulty of the FlexibleEFTHiggs approach is to make sure that
large logarithms cancel in the matching between the EFT and the UV model, as
required. Indeed, avoiding double counting leads to significant
complications in all hybrid 
calculations \cite{Athron:2017fvs,Bahl:2017aev,Bahl:2018ykj}.

In this paper we present an extension of the \feft\ hybrid
approach with a matching of the quartic Higgs coupling $\hat{\lambda}$
beyond 1-loop level (next-to-leading order, NLO) and apply it to
perform a state-of-the-art
hybrid calculation of the light CP-even Higgs boson mass in the real
MSSM.  Thereby our calculation incorporates several conceptual changes
and significant improvements:
\begin{itemize}
\item We parametrize the matching calculation at the high-energy scale
  in terms of parameters of the UV model (i.e.\ the MSSM).  This is in
  contrast to the usually chosen parametrization in terms of EFT
  parameters.  Our ``full-model parametrization'' has several
  significant advantages.  An important advantage is that the
  cancellation of large logarithmic corrections in the matching is
  more transparent.  Furthermore, our parametrization allows for 
  a computer algebraic implementation which is to a large
  extent independent of the chosen UV model. This fact enables
  the straightforward application 
  to a large class of SUSY models.  The
  detailed discussion of the different possible parametrizations is
  presented in \secref{sec:matching_general}.

\item In our application to the MSSM we include the state-of-the-art
  radiative corrections in the matching up to the 3-loop level at
  $\ord(\oneL + \twoloopfm + \thrloopfm)$ and perform
  renormalization-group running up to 4-loop level in QCD.  As a
  result, our calculation reaches a precision of \NCLO\ with a
  resummation of \NCLL, comparable with the calculation presented in
  ref.~\cite{Harlander:2019dge}.  The details of the matching of the
  MSSM to the SM are presented in \secref{sec:expansion_Master_eq}, and
  numerical results are shown in
  sections~\ref{sec:numerical_results}--\ref{sec:uncertainty}. 

\item The most important advantage of our new approach and the chosen
  full-model parametrization
  is the effective resummation of QCD-enhanced terms leading in the stop mixing
  parameter $X_t$, which is presented in \secref{sec:resummation}.
  More specifically, we show that the highest power contributions
  at $\ord(y_t^4g_3^{2n}, y_t^2g_{1,2}^2g_3^{2n} )$ for all $n>0$ are captured by our procedure.
    As a
  result, the perturbation expansion of the Higgs boson mass in terms
  of the MSSM parameters stabilizes significantly for large $X_t$,
  leading to a reduced theory uncertainty of the prediction.
\end{itemize}
We begin with a recap of the \SM\ and the \MSSM\ in
\secref{sec:model_intro}, introducing our conventions.
\secref{sec:matching_general} gives a general overview of the
implementation of the EFT approach, discussing in particular the role
of the parametrization.  Our new realization of the \feft\ approach
within a numerical code is discussed in
\secref{sec:expansion_Master_eq}.  In \secref{sec:resummation} we show
how our chosen parametrization in terms of MSSM parameters results in
a resummation of highest power $X_t$ contributions as described above.
 After a study of the numerical
results of our new calculation in \secref{sec:numerical_results}, we
perform a thorough analysis of the remaining theory uncertainty of our
calculation in \secref{sec:uncertainty}.

\section{Definition of the Standard Model and the MSSM}
\label{sec:model_intro}

In the following we will denote the Standard Model (SM) parameters,
defined in the \MSbar\ scheme, as
\begin{align}
  \hat{P} = \{\hat{g}_1, \hat{g}_2, \hat{g}_3, \hat{y}_t, \hat{y}_b, \hat{y}_\tau, \hat{\lambda}, \hat{v}\},
  \label{eq:SM_parameters}
\end{align}
where $\hat{g}_1 = \sqrt{5/3}~\hat{g}_Y$ and $\hat{g}_Y$, $\hat{g}_2$
and $\hat{g}_3$ denote the gauge couplings of the gauge groups
$U(1)_Y$, $SU(2)_L$ and $SU(3)_C$, respectively.  The Yukawa couplings
of the top quark, bottom quark and tau lepton are denoted as
$\hat{y}_t$, $\hat{y}_b$ and $\hat{y}_\tau$, respectively.  The
\nth{1} and \nth{2} generation Yukawa couplings as well as
CP-violation effects are neglected and we will set the CKM and PMNS
matrices to unity.  The quartic coupling $\hat\lambda$ of the SM
Higgs field $\Phi$ is defined by the Higgs potential
\begin{align}
  V(\Phi) = \hat{\mu}^2 |\Phi|^2 + \frac{\hat{\lambda}}{2} |\Phi|^4 .
\end{align}
We decompose the Higgs field as
\begin{align}
  \Phi =
  \begin{pmatrix}
    G^+ \\ \frac{1}{\sqrt{2}} (\hat{v} + h + i G^0)
  \end{pmatrix},
\end{align}
where $h$ is the SM Higgs particle,
$\hat{v}\equiv \sqrt{2} \langle \Phi \rangle$ is the Higgs
vacuum expectation value (VEV) (i.e.~the minimum of the 
SM effective potential) which satisfies $\hat{v}=
(-2\hat{\mu}^2/\hat{\lambda})^{1/2} \approx 246\GeV$ at tree level
and $G^{0,\pm}$ are the SM Goldstone bosons.
After spontaneous electroweak symmetry breaking, the \MSbar\
masses for the top, bottom and tau fermion and for the heavy physical
bosons are given by
\begin{align}
  \hat{m}_t   &= \frac{\hat{y}_t \hat{v}}{\sqrt{2}}, &
  \hat{m}_b   &= \frac{\hat{y}_b \hat{v}}{\sqrt{2}}, &
  \hat{m}_\tau&= \frac{\hat{y}_\tau \hat{v}}{\sqrt{2}}, \\
  \hat{m}_W   &= \frac{\hat{g}_2 \hat{v}}{2}, &
  \hat{m}_Z   &= \frac{\hat{v}}{2} \sqrt{\hat{g}_Y^2 + \hat{g}_2^2}, &
  \hat{m}_h^2 &= \hat{\lambda} \hat{v}^2 .
\end{align}
For convenience we define in addition the following symbols:
\begin{align}
  \athat &= \frac{\hat{y}_t^2}{4\pi}, &
  \ashat &= \frac{\hat{g}_3^2}{4\pi}, &
  \aemhat &= \frac{\hat{e}^2}{4\pi}, &
  \hat{e} &= \frac{\hat{g}_Y \hat{g}_2}{\sqrt{\hat{g}_Y^2 + \hat{g}_2^2}}.
\end{align}
We denote the corresponding relevant parameters of the ($R$-parity
conserving) Minimal Supersymmetric Standard Model (MSSM), defined in
the \DRbarPrime\ scheme, as
\begin{align}
  P = \{g_1, g_2, g_3, y_t, y_b, y_\tau, v\},
  \label{eq:MSSM_parameters}
\end{align}
where $g_1 = \sqrt{5/3}\, g_Y$ and $v = (v_u^2 + v_d^2)^{1/2}$,
whereas $v_u$ and $v_d$ denote the VEVs of the up- and down-type Higgs
fields which represent the minimum of the effective potential in the MSSM,
\begin{align}
  \langle H_u \rangle &= \frac{1}{\sqrt{2}}
  \begin{pmatrix} 0 \\ v_u \end{pmatrix}, &
  \langle H_d \rangle &= \frac{1}{\sqrt{2}}
  \begin{pmatrix} v_d \\ 0 \end{pmatrix}.
\end{align}
If not stated otherwise, we define $\tan\beta = v_u/v_d$.
After the spontaneous electroweak symmetry breaking in the MSSM, the
\DRbarPrime\ masses for the top, bottom and tau fermion as well as the
SM-like Higgs in the decoupling limit are given by
\begin{align}
  m_t &= \frac{y_t v_u}{\sqrt{2}}, \qquad\qquad\qquad
  m_b = \frac{y_b v_d}{\sqrt{2}}, \qquad\qquad\qquad
  m_\tau = \frac{y_\tau v_d}{\sqrt{2}}, \\
  m_h^2 &= \frac{1}{4} \left( g_Y^2 + g_2^2 \right) v^2 \cos^2 (2\beta).
\end{align}
We neglect inter-generation sfermion mixing, so the \DRbarPrime\
masses of the stops, sbottoms and staus are given by the eigenvalues
of the mass matrices
\begin{align}
  \mathsf{M}_t =
  \begin{pmatrix}
    m_t^2 + \msq^2 & m_t X_t \\
    m_t X_t & m_t^2 + \msu^2
  \end{pmatrix}, \\
  \mathsf{M}_b =
  \begin{pmatrix}
    m_b^2 + \msq^2 & m_b X_b \\
    m_b X_b & m_b^2 + \msd^2
  \end{pmatrix}, \\
  \mathsf{M}_\tau =
  \begin{pmatrix}
    m_\tau^2 + \msl^2 & m_\tau X_\tau \\
    m_\tau X_\tau & m_\tau^2 + \mse^2
  \end{pmatrix},
\end{align}
where $\msq^2$, $\msu^2$, $\msd^2$, $\msl^2$ and $\mse^2$ denote the
squared soft-breaking mass parameters of the left- and right-handed
\nth{3} generation squarks and sleptons and electroweak contributions
from $D$-terms have been omitted.\footnote{In the calculation of the
  2-loop and 3-loop matching corrections presented in the next
  sections, we neglect $D$-term contributions, as we work in the
  gauge-less limit. In the calculation of the 1-loop threshold
  correction $\Delta\lambda^\oneL$, all $D$-term contributions are
  taken into account.}  The sfermion mixing parameters $X_t$, $X_b$
and $X_\tau$ are defined as
\begin{align}
  X_t &= A_t - \mu\cot\beta, &
  X_b &= A_b - \mu\tan\beta, &
  X_\tau &= A_\tau - \mu\tan\beta,
\end{align}
where $A_f$ ($f=t,b,\tau$) are the trilinear Higgs--sfermion--sfermion
couplings and $\mu$ is a MSSM superpotential parameter.  For
convenience we define in addition the following symbols:
\begin{align}
  \at &= \frac{y_t^2}{4\pi}, &
  \as &= \frac{g_3^2}{4\pi}, &
  \aem &= \frac{e^2}{4\pi}, &
  e &= \frac{g_Y g_2}{\sqrt{g_Y^2 + g_2^2}}, &
  \MS^2 &= \mstop{1} \mstop{2},
\end{align}
where $\mstop{i}$ denotes the $i\textsuperscript{th}$ \DRbarPrime\
stop mass.

\section{Matching procedure in general}
\label{sec:matching_general}

We begin by recalling a few basic aspects of the effective field
theory approach to compute weak-scale observables, such as the pole
mass of the Higgs boson, $M_h$, in SUSY models in scenarios where the
SUSY scale $\Lambda$ is significantly larger than the weak scale.
This will help later in characterizing our approach and in comparing
it to other approaches.

\subsection{Basics of the effective field theory approach}

In SUSY models with very heavy new particles of mass $\Lambda\gg v$,
an observable $O$ can be expanded perturbatively in a three-fold way:
in terms of \emph{loops} (counted by a generic loop-counting parameter
$\alpha$),
\emph{large logarithms} of the large mass ratio
$L\equiv\log(\Lambda/v)$ and a \emph{mass suppression factor}
$v/\Lambda$.  For the particular  case of a dimensionless observable $O$ 
which has a tree-level
contribution of order $\alpha^0$, the leading
$n$-loop contribution ($n \ge 0$) is typically of the form
$ \alpha^n L^n$.  Subleading/higher-order contributions have more
powers of $\alpha$, fewer powers of $L$ and/or additional factors of
$v/\Lambda$.  Hence, one can write to all orders
\begin{align}
  O = \sum_{n=0}^\infty \sum_{l=0}^n\sum_{k=0}^\infty c_{nlk}~ \alpha^n  L^l  \left(\frac{v}{\Lambda}\right)^k ,
\end{align}
where the sum of the terms with $n=0$ represent the tree-level
contribution $O^{\tL}$ and the coefficients $c_{nlk}$ are constants
which may contain the parameters of the full model and logarithms of small
mass ratios.  An effective field theory calculation allows to include
all terms at the $m$-th subleading $\log$ level,
\begin{align}
  \alpha^n L^n, \ldots, \alpha^{n+m} L^n \quad \forall n \ge 0.
\end{align} 
Since terms of all loop orders are contained in the sum of these
terms, their inclusion is also called ``resummation of logarithms''.
Usually the resummation is achieved by performing the following three
steps (see \figref{fig:eft-approach}):
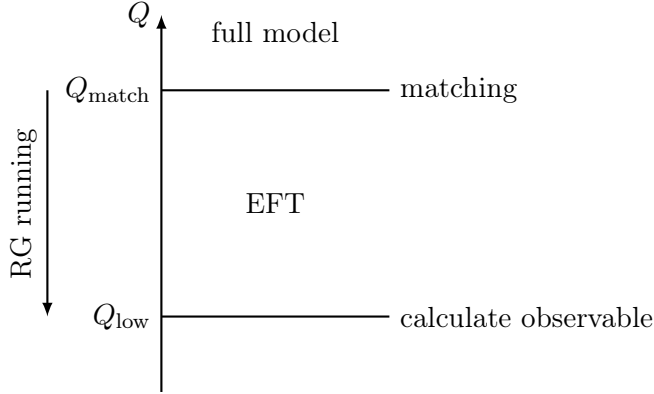
\begin{figure}[tbh]
  \centering
  \begin{tikzpicture}
    \path[arrow] (0,0) -- (0,5) node[left]{$Q$};
    \draw[thick] (0,4) node[left]{$\Qmatch$} -- node[above = 0.5cm]{full model} (3,4) node[right]{matching};
    \draw[thick] (0,1) node[left]{$\Qlow$} -- (3,1) node[right]{calculate observable};
    \draw[thick] (1.5,2.5) node[]{EFT};
    \path[arrow,latex-] (-1.5,1) -- node[above,rotate=90]{RG running} (-1.5,4);
  \end{tikzpicture}
  \caption{Calculation of an observable in an effective field theory
    of a full model.}
  \label{fig:eft-approach}
\end{figure}
\begin{enumerate}
\item Construct a Lagrangian of the effective theory and derive a
  relation between the running parameters of the full and the
  effective theory by a matching calculation at the $m$-loop level at
  the high scale $\Qmatch \approx \Lambda$.\footnote{Alternatively one
    may integrate out the heavy states and derive the Lagrangian of
    the EFT, from which the relation between the running parameters of
    the full and the effective theory can be read off.}
\item Use ($m+1$)-loop renormalization group running to evolve the
  parameters of the EFT from the scale $\Qmatch$ to the low-energy scale
  $\Qlow$.  In this process the large logarithms are resummed to the
  \NKLL{m} order.
\item Match the parameters of the EFT at the scale $\Qlow$ to observed
  quantities and compute the observable in question at $m$-loop level.
\end{enumerate}
It is not only crucial to take into account all $m$-loop terms,
but it is also important to consistently truncate the
perturbation expansion at the $m$-loop order.  In particular, it is
imperative not to include any spurious $(>m)$-loop terms enhanced by
large logarithms as this would spoil the correct resummation.  On the
other hand it is allowed to incorporate $m$-loop terms suppressed by
powers of
$v/\Lambda$ in the $m$-loop matching, i.e.\ to take into account fixed
order terms of the form ${\alpha^{m} L^{k} v/\Lambda}$.  In this way
the computation of low-energy observables can be improved by power
suppressed terms at fixed loop order.  Note, however, that as long as
only running of operators of mass dimension $\leq 4$ is used,
power-suppressed large logarithms of the form $L^k v/\Lambda$ are not
resummed \cite{Athron:2016fuq}, i.e.~terms of order
$\alpha^{n+m} L^{n+(k\leq m)} v/\Lambda$ with $n \geq 1$ are not
correctly predicted.
However, in ref.~\cite{Bagnaschi:2017xid} it was shown that this
effect is negligible for the purpose of Higgs pole mass prediction in
the relevant parameter space of the MSSM.

\subsection{Parametrization of the matching relations}
\label{sec:parametrization_choices}

In the following we discuss different possibilities to perform the
high-scale matching.  Specifically, for a matching at some given loop order,
one needs to consistently expand either in terms of the running
parameter of the fundamental theory $\alpha$ or of the EFT
$\hat{\alpha}$.  In principle both options are correct and
equivalent.  However, once perturbation theory is truncated it matters
whether truncation is done at the order $(\alpha)^m$ or
$(\hat{\alpha})^m$, because these two kinds of expansions differ by
higher-order terms.  We give a simple illustration using a 1-loop
toy example which is similar to the case of the Higgs pole mass
calculation.

We suppose the exact matching condition is given by the equality
\begin{align}\label{eq:0}
  \Gamma^{\eft}=\Gamma^{\full},
\end{align}
where $\Gamma$ is some Green function. In the full theory, the
1-loop expression is
\begin{align}
  \Gamma^{\full} = \alpha + \alpha^2 [\Delta_\gamma L + \Delta_c],
  \label{eq:1}
\end{align}
where $\Delta_\gamma$ and $\Delta_c$ are numerical coefficients.  In
the EFT, the 1-loop expression reads
\begin{align}
  \Gamma^{\eft}= \hat{\lambda} + \hat{\alpha}^2 [\Delta_\gamma L].
  \label{eq:2}
\end{align}
The coefficient $\Delta_\gamma$ of the large logarithm $L$ is the same
in both cases, because it must cancel in the matching condition.  We
assume that $\alpha$ and $\hat{\alpha}$ are related at 1-loop level
by
\begin{align}
  \hat{\alpha}= \alpha + \alpha^2 \Delta_\alpha .
\end{align}
The matching condition can now be solved perturbatively for
$\hat{\lambda}$ in terms of $\alpha$ or $\hat{\alpha}$.  At
tree-level one obtains
\begin{align}
  &\textbf{tree-level}:&&
  \hat{\lambda}=\alpha=\hat{\alpha}.
%
\intertext{At the 1-loop level one obtains in terms of $\alpha$:}
%
  &\textbf{full-model parametrization}~\oneL:&&
  \hat{\lambda} = \alpha + \alpha^2 \Delta_c
  \label{full_parametrization}
%
\intertext{and in terms of $\hat{\alpha}$:}
%
  &\textbf{EFT parametrization}~\oneL:&&\hat{\lambda} = \hat{\alpha}+\hat{\alpha}^2\left[ \Delta_c
- \Delta_\alpha\right].
  \label{EFT_parametrization}
\end{align}
Both expressions \eqref{full_parametrization} and
\eqref{EFT_parametrization} are valid possibilities for the 1-loop
matching relations, but the results for $\hat{\lambda}$ differ by
non-log-enhanced 2-loop terms.  In fact, this difference could be used
as an estimate of the theory uncertainty.  For the prediction of the
Higgs boson pole mass, the EFT parametrization is used in several
calculations such as \HSSUSY\footnote{According to
  ref.~\cite{Bagnaschi:2017xid}, the bottom Yukawa coupling
  inside the 2-loop threshold correction $\Delta\lambda$ are in the
  full-model parametrization for the reason of correct $\tan \beta$
  treatment, as will be discussed in \secref{sec:resummation}.}
\cite{Athron:2017fvs,Allanach:2018fif}, \MhEFT \cite{Lee:2015uza} and
\susyhd \cite{Vega:2015fna}, although further parametrizations have
been presented in refs.~\cite{Martin:2007pg,Draper:2013oza}.

We note that in an algorithmic implementation of the full-model
parametrization, the Green function $\Gamma^{\full}$ may be evaluated
numerically, while $\Gamma^{\eft}$ needs to be analytically expanded
in terms of $\alpha$ and truncated consistently at the 1-loop
level. Hence, an analytic manipulation of $\Gamma^{\eft}$ is needed.
Conversely, an algorithmic implementation of the EFT parametrization
would require an analytic expansion of $\Gamma^{\full}$ in terms of
$\hat{\alpha}$ and a consistent truncation of that expansion.

Finally, we note that one might be tempted to plug  the respective 1-loop
results \eqref{eq:1}--\eqref{eq:2} into the matching condition
\eqref{eq:0} to obtain
\begin{align}
  \hat{\lambda} + \hat{\alpha}^2\left[\Delta_\gamma L \right] =
  \alpha + \alpha^2\left[\Delta_\gamma L + \Delta_c
  \right]
\end{align}
and solve for $\hat{\lambda}$, e.g.~numerically.  One would then
obtain
\begin{align}
  &\textbf{incorrect:}&&\hat{\lambda}=\alpha + \alpha^2 \Delta_c + \alpha^3 2\Delta_\alpha \Delta_\gamma L + \mathcal{O}(\alpha^4)\,.
\end{align}
Here, a spurious $\log$-enhanced 2-loop term is generated.  If such an
implementation were used, the resummation of subleading logarithms
would be spoiled.  A problem of this kind appeared in
refs.~\cite{Athron:2016fuq,Staub:2017jnp} and a solution was first
discussed in ref.~\cite{Athron:2017fvs}.

\subsection{Matching of the quartic Higgs coupling}
\label{sec:matching_of_quartic}

In the following we will discuss the differences between the two
parametrizations in the context of predicting the quartic Higgs
coupling $\hat{\lambda}$ from a matching of the Standard Model to the
MSSM.

\paragraph{EFT (SM) parametrization.}

In this parametrization the quartic Higgs coupling $\hat{\lambda}$ is
expressed in terms of the \MSbar-renormalized SM parameters
$\{\hat{g}_1, \hat{g}_2, \hat{g}_3, \hat{y}_t, \hat{y}_b,
\hat{y}_\tau, \hat{v}\}$ at the matching scale $\Qmatch$.  In the
scenario with degenerate SUSY mass parameters and $\Qmatch = \MS$, the
1-loop contribution to $\hat{\lambda}$ from stops is given by
\begin{align}
  \textbf{EFT parametrization:}&&
  \left.\Delta\lambda^\oneL \right|_{\hat{y}_t^4} = \frac{1}{(4\pi)^2} \hat{y}_t^4 6\left[
  x_t^2 - \frac{x_t^4}{12}
  \right],
  \label{eq:DLambda_sm}
\end{align}
where $x_t = X_t/\MS$ is the dimensionless stop-mixing parameter in
the \DRbarPrime\ scheme.

\paragraph{Full-model (MSSM) parametrization.}

In this parametrization the MSSM parameters are treated as
fundamental.  At the matching scale the quartic Higgs coupling
$\hat\lambda$ is then fixed in terms of the MSSM \DRbarPrime\
parameters.  As a result, the 1-loop contribution to
$\Delta\lambda^\oneL$ reads at $\ord(y_t^4)$
\begin{align}
  \textbf{full-model parametrization:}&&
  \left.\Delta\lambda^\oneL \right|_{y_t^4}  = \frac{1}{(4\pi)^2} y_t^4 s_\beta^46\left[
  x_t^2 - \frac{x_t^4}{12}
  \right],
  \label{eq:DLambda_MS}
\end{align}
where $y_t$ denotes the MSSM top Yukawa coupling in the \DRbarPrime\
scheme.

With respect to the top Yukawa and strong gauge coupling, the
difference between the EFT and the full-model parametrization
\eqref{eq:DLambda_sm} and \eqref{eq:DLambda_MS} is of 2-loop order.
This can be seen by equivalently reparametrizing
eq.~\eqref{eq:DLambda_MS} in terms of the SM \MSbar top Yukawa
coupling $\hat{y}_t$, which leads to
\begin{align}\begin{split}
    \left.\Delta\lambda^\oneL \right|_{y_t^4} ={}& \frac{1}{(4\pi)^2} \hat{y}_t^4 6\left[
      x_t^2 - \frac{x_t^4}{12}
    \right]
    - \frac{1}{(4\pi)^4}\frac{8}{3}\hat{y}_t^4\hat{g}_3^2\left[
      x_t^5 + \propto x_t^{n\le 4}\right] +\ord(g_3^4y_t^4) .
  \end{split}
  \label{eq:DLambda_MSSM_rep}
\end{align}
Comparing the two versions of the threshold corrections
\eqref{eq:DLambda_sm} and \eqref{eq:DLambda_MSSM_rep} reveals several
important points.
We note first that by construction the 2-loop term on the r.h.s.\ of
eq.~\eqref{eq:DLambda_MSSM_rep} does not contain large logarithms, in
agreement with the effective field theory paradigm.  Clearly, the
2-loop difference between eqs.~\eqref{eq:DLambda_sm} and
\eqref{eq:DLambda_MSSM_rep} could be used as a measure of the theory uncertainty of the
1-loop prediction of $\hat{\lambda}$ at the matching scale. Finally note that
this reparametrization generates a 2-loop $x_t^5$ term on
the r.h.s.\ of eq.~\eqref{eq:DLambda_MSSM_rep}. This term is correct,
i.e.\ it  appears in the
explicit 2-loop calculation of ref.~\cite{Bagnaschi:2014rsa}.
In \secref{sec:resummation} we will show
that this is not an accident; the full-model
parametrization includes important terms correctly, which in the
EFT parametrization would require  higher-order calculations. It can
thus be used to improve the precision 
of Higgs pole mass prediction in the effective field theory approach.

\paragraph{Automatization of the matching beyond 1-loop level.}

Besides the higher precision, the full-model parametrization may also
be easier to implement in generic spectrum generators that use the
\feft\ approach \cite{Athron:2016fuq,Athron:2017fvs,Staub:2017jnp}.
In this approach the condition
\begin{align}
  (M_h^\MSSM)^2 = (M_h^\SM)^2
\end{align}
is numerically solved for $\hat{\lambda}$ at the matching scale.  As
discussed in ref.~\cite{Athron:2017fvs} and in
\secref{sec:parametrization_choices}, care has to be taken to avoid
the occurrence of spurious large logarithms of higher-order in the
matching.
A correct application of \feft\ approach beyond the 1-loop level using
the EFT parametrization requires an expansion of the full-model BSM Higgs
self-energy $\Sigma_\phi^\BSM(P)$ in terms of the full-model
 parameters $P$ and a following expansion of $P$ in
terms of the parameters of the \EFT\ (here the SM) $\hat{P}$,
including a truncation at some fixed order in $\hat{P}$.  This
expansion introduces ``implicit'' terms beyond 1-loop of the form
\begin{align}
  \textbf{EFT parametrization:}&&
  \Delta\lambda^\twoL \supset -\left(\frac{\partial}{\partial  P}
  \Sigma^{\BSM}_{\phi}\right) \Delta \hat{P},
  \label{eq:double-loop}
\end{align}
where $\Delta \hat{P} = P - \hat{P}$ is the threshold correction of
\SM-like parameters expressed through SM parameters.  Thus, the
inclusion of derivatives of the BSM Higgs self-energy w.r.t.\ SM-like
parameters becomes mandatory for the cancellation of large logarithms
in the matching beyond 1-loop.  The calculation of these derivatives
requires some extra computational effort, which must be performed for
each BSM model.  The application of this approach to arbitrary BSM
models thus requires some cost.

Within the full-model parametrization, the Higgs self-energy in the
EFT, $\Sigma^\SM_h(\hat{P})$, must be expanded in terms of the
parameters of the EFT, $\hat{P}$, which then must be expanded in terms
of the parameters of the full model, $P$.  As a result, 
2-loop structures of the following form are generated
\begin{align}
  \textbf{full-model parametrization:}&&
  \Delta\lambda^\twoL \supset \left(\frac{\partial}{\partial  \hat{P}}
  \Sigma^\SM_h\right) \Delta P,
  \label{eq:double-loop,t-d}
\end{align}
where $\Delta P = \hat{P} - P$ is the threshold correction of
\SM-like BSM parameters expressed through BSM parameters.  Thus, only
derivatives of the \EFT\ Higgs self-energy are required.  As long as
the employed EFT does not change, these derivatives can be computed
once and reused in the matching to arbitrary BSM models.  With respect
to computational effort and model independence, the full-model
parametrization is thus advantageous.  Due to the re-usability of the
appearing structures and the improved treatment of $x_t$ (and
$\tan\beta$) discussed later, we propose to use
the full-model parametrization instead of the
EFT parametrization used in \HSSUSY, \susyhd and the original \feft\
implementation \cite{Athron:2016fuq,Athron:2017fvs}.

\section{New FlexibleEFTHiggs matching procedure}
\label{sec:expansion_Master_eq}

In the following we apply the conclusions of the
previous section to the matching of the SM to the MSSM and describe a
new improved matching procedure of the \feft\ approach, which also
allows to extend the approach beyond the 1-loop level without
introducing spurious logarithms of higher order.

\subsection{FlexibleEFTHiggs matching conditions}

The \feft\ approach is based on the central matching condition
\begin{align}
  (M_h^\MSSM)^2=(M_h^\SM)^2,
  \label{eq:Higgs_pole_mass_match}
\end{align}
where $M_h^\MSSM$ denotes the pole mass of the SM-like Higgs as
predicted within the MSSM and $M_h^\SM$ the Higgs pole mass in the SM
as computed in terms of SM parameters.  Within the SM, the Higgs pole
mass is related to SM tree-level parameters and loop corrections as
\begin{align}
  (M_h^\SM)^2 &= s,
\end{align}
where
\begin{align}
\label{eq:SM_higgs_pole}
  0 &= s - \hat{m}_h^2 + \re\left[\Sigma_h^\SM(s)-\frac{t^\SM_h}{\hat{v}}\right],\\
  \hat{m}_h^2 &= \hat{\lambda} \hat{v}^2.
\end{align}
Here $\hat{v}$ is the minimum of the loop-corrected SM effective
potential and $\Sigma_h^\SM$, $t_h^\SM$ are the \MSbar-renormalized
self-energy and tadpole, respectively.  Within the MSSM, the Higgs
pole mass is related to MSSM tree-level parameters and \DRbarPrime\
renormalized loop corrections as
\begin{align}
  (M_h^\MSSM)^2 = s, &&\text{where}&&
  \label{eq:BSM_pole_mass_definition}
  0 = &~\text{det}\left[s\,\delta_{ij}-(m_\phi^2)_{ij} + 
  \re\left[\Sigma_{\phi,ij}(s) - \frac{t_{\phi,i}}{v_i}\delta_{ij}\right]\right]\,.
\end{align}
Here, the tree-level mass matrix $(m_\phi^2)_{ij}$ is parametrized
such that the soft-breaking Higgs-doublet mass parameters
$m_{H_{u,d}}^2$ are eliminated by employing the EWSB
equations at the loop level.  This elimination introduces the tadpoles
$t_{\phi,i}$ on the r.h.s.\ of eq.~\eqref{eq:BSM_pole_mass_definition},
which are of the same loop order as the momentum-dependent self-energy
matrix $\Sigma_{\phi,ij}(s)$ of the BSM model.  For later convenience
we introduce the abbreviations
\begin{align}
\label{eq:DsSM}
\Delta s_h^{\SM}(p^2)&= - \re\left[\Sigma_h^{\SM}(p^2) - \frac{t_h^\SM}{v}\right]\,,\\
\label{eq:DsBSM}
\Delta s_h^{\MSSM}&= s - m_h^2\,,
\end{align}
where $m_h^2$ is the SM-like tree-level mass eigenvalue of the matrix
$(m_\phi^2)_{ij}$.  Combining the previous expressions gives rise to
the following relation for the SM quartic coupling $\hat{\lambda}$:
\begin{align}
  \hat{\lambda} &= \frac{1}{\hat{v}^2} \left[
            (M_h^\MSSM)^2 -\Delta s_h^\SM((M_h^\MSSM)^2)\right]\,.
  \label{eq:lambda_from_Mh}
\end{align}
This is the master formula for the determination of $\hat{\lambda}$ in the
\feft\ approach; in principle it could be evaluated exactly and at
arbitrarily high orders.  In particular, it could be evaluated either in
the limit $v/M_{S}\rightarrow 0$ or by keeping power-suppressed
terms of order $v/M_{S}$.
The first option would correspond to the
pure EFT approach pursued e.g.~in \HSSUSY\ and \susyhd.  The second
option corresponds to the \feft\ approach.  For an extensive
discussion of this method we refer to
refs.~\cite{Athron:2016fuq,Athron:2017fvs}. As exemplified in appendix~A of
ref.~\cite{Athron:2016fuq} and in \appref{App:lambda_atat} of the present paper the two options indeed coincide
analytically in the limit $M_S\rightarrow \infty$.

In the following we evaluate the master formula
\eqref{eq:lambda_from_Mh} according to the following prescription:
\begin{itemize}
\item We use the \feft\ hybrid method introduced in
  ref.~\cite{Athron:2016fuq}, i.e.\ we evaluate
  eq.~\eqref{eq:lambda_from_Mh} as it stands, including
  power-suppressed terms of $\ord(v^2/M_S^2)$ arising in the
  self-energies and tadpoles.

\item Eq.~\eqref{eq:lambda_from_Mh} is evaluated in the full-model
  (MSSM) parametrization, which is rather easy to generalize to other
  SUSY models and allows for a resummation of leading $x_t$ and
  $\tan\beta$ contributions in the Yukawa couplings $y_t$ and $y_b$ as
  well as in the quartic coupling $\hat{\lambda}$.

\item The threshold correction for $\hat{\lambda}$ is calculated at
  \NCLO with all 1-loop corrections, 2-loop corrections in the
  gaugeless limit at $\order{\twoloopfm}$\footnote{Additional 2-loop
    corrections of $\ord(v^2y_b^4y_\tau^2+v^2y_b^2y_\tau^4)$ to the
    pole mass of the Higgs boson can be found in
    ref.~\cite{Allanach:2004rh}, but we don't include these corrections in the present study.} and
  3-loop corrections of $\ord(\thrloopfm)$.\footnote{The threshold
    corrections included in \HSSUSY\ for $\hat{\lambda}$ are of the
    same order \cite{Harlander:2018yhj}, but expressed in the EFT
    parametrization.}  By including the SM $\beta$-functions up to the
  4-loop level, this matching allows for a resummation of \NCLL\ at
  the considered order in the couplings.  The final Higgs mass
  prediction will include the complete series of power-suppressed
  $(v^2/\MS^2)^n$ terms at 1-loop and 2-loop level at the given
  orders.  However, 3-loop suppressed terms are not included in our
  calculation, because they are neither publicly available in the
  literature \cite{R.:2019ply} nor implemented in the \Himalaya\
  library \cite{Harlander:2017kuc}.
\end{itemize}

In addition to the master formula \eqref{eq:lambda_from_Mh}, the
matching conditions for the other SM parameters
$\{\hat{g}_1, \hat{g}_2, \hat{g}_3, \hat{y}_t, \hat{y}_b,
\hat{y}_\tau, \hat{v}\}$ are given by
\begin{subequations}
\begin{align}
  (M_V^\MSSM)^2 &= (M_V^\SM)^2, & V&=W,Z, \label{eq:MV_matching} \\
  M_f^\MSSM &= M_f^\SM, & f&=t,b,\tau, \label{eq:Mf_matching} \\
  \Gamma_{\bar f fA^\mu}^\MSSM&=\Gamma_{\bar f fA^\mu}^\SM, \\
  \Gamma_{\bar q qg^\mu_a}^\MSSM&=\Gamma_{\bar qqg^\mu_a}^\SM, \label{eq:as_matching}
\end{align}%
\label{eq:matching_conditions}%
\end{subequations}%
where $M$ denotes the pole mass of the corresponding particle and
$\Gamma$ is a Green function.  The symbols $A^\mu$ and $g^\mu_a$
denote the QED and QCD gauge fields, respectively.  Quarks are denoted
as $q$ and SM fermions with a non-vanishing electric charge are
referred to as $f$.

\subsection{Perturbative expansion of the matching conditions}

In this section we perform the explicit perturbative expansion of the
master formula \eqref{eq:lambda_from_Mh} and the matching conditions
\eqref{eq:matching_conditions} in the full-model
parametrization.  As a result we will obtain all building blocks
necessary for the 3-loop Higgs pole mass prediction in the improved
\feft\ approach.

We start by performing the matching at tree-level.  At this order one
has
\begin{subequations}
\begin{align}
  (M_h^\MSSM)^2 &= m_h^2, & \Delta s_h^\SM(p^2) &= 0, \\
  (M_V^\MSSM)^2 &= m_V^2, & (M_V^\SM)^2 &= \hat{m}_V^2, & V&=W,Z \\
  M_f^\MSSM &= m_f, & M_f^\SM &= \hat{m}_f, & f&=t,b,\tau, \\
  \Gamma_{\bar f fA^\mu}^\MSSM &= -e\gamma^\mu Q_f, &
  \Gamma_{\bar f fA^\mu}^\SM &= -\hat{e}\gamma^\mu Q_f,\\
  \Gamma_{\bar{q}qg_a^\mu}^\MSSM &= - g_3 \gamma^\mu T_a, &
  \Gamma_{\bar{q}qg_a^\mu}^\SM &= -\hat{g}_3 \gamma^\mu T_a.
\end{align}%
\label{eq:tree_level_matching}%
\end{subequations}%
Inserting eqs.~\eqref{eq:tree_level_matching} into the master formula
\eqref{eq:lambda_from_Mh} and into the matching conditions
\eqref{eq:matching_conditions} one obtains the \MSbar\ SM parameters
expressed in terms of \DRbarPrime\ MSSM parameters at tree level:
\begin{subequations}
\begin{align}
  \hat{\lambda}^{\tL} &= m_h^2/v^2, \\
  \hat{g}_i^{\tL} &= g_i, & i&=1,2,3, \\
  \hat{y}_t^{\tL} &= y_t s_\beta, \\
  \hat{y}_f^{\tL} &= y_f c_\beta, & f&=b,\tau, \\
  \hat{v}^{\tL} &= v.
\end{align}%
\label{eq:SM_parameters_tree_level}%
\end{subequations}%
For later convenience we denote the tree-level SM \MSbar\ parameters
on the l.h.s.\ of eqs.~\eqref{eq:SM_parameters_tree_level} as
$\hat{P}^\tL$.
At the 1-loop level we obtain accordingly
\begin{subequations}
\begin{alignat}{4}
  \hat{\lambda}^{\oneL} &{}= \hat{\lambda}^{\tL} + \Delta\lambda^{\oneL}&{}={}& m_h^2/v^2 + \Delta\lambda^{\oneL} , \\
  \hat{g}_i^{\oneL} &{}= \hat{g}_i^{\tL} + \Delta g_i^{\oneL}&{}={}& g_i + \Delta g_i^{\oneL} , &&{} i&=1,2,3, \\
  \hat{y}_t^{\oneL} &{}= \hat{y}_t^{\tL} + \Delta y_t^{\oneL}&{}={}& y_t s_\beta + \Delta y_t^{\oneL} , \\
  \hat{y}_f^{\oneL} &{}= \hat{y}_f^{\tL} + \Delta y_f^{\oneL}&{}={}& y_f c_\beta + \Delta y_f^{\oneL} , &&{} f&=b,\tau, \\
  \hat{v}^{\oneL} &{}= \hat{v}^{\tL} + \Delta v^{\oneL}&{}={}& v + \Delta v^{\oneL} ,
\end{alignat}%
\label{eq:SM_parameters_one_loop}%
\end{subequations}%
where the 1-loop threshold corrections on the r.h.s.\ of
eqs.~\eqref{eq:SM_parameters_one_loop} are expressed in terms of MSSM
\DRbarPrime\ parameters.  In the pure EFT limit $v\to0$ the 1-loop
threshold corrections can be found for example in
refs.~\cite{Draper:2013oza,Bagnaschi:2014rsa,Martin:2007pg}.  The
explicit calculation of $\Delta y_t^{\oneL}$ and
$\Delta\lambda^{\oneL}$ beyond the pure EFT limit will be exemplified
below.  For brevity we denote in the following the 1-loop
SM \MSbar\ parameters on the l.h.s.\ of
eqs.~\eqref{eq:SM_parameters_one_loop} generically as
$\hat{P}^{\oneL}$.  Similarly, the $n$-loop SM \MSbar\ parameters are
denoted as $\hat{P}^{\nL}$.  Furthermore we denote the general threshold correction as $\Delta P =\hat P- \hat P^{\tL}$ and specify
the notation of a generic $n$-loop threshold correction as
\begin{align}
  \Delta P^{\nL} &\equiv \hat{P}^{\nL} - \hat{P}^{(n-1)\ell}, \\
  \Delta P^{\alpha^n} &\equiv \left.\Delta P\right|_{\alpha^n} ,
\label{Pnotation}
\end{align}
which are expressed in terms of MSSM \DRbarPrime\ parameters, where $\left.\Delta P\right|_{\alpha^n}$ denotes all contributions to $\Delta P$ of order $\ord(\alpha^n)$ .

For the prediction of the SM-like Higgs pole mass in the MSSM with the
improved \feft\ approach up to the order
$\order{\twoloopfm +\thrloopfm}$, it is sufficient to determine all SM parameters at the
1-loop level, except for $\hat{\lambda}$ and $\hat{y}_t$, which must
be determined at a higher order.  For this reason we describe in the
following in more detail the calculation of the threshold corrections
$\Delta y_t^{\nL}$ and $\Delta\lambda^{\nL}$.  In order to
express these threshold corrections consistently in the full-model
parametrization, an extra expansion of the loop corrections in terms
of MSSM \DRbarPrime\ parameters must be performed.  We will refer to
this procedure as ``double loop expansion''.

\paragraph{Expansion of the top quark pole mass matching condition.}

The 2-loop threshold correction for the top Yukawa coupling,
$\Delta y_t^\twoL$, can be obtained from the top quark pole mass
matching condition eq.~\eqref{eq:Mf_matching} with
\begin{subequations}
	\begin{align}
	M_t^\SM ={}& \hat{m}_t + \Delta \hat{m}_t^{\oneL}(p=\hat m_t) + \Delta \hat{m}_t^{\twoL}, \\
	M_t^\MSSM ={}& m_t + \Delta m_t^{\oneL}(p= m_t) + \Delta m_t^{\twoL}.
	\end{align}%
	\label{eq:MtSM_and_MSSM}%
\end{subequations}%
The 1-loop corrections on the r.h.s.\ of eqs.~\eqref{eq:MtSM_and_MSSM} are given
by
\begin{subequations}
	\begin{align}
	\Delta \hat{m}_t^{\oneL}(\hat m_t) ={}& -\re\Sigma_{t,S}^{\SM,\oneL}
	- \hat{m}_t \left[
	\re\Sigma_{t,L}^{\SM,\oneL}
	+ \re\Sigma_{t,R}^{\SM,\oneL}
	+ \Delta \hat{m}_t^{\QCD,\oneL}
	\right], \\
	\Delta m_t^{\oneL}(m_t) ={}& -\re\Sigma_{t,S}^{\MSSM,\oneL}
	- m_t \left[
	\re\Sigma_{t,L}^{\MSSM,\oneL}
	+ \re\Sigma_{t,R}^{\MSSM,\oneL} 
	+ \Delta m_t^{\QCD,\oneL}
	\right], \\
	\Delta \hat{m}_t^{\twoL} ={}& - \hat{m}_t \Delta \hat{m}_t^{\QCD,\twoL}, \\
	\Delta m_t^{\twoL} ={}& - m_t \Delta m_t^{\QCD,\twoL},
	\end{align}
\end{subequations}
where $\Sigma_{t,\{S,L,R\}}^{\{\SM,\MSSM\},\oneL}$ denote the
renormalized scalar, left- and right-handed components of the 1-loop
top self-energy evaluated at momentum $p=\hat m_t$ and $p= m_t$ in the
SM and MSSM, respectively, without the QCD contributions.  The SM
self-energies are renormalized in the \MSbar\ scheme and the MSSM
self-energies are renormalized in the \DRbarPrime\ scheme.  In the
degenerate SUSY mass limit the 1- and 2-loop QCD contributions are
given by \cite{Fleischer:1998dw}
\begin{subequations}
	\begin{align}
	\Delta m_t^{\QCD,\oneL} &= -\frac{g_3^2}{(4\pi)^2} \frac{4}{3} \left[5 - 3 \;\barlog(t) - x_t + \barlog(M_S^2)\right], \label{eq:Mtpole_1L_MSSM} \\
	\begin{split}
	\Delta m_t^{\QCD,\twoL} &=  \frac{g_3^4}{54 (4\pi)^4} \big[
	1745-640 x_t + 4\; \barlog(M_S^2)\left(677 -16x_t +93\; \barlog(M_S^2)\right)\\
	&\phantom{=  \frac{g_3^4}{54 (4\pi)^4} \big[}
	+288\; \barlog(t) \left(x_t - 4\; \barlog(M_S^2)\right)
	\big]
	\end{split}
	\label{eq:Mtpole_2L_MSSM} \\
	\Delta \hat{m}_t^{\QCD,\oneL} &= -\frac{\hat{g}_3^2}{(4\pi)^2} \frac{4}{3} \left[4 - 3 \;\barlog(\hat{t})\right], \label{eq:Mtpole_1L_SM} \\
	\Delta \hat{m}_t^{\QCD,\twoL} &=
	-\frac{\hat g_3^4}{18 (4\pi)^4}
	\left[396 \;\barlog^2(\hat{t}) - 1452 \;\barlog(\hat{t}) - 48 \zeta_3 + 2053 + 16 \pi^2 (1+\log 4)\right],
	\label{eq:Mtpole_2L_SM}
	\end{align}
\end{subequations}
where $\hat{t} = \hat{m}_t^2$, $t = m_t^2$ and
$\barlog(x) \equiv \log(x/Q^2)$.
The 2-loop MSSM QCD contribution $\Delta m_t^{\QCD,\twoL}$ for
non-degenerate SUSY mass parameters can be found in
refs.~\cite{Bednyakov:2007vm,Bednyakov:2002sf,Bednyakov:2005kt}.  Note
that the SM QCD contributions have already been evaluated at
$p=M_t^\SM$ up to order $\ord(\hat{m}_t \hat{g}_3^4)$, while the MSSM
QCD contributions have been evaluated at $p=M_t^\MSSM$ up to order
$\ord(m_t g_3^4)$.  Thus, the 2-loop contributions
$\Delta m_t^{\twoL}$ and $\Delta \hat{m}_t^{\twoL}$ contain terms
stemming from momentum iteration out of $\Delta {m}_t^{\oneL}$ and
$\Delta \hat{m}_t^{\oneL}$, respectively.

To obtain the 1- and 2-loop threshold corrections for the top Yukawa
coupling in the full-model parametrization, the top quark pole masses
\eqref{eq:MtSM_and_MSSM} are inserted into the matching condition
\eqref{eq:Mf_matching}, where both sides must be evaluated at
$p = M_t^\MSSM$. The subsequent expansion of the matching condition
in terms of MSSM \DRbarPrime\ parameters (double loop expansion) 
is equivalent to taking the expressions in eqs.~\eqref{eq:MtSM_and_MSSM}
and expanding up to $\ord(y_tg_3^4)$, which yields
\begin{subequations}
	\begin{align}
	\Delta y_t^\oneL &= \frac{\sqrt{2}}{v} \left[
	\Delta m_t^{\oneL}(m_t) - \Delta \hat{m}_t^{\oneL}(m_t)
	 - \frac{m_t}{v} \Delta v^{\oneL}\right],
	 	\label{eq:yt_from_Mtpole1l} \\
	\Delta y_t^\twoL &= \frac{\sqrt{2}}{v} \left[
	\Delta m_t^{\twoL} - \Delta \hat{m}_t^{\twoL}
	+ \sum_{P}\left(\frac{\partial}{\partial  P} \hat{m}_t \Delta \hat{m}_t^{\QCD,\oneL}\right) \Delta P \right],
	\label{eq:yt_from_Mtpole2l}
	\end{align}
        \label{eq:yt_from_Mtpole}
\end{subequations}
with $P\in\{\hat y_t, \hat g_3\}$.

\paragraph{Expansion of the master formula.}

The $n$-loop threshold correction $\Delta\lambda^{\nL}$ is obtained
from the master formula eq.~\eqref{eq:lambda_from_Mh}.  To derive the
necessary building blocks to express $\Delta\lambda^{\nL}$ in the
full-model parametrization, the following three expansions must be
performed:
\begin{itemize}
\item The prefactor $1/\hat{v}^2$ on the r.h.s.\ of
  eq.~\eqref{eq:lambda_from_Mh} must be expressed in terms of MSSM
  parameters, which yields%
  \footnote{Threshold contributions to $v$ at 2-loop  would
  induce 2-loop contributions in eq.~\eqref{eq:lambda_from_Mh} beyond the gaugeless limit.}
  \begin{align}
    \frac{1}{\hat{v}^2}=\frac{1}{v^2}\left[1-2\frac{\Delta v^\oneL}{v}\right]\,.
  \end{align}

\item The squared Higgs pole mass in the MSSM, $(M_h^\MSSM)^2$, on the
  r.h.s.\ of eq.~\eqref{eq:lambda_from_Mh} is naturally expanded in
  terms of MSSM parameters as
  \begin{align}
    (M_h^\MSSM)^2 &= m_h^2 + \Delta s_h^{\MSSM,\oneL} + \Delta s_h^{\MSSM,\twoL} +
                    \Delta s_h^{\MSSM,\thrL}, \\
    \Delta s_h^{\MSSM,\nL} &\equiv s^{\MSSM,\nL} - s^{\MSSM,(n-1)\ell},
  \end{align}
  where $s_h^{\MSSM,\nL}$ is obtained from eq.~\eqref{eq:DsBSM} with
  all corrections included up to $n$-loop level.\footnote{Note that
    due to the non-linearity of the determinant
    \eqref{eq:BSM_pole_mass_definition}, the contribution
    $\Delta s_h^{\MSSM,\nL}$ contains products of self-energies of
    lower loop order, which are, however, suppressed by factors of
    $v^2/M_S^2$ as discussed in refs.~\cite{Espinosa:2000df,
      Bahl:2017aev}.}  The
  1-loop self-energy and tadpoles can be obtained from \sarah\ and we
  expand the self-energy as
  \begin{align}\label{eq:mom_iter_MSSM}
    \Sigma^\oneL_{\phi,ij}((M_h^2)^{\MSSM}) &= \Sigma^\oneL_{\phi,ij}(m_h^2)+
    \left(\frac{\partial}{\partial p^2}\Sigma^{\oneL}_{\phi,ij}(0)\right) 
    \Delta s_h^{\MSSM,\oneL}.
  \end{align}
  Note, that the last term on the r.h.s.\ of
  eq.~\eqref{eq:mom_iter_MSSM} contributes pure 2-loop Yukawa terms
  and thus must be taken into account for a consistent Higgs pole mass
  prediction at the 2-loop order
  $\order{v^2((y_t^2 + y_b^2)^3 + (y_t^2 + y_\tau^2)^3)}$, together with the
  corresponding explicit 2-loop self-energy and tadpole contributions.
  The explicit 2-loop corrections to $\Delta s_h^{\MSSM,\twoL}$ read
  \begin{align}
    \Sigma^{\twoL}_{\phi,ij}(0) - \frac{t_{\phi,i}^{\twoL}}{v_i}\delta_{ij},
  \end{align}
  which we take in the gaugeless limit at the order
  $\ord(v^2(g_3^2(y_t^4+y_b^4)+(y_t^2+y_b^2)^3 +y_\tau^6))$
  \cite{Degrassi:2001yf, Brignole:2001jy, Dedes:2002dy,
    Brignole:2002bz, Dedes:2003km}.  At 3-loop level we include the
  known MSSM contributions of $\ord(v^2 y_t^4g_3^4)$
  \cite{Harlander:2008ju,Kant:2010tf} to the Higgs pole mass,
  \begin{align}
    \Sigma^{\thrL}_{\phi,ij}(0) - \frac{t_{\phi,i}^{\thrL}}{v_i}\delta_{ij},
  \end{align}
  which we take from the \Himalaya\ library \cite{Harlander:2017kuc}.

\item The pure SM contributions $\Delta s_h^{\SM}(p^2)$ from
  eq.~\eqref{eq:DsSM} must also be expressed in terms of MSSM
  parameters, which we achieve by a double loop expansion:
  \begin{align}
    \Delta s_h^{\SM}((M_h^{\MSSM})^2) &= \Delta s_h^{\SM,\oneL}(m_h^2)+\Delta s_h^{\SM,\twoL}	+\Delta s_h^{\SM,\thrL},
  \end{align}
  where
  \begin{subequations}
  \begin{align}
    \Delta s_h^{\SM,\oneL}(p^2) &= -\re\left[\Sigma_h^{\SM,\oneL}(p^2) - \frac{t_h^{\SM,\oneL}}{v}\right],\\
    \label{eq:iter_2l}
    \Delta s_h^{\SM,\twoL} &= -\re\left[\Sigma_h^{\SM,\twoL}(0) - \frac{t_h^{\SM,\twoL}}{v}\right]
                             +\sum_P\left(\frac{\partial}{\partial P }\Delta s^{\SM,\oneL}_h(0)\right) \Delta P^{\oneL},\\
    \label{eq:iter_3l}
    \Delta s_h^{\SM,\thrL} &= -\re\left[\Sigma_h^{\SM,\thrL}(0) - \frac{t_h^{\SM,\thrL}}{v}\right]+
                             \sum_{n\cdot q+m=3}
                             \left(\frac{\partial^n}{\partial P^n}\Delta s^{\SM,m\ell}_h(0)\right) (\Delta P^{q\ell})^n.
  \end{align}
  \end{subequations}
  The sum on the r.h.s.\ of eq.~\eqref{eq:iter_2l} runs over
  $P\in \{p^2, \hat y_t, \hat y_b, \hat y_\tau, \hat g_3, \hat v\}$,
  where the $\Delta p^2$ contribution accounts for the fact that the
  momentum inserted in the 1-loop SM Higgs mass correction of
  eq.~\eqref{eq:lambda_from_Mh} is evaluated at $p^2=(M_h^{\MSSM})^2$.
  Hence the relation
  \begin{align}
  \label{eq:Delta_p}
  	\Delta p^2 = \Delta s_h^{\MSSM,\oneL}
  \end{align}
  includes corrections at 1-loop in the gaugeless limit.
  The sum on the r.h.s.\ of eq.~\eqref{eq:iter_3l} runs over
  $P\in\{\hat y_t, \hat g_3\}$ with
  $(\Delta P^{q\ell})^n\in\{\Delta y_t^{\oneL}, 1/2(\Delta
  y_t^{\oneL})^2,\Delta y_t^{\twoL},\Delta
  g_3^{\oneL}\}$.\footnote{Note that the threshold correction
    $(\Delta P^{q\ell})^n$ includes QCD corrections only.}  Mixed
  derivatives and products of threshold corrections do not appear in
  eq.~\eqref{eq:iter_3l}, since they would contribute beyond the
  considered $\ord(y_t^4g_3^4)$ in $\Delta \lambda$.  The explicit 2-loop corrections
  in the effective potential approach at $\ord(\hat v^2\hat y_t^4 \hat g_3^2)$ are taken from
  refs.~\cite{Athron:2017fvs,Degrassi:2012ry}. 
   The 2-loop corrections at
  $\ord(\hat v^2((\hat y_t^2+ \hat y_b^2)^3 +\hat y_\tau^6))$ are
  presented in \secref{sec:expl_results}.
  At 3-loop level, we include contributions of
  $\ord(\hat v^2 \hat y_t^4 \hat g_3^4)$ from ref.~\cite{Martin:2014cxa}.
\end{itemize}
With these ingredients, the expansion of the master formula
\eqref{eq:lambda_from_Mh} is given by
\begin{subequations}\label{eq:lambdamatchinggeneral}
\begin{align}
\hat{\lambda} &= \lambda + \Delta \lambda^\oneL +\Delta \lambda^\twoL +\Delta \lambda^\thrL\,,\label{eq:hat-lambda} \\
 \Delta\lambda^\oneL &= \frac{1}{v^2}\left[-2 m_h^2 \frac{\Delta v^{\oneL}}{v}
 +	\Delta s^{\MSSM,\oneL}_h-\Delta s^{\SM,\oneL}_h( m_h^2 )\right]\,,\\
 \label{eq:double-loop-exp-2L}
  \Delta\lambda^\twoL &= \frac{1}{v^2}\left[2\frac{\Delta v^{\oneL}}{v} 
  \left(\Delta s^{\SM,\oneL}_h(0)-\Delta s^{\MSSM,\oneL}_h\right)
 +	\Delta s^{\MSSM,\twoL}_h-\Delta s^{\SM,\twoL}_h\right]\,,\\
  \label{eq:double-loop-exp-3L}
  \Delta\lambda^\thrL &= \frac{1}{v^2}\left[
 	\Delta s^{\MSSM,\thrL}_h-\Delta s^{\SM,\thrL}_h\right]\,.
\end{align}
\end{subequations}
Note that for consistency the parameter shifts $\Delta P$, which
contribute to $\Delta\lambda^\twoL$ and $\Delta\lambda^\thrL$, must be
evaluated in the gaugeless limit.  Note also that the double loop
expansion for the 3-loop threshold correction $\Delta\lambda^\thrL$ is
relatively simple, because only a small sub-set of implicit corrections
contribute at order $\ord(y_t^4g_3^4)$.

\subsection{Explicit results and comparison to the literature}
\label{sec:expl_results}

In this subsection we
  present several explicit results and analytic expressions for  the
  threshold corrections at the 1-loop, 2-loop and 3-loop level. The
  first purpose is to demonstrate the internal consistency of
  the new way of setting up the
  threshold corrections by checking that all explicit large logarithms
cancel in   eqs.~\eqref{eq:yt_from_Mtpole} and
\eqref{eq:lambdamatchinggeneral}. A second purpose is to verify the
correctness of the results by comparing to results presented in the
literature, appropriately reparametrized. Finally, we also present
several new analytic results at the 2-loop level.

All results in this subsection are provided in the EFT limit $v^2\ll
M_S^2$ and for the degenerate
mass case and non-trivial stop mixing.  

\subparagraph{\underline{$\Delta y_t$ at $\ord(y_tg_3^4)$}}
\label{yt_2-loop}
We start with the derivation of the 1- and 2-loop threshold
corrections for the top Yukawa coupling, $\Delta y_t^\oneL$ and
$\Delta y_t^\twoL$, at $\ord(y_t g_3^{2n})$ from
eqs.~\eqref{eq:yt_from_Mtpole}.  The exact 2-loop pole mass
contribution in the MSSM (self-energy + momentum iteration) is
obtained from refs.~\cite{Bednyakov:2002sf,Bednyakov:2005kt}.  The
2-loop pole mass contribution in the SM is taken from
ref.~\cite{Fleischer:1998dw}. Using the notation as introduced in
eq.~\eqref{Pnotation} and  for the case of degenerate SUSY mass
parameters the corrections on the r.h.s.\ of
eqs.~\eqref{eq:yt_from_Mtpole1l} and \eqref{eq:yt_from_Mtpole2l}
evaluate to
\begin{subequations}
\begin{align}\label{eq:Dyuk_MSSM_2L}
  \left.\hat{y}_t \right|_{y_t \as^{\leq 2}}  &= y_t s_\beta + \Delta y_t^{\as} + \Delta y_t^{\as^2},\\
  \Delta y_t^{\as} &=  y_t s_\beta\frac{g_3^2 }{(4\pi)^2}\frac{4}{3}
                     \left[1-x_t + \barlog(M_S^2)\right],
                     \label{eq:Dyuk_1l_MSSM} \\
  \Delta y_t^{\as^2} &= y_t s_\beta\frac{g_3^4 }{(4\pi)^4}\frac{1}{54}
                     \left[2099-832x_t + (1748 - 64 x_t) \;\barlog(M_S^2) + 372 \;\barlog^2(M_S^2)\right].
                     \label{eq:2l_yt_td}
\end{align}%
\label{eqs:2l_yt}%
\end{subequations}
By squaring eq.~\eqref{eq:Dyuk_MSSM_2L} we obtain perfect analytic
agreement with the 2-loop threshold correction presented in
ref.~\cite{Harlander:2018yhj}.  Note that the presented threshold
corrections $\Delta y_t^{\as}$ and $\Delta y_t^{\as^2}$ are linear in
$x_t$, which will be relevant for the determination of $\hat\lambda$ in
the full-model parametrization below.

For future use we record here the corresponding result in EFT
parametrization in an analogous notation as in eq.~\eqref{eqs:2l_yt}:
\begin{subequations}
\begin{align}
  \label{eq:yt_bu}
  \left. y_t \right|_{\hat y_t \ashat^{\leq 2}}  &= \frac{\hat{y}_t}{s_\beta} + \Delta \hat{y}_t^{\ashat} + \Delta \hat{y}_t^{\ashat^2} ,\\
  \Delta \hat{y}_t^{\ashat} &=  \frac{\hat{y}_t}{s_\beta}\frac{\hat{g}_3^2 }{(4\pi)^2}\frac{4}{3}
  \left[-1+x_t - \barlog(M_S^2)\right] ,\\
  \Delta \hat{y}_t^{\ashat^2} &=  \frac{\hat{y}_t}{s_\beta}\frac{\hat{g}_3^4}{(4\pi)^4}\frac{1}{54}
  \left[-2075+712x_t+96x_t^2
  - (1340 + 416 x_t) \;\barlog(M_S^2) + 12 \;\barlog^2(M_S^2)\right]\,.
  \label{eq:2l_yt_bu}
\end{align}
\end{subequations}
A noteworthy difference is the term  $\propto g_3^4x_t^2$, which
appears in the EFT parametrized expression
in eq.~\eqref{eq:2l_yt_bu}, but not in the full-model parametrization
in eq.~\eqref{eq:2l_yt_td}. This term
originates from an implicit (conversion) correction.  Its origin can
be traced back to eq.~\eqref{eq:yt_from_Mtpole2l}, which 
contains a derivative term of the form
\begin{align}
  \label{eq:implicit_two_loop}
  \left(\frac{\rd}{\rd m_t}\Sigma_t^{\MSSM,\as}(m_t, m_t) \right)
  \frac{v}{\sqrt{2}}\Delta y_t^{\as} ,
\end{align}
when evaluated in the EFT parametrization.  The 1-loop correction
$\Delta y_t^{\as}$ and the 1-loop MSSM top quark self-energy
$\Sigma_t^{\MSSM,\as}$ each contain terms $\propto x_t$, which results
in a contribution $\propto x_t^2$ in eq.~\eqref{eq:implicit_two_loop}.
As will be discussed below, this 2-loop $x_t^2$ term is also present
in the MSSM-parametrized threshold correction, where it is
\emph{implicitly} taken into account by the 1-loop threshold of
eq.~\eqref{eq:Dyuk_1l_MSSM}.

\subparagraph{\underline{$\Delta\lambda$ at $\ord(g_3^2 y_t^4)$}}

We continue and show that eq.~\eqref{eq:double-loop-exp-2L} in that
form leads to the known expression of $\Delta\lambda^\twoL$ at
$\ord(g_3^2 y_t^4)$, presented in ref.~\cite{Bagnaschi:2014rsa}.  The
calculation of this correction from eq.~\eqref{eq:double-loop-exp-2L}
at 2-loop level requires the explicit 2-loop corrections of
$\ord(v^2g_3^2 y_t^4)$ for both the SM \cite{Degrassi:2012ry} and MSSM
\cite{Degrassi:2001yf} Higgs pole mass.  Since the threshold
correction to the VEV $\Delta v$ does not contribute at this order,
eq.~\eqref{eq:double-loop-exp-2L} simplifies for $Q=M_S$ to
\begin{subequations}
\begin{align}\label{eq:Delta_as*at^2_pre}
  \left.\Delta\lambda^\twoL\right|_{g_3^2 y_t^4} &= \frac{1}{v^2} \left[\Delta s^{\MSSM,\twoL}_h-\Delta s^{\SM,\twoL}_h\right]\\
  &= \frac{1}{v^2} \left[ \Delta s_h^{\MSSM,y_t^4g_3^2} +
  \re\left[\Sigma_h^{\SM,y_t^4g_3^2}(0) - \frac{t_h^{\SM,y_t^4g_3^2}}{v}\right]-
  \left(\frac{\partial}{\partial \hat y_t }\Delta s^{\SM,\hat y_t^4}(0)\right) \Delta y_t^{\as} \right]\\
  &= \frac{g_3^2y_t^4s_{\beta}^4}{(4\pi)^4}
  \frac{4}{3}x_t\left[ -24 + 12 x_t +4x_t^2 -x_t^3\right],
  \label{eq:Delta_as*at^2}
\end{align}
\end{subequations}
where we have used $\Delta y_t^\oneL$ from
eq.~\eqref{eq:Dyuk_1l_MSSM}.  In contrast to the result presented in
ref.~\cite{Bagnaschi:2014rsa}, eq.~\eqref{eq:Delta_as*at^2} is
expressed in the full-model parametrization.  To compare our result
with eq.~(36) of ref.~\cite{Bagnaschi:2014rsa}, which is presented in
the EFT parametrization, one has to express the SM top Yukawa coupling
$\hat{y}_t$ in $\Delta \lambda^{\oneL}$ from eq.~\eqref{eq:DLambda_sm}
in terms of the MSSM top Yukawa coupling $y_t$.  After truncation at
the 2-loop order $\ord(g_3^2 y_t^4)$, the combined expression
is identical to eq.~\eqref{eq:Delta_as*at^2}.

\subparagraph{\underline{$\Delta\lambda$ at $\ord(y_t^6)$}}

At 2-loop $\ord(y_t^6)$ we can perform an even stricter consistency
test of eq.~\eqref{eq:double-loop-exp-2L}, because the implicit
corrections at this order have all non-trivial contributions from
eqs.~\eqref{eq:mom_iter_MSSM}, \eqref{eq:iter_2l} and
\eqref{eq:double-loop-exp-2L}.  Since the explicit 2-loop
contributions in the MSSM \cite{Espinosa:2000df} and SM
\cite{Vega:2015fna} are known, the only missing contributions in
eq.~\eqref{eq:double-loop-exp-2L} are the implicit corrections.  The
1-loop threshold corrections of $\ord(\at)$ to $P \in \{p^2,y_t,v\}$
are obtained from ref.~\cite{Draper:2013oza}.  In
\appref{App:lambda_atat} we present all contributions and the explicit
derivation in an expansion of $ct_\beta \equiv 1/ \tan \beta$ up to
the second order for $Q=M_S$.  Besides this expansion we have also
proven the equivalence between our threshold correction and the one
presented in ref.~\cite{Vega:2015fna} for the general $\tan\beta$ dependence.
Since the expression from ref.~\cite{Vega:2015fna}
does (by construction) not contain large logarithmic contributions, we
have shown that all large logarithmic contributions vanish in
eq.~\eqref{eq:double-loop-exp-2L} at $\ord(y_t^6)$.  Our explicit
result for the 2-loop threshold correction in the full-model
parametrization reads:
\begin{align}
  \label{eq:Delta_at^3_alt}
  \begin{split}
    \left.\Delta\lambda^\twoL\right|_{y_t^6}
    = \frac{y_t^6}{4(4\pi)^4} \big[
    & - 4 x_t^6 + 53.751 x_t^4 - 192.254 x_t^2 - 109.503 \\
    & + ct_\beta \left(-0.497 x_t^5+9.99x_t^3 -9.015 x_t\right)\\
    & + ct_{\beta}^2 \left(11.751 x_t^6 -145.508  x_t^4+592.02 x_t^2-317.702\right)\\
    & + \ord(ct_{\beta}^3)\big],
  \end{split}
\end{align}
where the analytical result is shown in eq.~\eqref{eq:Delta_at^3}.

\subparagraph{\underline{$\Delta\lambda$ at $\ord(y_t^4 y_\tau^2)$}}
At $\ord(y_t^4 y_\tau^2)$ an interesting comparison of threshold
contributions in both parametrizations can be made.  In a fixed-order
calculation there are no 2-loop diagrams which explicitly contribute
to the squared Higgs pole mass $M_h^2$ at $\ord(v^2 y_t^4 y_\tau^2)$.
However, in a fixed-order calculation contributions of this order are induced by
momentum iteration.  In an EFT calculation based on EFT
parametrization, the contributions of $\ord(\hat y_t^4 \hat y_\tau^2)$
to $\Delta\lambda^\twoL$ vanish, see the remarks in section 2 of
ref.~\cite{Bagnaschi:2017xid}.  However, in our matching procedure,
which is based on the full-model parametrization, terms of this order are
explicitly generated.  This can be seen as follows.  Evaluating the
implicit correction on the r.h.s.\ of eq.~\eqref{eq:iter_2l} of
$\ord(y_t^4 y_\tau^2)$ yields
\begin{subequations}
\begin{align}
\Delta s^{\SM,y_t^4 y_{\tau}^2}_h&=\left.\sum_P\left(\frac{\partial}{\partial P }\Delta s^{\SM,\oneL}_h(0)\right) \Delta P^{\oneL}\right|_{v^2y_t^4 y_\tau^2} \\
&=
\left(
\frac{\partial}{\partial \hat y_t}  \Delta s_h^{\SM, \hat y_t^4}\right)
\Delta y_t^{\atau}
+\left(\frac{\partial}{\partial \hat v}\Delta s_h^{\SM,\hat y_t^4} \right)
\Delta v^{\atau}
 +\left(\frac{\partial}{\partial p^2 }\Delta s^{\SM,\hat y_{\tau}^4}\right) \Delta (p^2)^{\at}.
 \label{eq:implicit_deriv_atau_at}
\end{align}
\end{subequations}
Since the tau lepton and stau slepton do not explicitly
contribute to the top quark self-energy at 1-loop level, the threshold
corrections $\Delta y_t^{\atau}$ and $\Delta v^{\atau}$ are related as
\begin{align}
\Delta y_t^{\atau} = - \sqrt{2}\frac{m_t}{v^2} \Delta v^{\atau} = -y_t s_\beta\frac{y_\tau^2}{(4\pi)^2} \frac{c_\beta^2 x_\tau^2}{12}, 
\label{eq:yt_thresh_atau}
\end{align}
where the threshold correction $\Delta v^{\atau}$ can be obtained from
the wave function renormalization of the Higgs, analogously to
$\Delta v^{\at}$ out of ref.~\cite{Draper:2013oza}.
Eventually, in the degenerate mass case and $Q=M_S$ the threshold
correction evaluates to
\begin{subequations}
	\begin{align}\label{eq:Delta_as*at*atau_pre}
	\left.\Delta\lambda^\twoL\right|_{y_t^4 y_\tau^2} &= \frac{1}{v^2} \left[
	2\frac{\Delta v^{\atau}}{v} 
	\left(\Delta s^{\SM,y_t^4}_h(0)-\Delta s^{\MSSM,y_t^4}_h\right)+
	\Delta s^{\MSSM,y_t^4 y_{\tau}^2}_h-\Delta s^{\SM,y_t^4 y_{\tau}^2}_h\right]\\
	&= 2\frac{y_t^4 y_\tau^2 }{(4\pi)^4}s_{\beta}^4 c_{\beta}^2
	x_t^2x_\tau^2\left[ -1 + \frac{x_t^2 }{12} \right],
	\label{eq:Delta_as*at*atau}
	\end{align}
\end{subequations}
where $\Delta s^{\MSSM,y_t^4 y_\tau^2}$ has been computed similarly to
eq.~\eqref{eq:split_at^2_contr}.  We note that large logarithmic
contributions within the terms of eq.~\eqref{eq:Delta_as*at*atau_pre}
cancel against each other.  In order to validate our result from
eq.~\eqref{eq:Delta_as*at*atau}, we have computed the analogous
expressions in EFT parametrization by performing a reparametrization
of the 1-loop correction in eq.~\eqref{eq:DLambda_sm}.  We find that
the EFT-parametrized threshold correction vanishes, which is in line
with the remarks in ref.~\cite{Bagnaschi:2017xid}.  The discussion of
the correction of $\ord(y_t^2 y_\tau^4)$ is analogous to
$\ord(y_t^4 y_\tau^2)$.
 
\subparagraph{\underline{Higgs mass loop corrections at $\ord(\hat v^2 ((\hat y_t^2+ \hat y_b^2)^3 +\hat y_\tau^6))$ in the SM}}

\begin{figure}[tbh!]
\centering
\begin{tikzpicture}[line width=1.5 pt, scale=1.3]
\begin{scope}[shift={(.175,0)} ,rotate=5]
\draw[fermion] (0,1) arc (270:90:1);
\draw[fermion] (0,3) arc (90:-90:1);
\end{scope}
\draw[scalarnoarrow]  (0,1)--(0,3);
\node at (-1.35,2){$t$};
\node at (1.3,2){$b$};
\node at (0.35,2){$G^\pm$};
\end{tikzpicture}
\caption{Vacuum bubble diagram in the SM, containing a top quark, a
  bottom quark and a charged Goldstone boson, which gives rise to
  2-loop radiative corrections to the Higgs mass at
  $\ord(\hat v^2  \hat y_t^4 \hat y_b^2 \log^2(\hat v^2/Q^2))$.}
\label{fig:Veff_atab}
\end{figure}
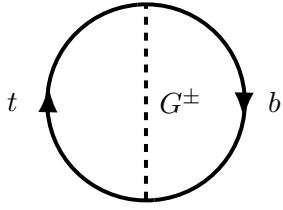

In the MSSM the 2-loop corrections of $\ord(v^2 (y_t^2+ y_b^2)^3 )$ to
the CP-even Higgs pole mass are known by ref.~\cite{Dedes:2003km} and
are included in several public codes to calculate the CP-even
Higgs pole masses in a fixed-order calculation.  In the SM, however, the
corresponding 2-loop corrections are not available in a simple and
explicit form in the literature to our knowledge.\footnote{The 2-loop
  corrections of that order can in principle be calculated from the
  generic 2-loop effective potential for general renormalizable
  theories \cite{Martin:2001vx}.}  In order to include the 2-loop
threshold corrections of $\ord((y_t^2 + y_b^2)^3)$ into the quartic
Higgs coupling $\hat\lambda$, our approach requires the corresponding
fixed-order corrections to be available for both the MSSM and the SM
separately.  If the contributions in the SM would be omitted, wrong
logarithmic enhanced terms of $\ord(\hat v^2\hat y_t^4 \hat y_b^2 \log^2(M_S^2/ \hat m_t^2))$
would propagate into the expression for the Higgs pole mass.  Thus, we
calculate here the contributions to the 2-loop effective potential of
the SM in the gaugeless limit for a non-vanishing bottom Yukawa
coupling.  The relevant diagram contributing to the mixed
contributions of $\ord( \hat v^4\hat y_t^4 \hat y_b^2)$ is shown in
\figref{fig:Veff_atab}.

Following the approach in ref.~\cite{Dedes:2002dy}, we compute the
2-loop bubble diagrams $V_\twoL^\SM$ and expand the 1-loop effective
potential around the \MSbar-renormalized masses of the top and bottom quark,
\begin{align}
\label{eq:Vatab}
\hat V_\twoL^\SM\big|_{(\hat y_t^2+ \hat y_b^2)^3 \phi^4} =
V_\twoL^\SM\big|_{( \hat y_t^2+ \hat y_b^2)^3 \phi^4} +
\frac{\partial (V^\SM_{\oneL})^\epsilon}{\partial \hat m_b^2} \delta \hat m_b^2
+  \frac{\partial (V^\SM_{\oneL})^\epsilon}{\partial \hat m_t^2} \delta \hat m_t^2,
\end{align}
where $(V^\SM_{\oneL})^\epsilon$ represents the part of the 1-loop
effective potential which is proportional to $(4-D)/2=\epsilon$ and
$\phi$ is a background field.  We have checked that
$\hat V^\SM_\twoL|_{(\hat y_t^2+ \hat y_b^2)^3 \phi^4}$ is reproduced by using
$V^\SM_\twoL|_{(\hat y_t^2+ \hat y_b^2)^3 \phi^4}$ with the subtracted integrals
$\hat I$ and $\hat J$ instead of $I$ and $J$, which have been introduced in  ref.~\cite{Ford:1992pn}.  After expanding the 2-loop integrals
around the renormalized Goldstone mass parameter and taking only
Yukawa coupling enhanced contributions into account, the finite result
expressed in SM \MSbar\ parameters reads
\begin{align}
\label{eq:at_ab_eff_pot}
\begin{split}
\hat V_{\twoL}^\SM|_{(\hat y_t^2+ \hat y_b^2)^3\phi^4}=
\frac{1}{(4\pi)^4} \frac{3}{2}  \bigg[&\phi^2 \hat y_b^4 (\hat I_{0TB}+2 \hat I_{BB0})+\phi^2
\hat y_t^4 (\hat I_{0TB}+2 \hat I_{TT0})\\
&+2 \hat y_t^2 \left(-\phi^2
\hat I_{0TB} \hat y_b^2+\hat J_{TB}+\hat J_{TT}\right)+2 \hat y_b^2
(\hat J_{BB}+\hat J_{TB})
\bigg].
\end{split}
\end{align}
The squared top and bottom quark mass parameters
$T= \hat y_t^2\phi^2/2$ and $B = \hat y_b^2\phi^2/2$ are expressed in
terms of the background field $\phi$.  In the MSSM, the corresponding
SM-like contributions are included in eqs.~(3.39) and (3.40) of
ref.~\cite{Martin:2002iu} and we have checked that the effective
potential in eq.~\eqref{eq:at_ab_eff_pot} reproduces these results
when omitting the corrections from BSM Higgs bosons.  The loop
corrections to the Higgs pole mass are derived from the effective
potential by differentiating w.r.t.~to the background field $\phi$ as
\begin{subequations}
\begin{align}
  \Delta m^{2,\SM}_{h,\EP}&=
  \re \left[\frac{t_h^{\SM,\twoL}}{v}- \Sigma_h^{\SM,\twoL}(0)\right]\\
  &=\left. \left(-\frac{1}{\phi}\frac{\partial}{\partial\phi}+\frac{\partial^2}{\partial\phi^2}\right)\hat V_{\twoL}^\SM\right|_{\phi=\hat v}.
  \label{eq:derivative_V}
\end{align}
\end{subequations}
Because the implicit corrections of eq.~\eqref{eq:iter_2l} are analogous to
eq.~\eqref{eq:implicit_deriv_atau_at}, 
the shift $\Delta s_h^{\SM,\twoL}$ 
at $\ord(\hat v^2(\hat y_b^2+\hat y_t^2)^2)$ is
completed by derivatives of eq.~\eqref{eq:at_ab_eff_pot}.
Neglecting terms of order $\mathcal O(\hat{y}_b^4 \hat b^2/\hat t)$, the loop
corrections read
\begin{align}
\label{eq:_DMh_ab_at}
\begin{split}
\left.\Delta m^{2,\SM}_{h,\EP}\right|_{(\hat y_t^2+ \hat y_b^2)^3 \hat{v}^2}
=\frac{1}{(4\pi)^4}\bigg[
&-2\hat y_t^4 \hat t \left(6+\pi^2- 21\;\barlog(\hat t) +  9\;\barlog^2(\hat t) \right)\\
&+6 \hat y_t^2\hat y_b^2\hat t\left(\pi^2 +\;\barlog(\hat t) + 3\;\barlog^2(\hat t) \right)\\
&+3 \hat y_b^4 \hat t \bigg(-15 -2\pi^2+2\;\barlog(\hat t) + 6\;\barlog^2(\hat t)
\\
&\phantom{+\hat y_b^4\hat b^2\frac13\bigg(}
+4\log x_{bt}(2+3\;\barlog(\hat t))
\bigg)\\
&+\frac{\hat y_b^4 \hat b}{3}\bigg(
49+6\pi^2+12\log x_{bt}(5-9\;\barlog(\hat t))
\\
&\phantom{+y_b^4m_b\bigg(}-36\log^2 x_{bt}
+18(7-3\;\barlog(\hat t))\;\barlog(\hat t)
\bigg) \bigg],
\end{split}
\end{align}
with $\hat{t} \equiv \hat{m}_t^2$, $\hat{b} \equiv \hat{m}_b^2$,
$x_{bt} \equiv \hat b/\hat t$ and 
$\barlog(\hat{t}) \equiv \log(\hat{t}/Q^2)$.  We note that the
first line in eq.~\eqref{eq:_DMh_ab_at} corresponds to the loop
corrections of $\ord(\hat y_t^6 \hat v^2)$, which can be found in
ref.~\cite{Vega:2015fna}, and which we reproduce here.  In the same
way as prescribed in eq.~(2.49) of ref.~\cite{Martin:2014cxa}, we
checked the renormalization scale invariance of the Higgs pole mass at
$\ord(\hat v^2(\hat y_t^2+ \hat y_b^2)^3)$ with the contributions of
eq.~\eqref{eq:_DMh_ab_at}, which is a non-trivial confirmation of our
result.

At $\ord(\hat y_\tau^6 \hat v^2)$ we repeated the calculation in eqs.~\eqref{eq:Vatab} and \eqref{eq:derivative_V} 
for the tau and tau neutrino contributions.
For massless neutrinos the result is 
analogous to the first line of eq.~\eqref{eq:_DMh_ab_at},
\begin{align}
	\left.\Delta m^{2,\SM}_{h,\EP}\right|_{\hat y_\tau^6 \hat{v}^2}
	=-\frac{2\hat y_\tau^4 \hat \tau }{(4\pi)^4}
	&\left(2+\frac{\pi^2}{3}- 7\;\barlog(\hat \tau) +  3\;\barlog^2(\hat \tau) \right),
\end{align}
with $\hat{\tau} \equiv \hat{m}_\tau^2$ and 
$\barlog(\hat{\tau}) \equiv \log(\hat{\tau}/Q^2)$.

\subparagraph{\underline{$\Delta s_h^{\SM,\thrL}$ at $\ord(g_3^4y_t^4)$}}

In order to check the consistency of the 3-loop expression of
eq.~\eqref{eq:double-loop-exp-3L}, we derive the second term on the
r.h.s., $\Delta s_h^{\SM,\thrL}$, at $\ord(v^2g_3^4y_t^4)$ in the full-model
parametrization and compare it to the result presented in
ref.~\cite{Harlander:2018yhj}.
At 3-loop $\ord(v^2 g_3^4y_t^4)$, the SM Higgs mass correction of
eq.~\eqref{eq:iter_3l} receives an explicit self-energy and tadpole
contribution, which we take from ref.~\cite{Martin:2014cxa}.  We
determine the implicit (derivative) contributions of
eq.~\eqref{eq:iter_3l} using the 1-loop threshold correction
$\Delta g_3^{\as}$ from ref.~\cite{Pierce:1996zz} and the 2-loop
threshold correction $\Delta y_t^{\as^2}$ form
eq.~\eqref{eq:2l_yt_td}.\footnote{Note that in this section we ignore
  contributions of $\ord(v^2/M_S^2)$ for brevity and for
  cross-checking against expressions from the literature.  In our
  actual implementation of the corrections, presented later, we take
  all available terms of $\ord(v^2/M_S^2)$ into account.
} In the EFT limit and for degenerate SUSY mass parameters our 3-loop
Higgs pole mass contribution, expressed in terms of MSSM parameters,
reads
\begin{align}
  \label{eq:delta_s_SM_3l}
  \begin{split}
  \Delta s_h^{\SM,y_t^4 g_3^4}={}&
  \frac{4}{405} \frac{g_3^4 y_t^2}{(4\pi)^4}t s_{\beta }^2 
  \bigg[-540 \;\barlog(M_S^2)^2 \left(86 \;\barlog(t)+59\right)\\
  &+180
  \;\barlog(M_S^2) \left(2 \;\barlog(t) \left(252 \;\barlog(t)+88 x_t-281\right)+184
  x_t-557\right)
  \\&-32 \left(360 \;\text{Li}_2\left( \frac{1}{2}\right)^2+120 \pi^2
  \;\text{Li}_2\left( \frac{1}{2}\right)-29 \pi ^4\right)
  \\&
  -45 \left(1536 \;\text{Li}_4\left( \frac{1}{2}\right)-720
  \zeta_3 + 3187\right)+59040 x_t
  \\&-360
  \bigg(\barlog(t) \left[4 x_t \left(9 x_t-22\right)-36 \zeta_3
  +185\right]
  \\&+18 \;\barlog(t)^2 \left(8 x_t-9\right)+207
  \;\barlog(t)^3+42 x_t^2\bigg)
  \bigg],
  \end{split}
\end{align}
with $t\equiv m_t^2$ and $\barlog(x) \equiv \log(x/Q^2)$.
Our result agrees with eq.~(27) of ref.~\cite{Harlander:2018yhj}.
Furthermore, as a validation of eq.~\eqref{eq:double-loop-exp-3L}, we
checked numerically, that inserting eq.~\eqref{eq:delta_s_SM_3l} and
the 3-loop MSSM Higgs pole mass contribution from the \Himalaya\
library into eq.~\eqref{eq:double-loop-exp-3L} reproduces eq.~(43)
from \cite{Harlander:2018yhj}.

\section{Resummation of leading squark mixing contributions}
\label{sec:resummation}

In the introductory example of \secref{sec:parametrization_choices}, 
differences between the full-model and EFT parametrization were discussed.  It
was shown that both approaches are equivalent up to higher-order
terms, which contribute numerically to the difference of both
approaches.  From a technical point of view, implementing the
full-model parametrized matching is easier to achieve; in this
section we present a second, more important argument in favor of this
approach: the
resummation of higher-order contributions of the full-model squark
mixing parameter $x_f \equiv X_f/M_S$ in the context of Higgs mass
predictions.

The resummation is analogous to the resummation of large $n$-loop
$(\tan\beta)^n$-corrections to $m_b$ of refs.~\cite{Hall:1993gn,Carena:1994bv,Carena:1999py,Hofer:2009xb},
suitably generalized.\footnote{
At this point we want to note that the value of $x_t$ is bounded by the necessity of avoiding charge and color breaking minima
\cite{Bagnaschi:2014rsa}. For large $\msq \approx \msu$ the
  absolute value of the dimensionless stop-mixing parameter $|x_t|$ is
  restricted to be less than 3, whereas $\tan \beta$ can be as large
  as $50$--$60$.}
We begin here by recalling main features of the
$\tan\beta$-resummation in $m_b$, rephrase it in the appropriate
language and then present the generalization. More details and further
generalizations will be presented elsewhere \cite{KSTheorem}.

\subparagraph{$m_b$-matching and $\tan\beta$-resummation:}

First we review the resummation of all-order $\tan\beta$-enhanced
contributions in the \DRbarPrime-renormalized MSSM parameter $m_b$.
The resummation relies on the following theorem proven in  ref.~\cite{Carena:1999py}:
\begin{center}
\begin{minipage}[c]{0.9\textwidth}
  \textit{There are no contributions to $\Delta m_b$ of the order
    $\ord((\as\,\tan\beta\,\mu/M_3)^n)$ for $n\ge 2$.}
\end{minipage}
\end{center}
Here $\Delta m_b$ is the loop correction between the $b$-quark pole
mass and \DRbarPrime\ running mass,
    $M_b = m_b (1 + \Delta m_b)$. The quantity $\Delta m_b$ contains a
1-loop term of the order $\as\tan\beta$, terms with lower orders
in $\tan\beta$ and terms governed by other couplings, but no
higher-loop terms of the orders given in the theorem. The theorem only
holds for ``unsuppressed'' terms, i.e.\ terms not suppressed by powers
of $v/\MS$.

The theorem can be equivalently formulated in the language of
full-model versus EFT parametrization of the $b$-quark mass matching
between the MSSM and the SM. Full-model parametrization means to
express the SM
\MSbar\ bottom quark mass $\hat{m}_b$ as a perturbative series in
the full-model MSSM parameters $m_b$, $\as$, \ldots, and
truncating the series at some desired order. The full-model
parametrized relation between $\hat{m}_b$ and $m_b$, truncated at
order $\as$ then reads%
\footnote{As implicitly indicated by ref.~\cite{Carena:1999py}, a
  stronger restriction can be formulated which forbids unsuppressed
  terms of $\ord(\as^n \tan^{>1} \beta)$ in the threshold corrections
  to the bottom mass matching between the THDM and MSSM.}
\begin{align}
  \label{eq:tanb_resum}
  \hat{m}_b|_{m_b\as^{\leq 1}} &= m_b\left(1 + \Delta m_b^{\as}\right) + \cdots\,.
\end{align}
Here the notation of eqs.~\eqref{Pnotation} and \eqref{eq:Dyuk_MSSM_2L}
has been used, and the
subscript $m_b\as^{\leq 1}$ refers to the full-model parametrization and the
chosen truncation order. The dots denote terms irrelevant for the present
discussion (containing less powers of
$\tan\beta$ and/or power suppressions). In a numerical code, this
equation is often numerically solved for $m_b$, effectively
giving
\begin{align}
  m_b|_{m_b\as^{\leq 1}} = \frac{\hat{m}_b}{1 + \Delta m_b^{\as}} + \cdots\,.
  \label{eq:mb_resum}
\end{align}
The point of the theorem is that eq.~\eqref{eq:tanb_resum} is
1-loop exact with respect to $(\as\tan\beta)^n$-terms and
eq.~\eqref{eq:mb_resum} correctly takes into account (``resums'') all
terms of these orders.\footnote{%
  We note that all codes mentioned in the present paper resum the
$m_b(\as\tan\beta)^n$ corrections in this way, even if they otherwise
do not use full-model parametrization.}

On the other hand, in a calculation using the EFT parametrization, $m_b$
would be expressed as a perturbative series in terms of $\hat{m}_b$,
$\ashat$, \ldots,\footnote{%
  For the purpose of the present section the distinction between the SM
  parameter $\ashat$ and the MSSM parameter $\as$ is not relevant,
  since the two parameters differ by terms which do 
  not depend on $\tan\beta$ or $x_f$.}
truncated at some desired order. E.g.\ at order $\ashat$,
\begin{align}
  \left.m_b\right|_{\hat{m}_b\ashat^{\leq 1}} = \hat{m}_b (1 - \Delta m_b^{\ashat}) + \cdots\,,
  \label{eq:mb_non_resum}
\end{align}
which contains the correct $(\as\tan\beta)^n$ term only for $n=1$ but
misses all higher-order terms. Accordingly, we can evaluate the
difference
\begin{align}
  \label{eq:tanb_resum_explicit}
  \left.m_b\right|_{m_b\as^{\leq 1}} =
  \left.m_b\right|_{\hat{m}_b\ashat^{\leq 1}} + \sum_{n\ge 2} \left(-\Delta m_b^{\as} \right)^n + \cdots\,.
\end{align}

\subparagraph{$y_f$-matching and $x_f$-resummation:}

An analogous resummation is possible for the $\as x_f$-enhanced
contributions to the Yukawa matching for all colored fermions. We
note that the factor $\mu\tan\beta$ in the previous theorem arises
via $X_b=(A_b - \mu^*\tan\beta)$.
Hence the above theorem generalizes to
\begin{center}
\begin{minipage}[c]{0.9\textwidth}
  \textit{There are no unsuppressed contributions to the threshold
    correction $\Delta y_f$ at $\ord(\as^nx_f^{> 1})$ for $n\ge 1$
    in full-model parametrization.}
\end{minipage}
\end{center}
In the following we specialize to the case of the top quark. If we
express the SM coupling $\hat{y}_t$ as a 
perturbative series in $y_t$, $\as$, \ldots\ and truncate at the
order $\as$, we obtain the full-model parametrized expression
\begin{align}\label{eq:y_with_X}
  \hat{y}_t|_{y_t\as^{\leq 1}} = y_t s_\beta + \Delta y_t^{\as} .
\end{align}
The theorem then implies that this relation is 1-loop exact, i.e.\ there
are no unsuppressed higher-order corrections to $\Delta y_t^{\as}$ of
$\ord(\as^nx_t^{> 1})$ for $n\ge 2$.
In a numerical code such as \feft, eq.~\eqref{eq:y_with_X} is solved
numerically for $y_t$, giving
\begin{align}\label{eq:y_with_Xsolved}
y_t s_\beta|_{y_t\as^{\leq 1}}   = \frac{\hat{y}_t}{1+\frac{\Delta
    y_t^{\as}}{y_t s_\beta}},
\end{align}
where the chiral symmetry ensures that for QCD corrections $\Delta y_t^{\as} \propto y_t$, such that the denominator has the structure $1+ \ord(\as x_t) +\cdots$.
This expression correctly ``resums'' in particular all terms of the
orders $\ord((\as x_t)^n)$. In the EFT parametrization, however, $y_t$
is expressed as a perturbative series in $\hat{y}_t$, $\ashat$,
\ldots, truncated at some desired order.\footnote{%
  The previous version \feft\ \oneL\ from ref.~\cite{Athron:2017fvs},
  implemented in \fs\ 2.0, truncates at 1-loop, while \HSSUSY 
  truncates at 2-loop order.}  The resulting
expression is then correct up to that order but misses all
higher-order terms of the form $(\as x_t)^n$.  Similar to
eq.~\eqref{eq:tanb_resum_explicit}, we can express the difference
between the two parametrizations as
\begin{align}\label{eq:y_reparam}
  y_t|_{y_t\as^{\leq 1}} 
      &= y_t|_{\hat{y}_t\ashat^{\leq n}} + 
        \sum_{k>n} a_k\hat{y}_t(\ashat x_t)^k + \cdots\,.
\end{align}
Here the subscript on the r.h.s.\ denotes the truncation of the
EFT parametrization at $n$-loop level, and the dots denote terms
irrelevant for the present discussion. The main point is that the
difference contains terms of the orders $(\as x_t)^k$ with $k>n$, and
all these terms are correctly contained in the full-model
parametrization but missing in the EFT parametrization, and the
coefficients $a_k$ can be analytically computed at all orders.
As an example of this discussion, we refer to the analysis on
the $y_t$ matching given in \secref{yt_2-loop}:
In \eqref{eq:2l_yt_bu} the explicit 
2-loop threshold correction expressed in SM parameters contains
the term $\propto \hat y_t( \ashat x_t)^2$. According to eq.~\eqref{eq:y_reparam},
the MSSM  Yukawa coupling in eq.~\eqref{eq:yt_bu} misses terms
$\propto \hat y_t(\ashat x_t)^{\ge 3}$ which, however, are implicitly taken into 
account in the 1-loop correction of eq.~\eqref{eq:Dyuk_MSSM_2L}.

\subparagraph{$\hat\lambda$-matching and $x_f$-resummation:}

Now we turn to the resummation of $\ord(g_3^2 x_t)$-contributions in the
matching of $\hat\lambda$ in the full-model parametrization.
In \secref{sec:matching_of_quartic} we already presented an example where the
1-loop matching of $\hat\lambda$ and $y_t$ in full-model
parametrization leads to a correct 2-loop term leading in
$x_t$.  This example illustrates a more general property; however the
resummation within $\hat\lambda$ is slightly more complicated than the two
cases discussed above. We first recall that the threshold correction
for $\hat\lambda$ in
full-model parametrization, truncated at 1-loop order, contains the
terms
\newcommand{\gx}{g_x}
\newcommand{\gxhat}{\hat{g}_x}
\begin{align}
  \Delta\lambda^{\oneL}
  &= c_{0t} y_t^4 x_t^4 + c_{01}y_t^2g_1^2 x_t^2+c_{02}y_t^2g_2^2x_t^2 + \cdots\,,
  \label{lambda1Lthresholdstructure}
\end{align}
where the $c_{0x}$ are coefficients and the dots denote terms
irrelevant for the present discussion. The terms leading in $x_t$ are
thus of the general form
$y_t^2 \gx^2x_t^m$ with $\gx\in\{y_t, g_1, g_2\}$ and
$m\in\{4,2,2\}$. The resummation of higher-order terms governed by
$(g_3^2 x_t)^n$ relies on the resummation within the Yukawa coupling
discussed before and on the following theorem for the explicit
contributions to the threshold corrections:
\begin{center}
\begin{minipage}[c]{0.9\textwidth}
  \textit{There
    are no unsuppressed contributions to $\Delta\lambda$ at
    $\ord(y_t^2\gx^2g_3^{2n} x_t^{>m})$ for $n>0$ in full-model
    parametrization.}
\end{minipage}
\end{center}
Again the theorem means that the full-model expression
\eqref{lambda1Lthresholdstructure} is 1-loop exact with respect to the
leading $\ord(g_3^2 x_t)$ terms. 
The proof of the theorem is analogous to the proof in ref.~\cite{Carena:1999py} 
and relies on the large mass expansion and inspection of individual
contributions. For 
details and generalizations we refer to ref.~\cite{KSTheorem}. 
If the Yukawa coupling in
eq.~\eqref{lambda1Lthresholdstructure} is replaced by $\hat{y}_t$ via
eq.~\eqref{eq:y_with_Xsolved}, we
obtain equations of the form
\begin{align}\label{eq:resummationgeneral}
  \left.\Delta \lambda\right|_{y_t^2 \gx^2} =
  \hat{c}_{0x}\hat{y}_t^2\hat{g}_x^2 x_t^m
+  \sum_{n\ge1}  \hat{c}_{nx}\hat{y}_t^2\hat{g}_x^2 \hat g_3^{2n} x_t^{m+n}
  + \cdots\,,
\end{align}
with appropriately modified coefficients $\hat{c}_{0x}$ from
tree-level matching and higher-order coefficients
$\hat{c}_{nx}$. In a numerical code based on the full-model
parametrization these higher-order terms are fully taken into account;
the coefficients are also  fully calculable analytically.
All the explicit higher-order terms in
eq.~\eqref{eq:resummationgeneral} are correct. In 
this sense the full-model parametrization resums these terms of
$\ord(\hat{y}_t^2 \gxhat^2 \hat{g}_3^{2n} x_t^{m+n})$.

On the other hand, if the $\hat\lambda$-matching is done in
EFT parametrization, $\Delta\lambda$ is expanded in terms of SM
parameters and truncated at some desired order. In that case, leading
$\ord(\hat{g}_3^2 x_t)$ terms are only taken into account up to that order. The
difference between the two versions of the threshold corrections can
be written as
\begin{align}\label{eq:resummationgeneraldifference}
  \left.\Delta \lambda\right|_{y_t^2 \gx^2} = 
  \left.\Delta \lambda\right|_{\hat{y}_t^2
    \gxhat^2\hat g_3^{\leq 2n}}
  +
  \sum_{k>n}  \hat{c}_{kx}\hat{y}_t^2\hat{g}_x^2 \hat g_3^{2k}x_t^{m+k}
  + \cdots\,.
\end{align}
All the explicitly given terms $\propto \hat{c}_{kx}$ on the r.h.s.\
of eq.~\eqref{eq:resummationgeneraldifference} are taken into account
correctly in the full-model parametrization but are missing in the EFT
parametrization.
In other words all these higher-order terms in
eq.~\eqref{eq:resummationgeneraldifference} do not arise from
explicit multi-loop diagrams; instead they only arise via the
reparametrization, and the values of the coefficients $\hat{c}_{kx}$
can be deduced from the 1-loop terms leading in $x_t$.
As an example, we reconsider the concrete calculation in eq.~\eqref{eq:DLambda_MSSM_rep}:
The 1-loop matching of $y_t$ at $\ord(g_3^2)$ in combination with
$\Delta\lambda$ at $\ord(y_t^4)$ captures the leading
$\ord(\hat{y}_t^4 \hat{g}_3^{2} x_t^5)$ contribution of
$\Delta\lambda^{\twoL}$ in EFT parametrization.  Likewise, the
3-loop term of $\ord(\hat{y}_t^4 \hat{g}_3^{4} x_t^{6})$ is
captured, too.

As mentioned before, the theorem from above can be formulated in a more
general way, including for example the known structure of
$\Delta\lambda^\twoL$ at $\ord(y_t^6)$ and $\Delta y_t^\twoL$ at
$\ord(y_t^3g_3^2)$, which allows for a resummation of further
higher-order corrections \cite{KSTheorem}.
However, we want to emphasize that  not all terms with a high power in $x_t$
can be resummed. For example, the pure Yukawa
$(n+1)$-loop contributions of 
$\ord(\hat y_t^{4+2n} x_t^{4+2n})$ cannot be captured by the application
of neither full-model nor EFT parametrization.
The reason is that there exist genuine $(n+1)$-loop diagrams
which provide unsuppressed contributions of this order.
The numerical impact of missing $\ord(\hat y_t^{4+2n} x_t^{4+2n})$ terms for $n=2$
is discussed in \secref{sec:numfurtherdetails} and \secref{sec:est_HS_uncer}.

As an important application and check we consider mixed QCD--electroweak
contributions. In codes such as \feft, \HSSUSY\ or \susyhd, the 2-loop
and 3-loop Higgs self-energy is only computed in the limit of
vanishing electroweak gauge couplings. But the resummation now allows
to compute the analytic form of the leading $x_t$-terms of the 2-loop
and 3-loop mixed QCD--electroweak contributions of the orders
$\ord(\hat{y}_t^2 \hat{g}_{1,2}^2 \hat g_3^2 + \hat{y}_t^2 \hat{g}_{1,2}^2
\hat g_3^4)$. The
1-loop threshold correction $\Delta\lambda^\oneL$ at
$\ord(y_t^2g_{1,2}^2)$ originates from $D$-term contributions and
involves a maximum $x_t$-dependence of order $x_t^2$.
Thus, our
prediction for the leading $x_t^3$ contribution at 2-loop mixed
electroweak order is
\begin{align}
  \label{eq:xt_pred}\begin{split}
    \left.\Delta \lambda\right|_{\hat{y}_t^2 \hat{g}_{1,2}^2 \hat{g}_3^2}={}& \frac{\hat{g}_3^2\hat{y}_t^2}
    {(4\pi)^4}\frac{X_t^3}{M_3 \msq\msu}\frac{2c_{2\beta}}{3} \tilde F_9
      \left(\frac{\msq}{M_3}, \frac{\msu}{M_3}\right)\\
    &\times\left[3 \left(\tilde F_4(x_{QU}) \hat{g}_2^2 +\tilde F_3(x_{QU}) \hat{g}_Y^2\right) -c_{2\beta} \tilde F_5(x_{QU})\left(\hat{g}_2^2+\hat{g}_Y^2\right)  \right],
\end{split}
\end{align}
with the functions $\tilde F_{3,4,5,9}$ defined in
ref.~\cite{Bagnaschi:2014rsa} with
$\tilde F_{3,4,5}(1) = \tilde F_{9}(1,1) = 1$ and
$x_{QU} = \msq/\msu$.
Eq.~\eqref{eq:xt_pred} is equal to the result obtained in
eq.~(27) of ref.~\cite{Bagnaschi:2019esc} by an explicit computation
of this order in EFT 
parametrization. The analogous prediction for the leading $x_t^4$-term of the
3-loop mixed QCD--electroweak contributions is
\begin{align}
  \label{eq:xt_pred_3L}\begin{split}
\left.\Delta \lambda\right|_{\hat{y}_t^2 \hat{g}_{1,2}^2 \hat{g}_3^4}={}& \frac{\hat{g}_3^4\hat{y}_t^2}
{(4\pi)^6}\frac{X_t^4}{M_3^2 \msq\msu}\frac{4c_{2\beta}}{9} \tilde F_9^2
\left(\frac{\msq}{M_3}, \frac{\msu}{M_3}\right)\\
&\times\left[3 \left(\tilde F_4(x_{QU}) \hat{g}_2^2 +\tilde F_3(x_{QU}) \hat{g}_Y^2\right) -c_{2\beta} \tilde F_5(x_{QU})\left(\hat{g}_2^2+\hat{g}_Y^2\right)  \right].
\end{split}%
\end{align}%
All these terms and the corresponding terms of $\ge4$-loop order are
automatically taken into account by the new \feft\ calculation based
on full-model parametrization.

\section{Running and matching procedure at the electroweak scale}
\label{sec:matching_low_scale}

In this section we describe the computations which are performed after
the SM parameters are obtained at the high scale by the matching
characterized in \secref{sec:expansion_Master_eq}.  As illustrated in
\figref{fig:eft-approach}, the subsequent calculation involves two
further steps. First the SM parameters are run down to the low-energy electroweak scale
by solving the RGEs (\subsecref{sec:rundown}). Second, the
low-scale parameters are related to input values of observables, and
the final prediction for the Higgs boson mass is computed. The
corresponding low-scale matching procedure is described in
\subsecref{sec:lowscalematching}.

\subsection{Running to the electroweak scale}\label{sec:rundown}

In order to relate the high-scale SM parameters with low-scale SM
parameters, the renormalization group equations of the SM are solved
numerically.  Our new \feft\ calculation uses the SM $\beta$ functions
of refs.~\cite{Mihaila:2012fm,Bednyakov:2012rb,
  Bednyakov:2012en,Chetyrkin:2012rz,Bednyakov:2013eba,Chetyrkin:2016ruf,Martin:2015eia,Bednyakov:2015ooa},
which include up to 4-loop corrections. The RGEs within the MSSM are
not needed for the actual Higgs mass computation. They are only needed
if the input scale of MSSM $\DRbarPrime$ parameters does not coincide
with the matching scale. In this case our new \feft\ calculation uses
the MSSM 3-loop $\beta$ functions of
refs.~\cite{Jack:2003sx,Jack:2004ch}, see also
ref.~\cite{Athron:2017fvs}.

\subsection{Matching of SM couplings to
  observables}\label{sec:lowscalematching}
  
In order to express the prediction for the Higgs pole mass in terms of
physical quantities, the running SM \MSbar\ parameters have to be
related to observables.  There are eight SM \MSbar\ parameters
relevant for the Higgs mass prediction (cf.\
eq.~\eqref{eq:SM_parameters}):
\begin{align}
  \hat{P} = \{\hat{g}_1, \hat{g}_2, \hat{g}_3, \hat{y}_t, \hat{y}_b, \hat{y}_\tau, \hat{\lambda}, \hat{v}\}.
\end{align}
Among these eight parameters, $\hat{\lambda}$ is fixed by the matching
to the MSSM at the SUSY scale, while the other seven parameters are
fixed by low-energy observables.  Following the approach described
in ref.~\cite{Athron:2017fvs}, we fix these seven parameters at the
scale $Q=M_Z$ by relating them to the seven low-energy quantities
shown in \tabref{tab:obs}.
\begin{table}[tbh]
  \centering
  \begin{tabularx}{0.9\textwidth}{lX}
    \toprule
    Quantity & Description \\
    \midrule
    $M_t$ & top quark pole mass \\
    $\mbmb$ & \MSbar\ bottom quark mass at the scale $Q=\hat{m}_b$ in the SM with five active quark flavors \\
    $M_\tau$ & $\tau$ lepton pole mass \\
    $M_Z$ & $Z$ boson pole mass \\
    $G_F$ & Fermi constant \\
    $\asMZ$ & \MSbar\ strong coupling in the SM with five active quark flavors at the scale $Q=M_Z$ \\
    $\aemMZ$ & \MSbar\ electromagnetic coupling in the SM with five active quark flavors at the scale $Q=M_Z$ \\
    \bottomrule
  \end{tabularx}
  \caption{Low-energy quantities for the determination of SM \MSbar\ parameters.}
  \label{tab:obs}
\end{table}

\paragraph{Matching procedure.}

Before continuing  the discussion about the included loop
corrections at the electroweak scale, we want to emphasize a qualitative
difference of our low-scale matching procedure w.r.t.\ the procedure
described in ref.~\cite{Athron:2017fvs}.
In our new \feft\ matching approach we consider the
full-model (MSSM) \DRbarPrime\ parameters at the SUSY scale to be
fundamental.  This includes the SUSY parameters and the SM-like
full-model parameters $P$ from eq.~\eqref{eq:MSSM_parameters}.  As a
consequence, we eventually express all observables, including the
low-energy observables from \tabref{tab:obs} as well as the predicted
Higgs pole mass in terms of the full-model \DRbarPrime\ parameters.
Technically, this is achieved by $(i)$ converting the (fundamental)
full-model \DRbarPrime\ parameters $P(M_S)$ to \MSbar\ EFT parameters
$\hat{P}(M_S)$ using the matching conditions \eqref{eq:lambda_from_Mh}
and \eqref{eq:matching_conditions}, $(ii)$ renormalization group running
of the EFT parameters from the scale $M_S$ to the low-energy (electroweak) scale
$\Qlow$ and $(iii)$ predicting the low-energy quantities
$O^{\text{pred}}_i$ from \tabref{tab:obs}. In the most
direct approach the relation between the renormalized SM \MSbar\ parameters and
the observables at the $n$-loop level is constructed as
\begin{align}\label{top-down_relation}
  O^{\text{pred}}_i = f_i(\hat{P}(\Qlow)) + \ord(\hbar^{n+1}) .
\end{align}
In eq.~\eqref{top-down_relation} $f_i(\hat{P})$ denotes the function
that calculates the observable $O^{\text{pred}}_i$ as a function of
the EFT parameters $\hat{P}$.  For a given set of SUSY parameters the
SM-like full-model parameters $P$ are adapted such that the predicted
observables $O^{\text{pred}}_i$ agree with the observed values
$O^{\text{input}}_i$ up to a sufficiently high precision
$\epsilon \ll 1$,
\begin{align}
\label{eq:pred_observ}
  \left|O^{\text{pred}}_i - O^{\text{input}}_i\right| < \epsilon \qquad \forall i.
\end{align}
In contrast, in spectrum generators working in the EFT parametrization
(e.g.\ \HSSUSY, \susyhd or \MhEFT), the low energy EFT \MSbar\
parameters $\hat{P}(\Qlow)$ (except for $\hat\lambda$) are not determined
from the full-model parameters, but rather they are directly extracted
from the observables as
\begin{align}\label{bottom-up_relation}
  \hat{P}_i(\Qlow) = h_i(O^{\text{input}}) + \ord(\hbar^{n+1})
\end{align}
with some function $h_i$ denoting the calculation.  The difference
between the EFT parameters $\hat{P}$ from both approaches
\eqref{top-down_relation} and \eqref{bottom-up_relation} is of higher
order.  The different higher-order terms depend dominantly on the
values of $\hat{P}(\Qlow)$. However, eqs.~\eqref{top-down_relation}
and \eqref{bottom-up_relation} can be modified by higher orders
in such a way that the differences expressed in terms of
$\hat{P}(\Qlow)$ vanishes, e.g.~see the discussion below eq.~\eqref{eq:mt_MSbar}.
Even if the conditions \eqref{top-down_relation} and
\eqref{bottom-up_relation} coincide in their inclusion of loop corrections,
a remaining difference in the numerical value of $\hat{P}(\Qlow)$ 
may occur depending on the parametrization of the high-scale
matching condition for $\hat \lambda$.
Though the difference, which originates from reparametrization effects 
at the high scale, depends only subdominantly on the SUSY
parameters, e.g.\ $\tan\beta$.%
\footnote{Considering the difference of the threshold correction
  $\Delta\lambda$ between the EFT and the full-model parametrization,
  the reparametrization effects do explicitly depend on SUSY
  parameters. Inserted into the $\beta$ functions of the SM, the
  reparametrization terms receive further loop suppressions.
  Therefore, the higher-order reparametrization effects which contain
  MSSM-specific parameters propagate into all EFT parameters at the
  electroweak scale with additional loop factors.}

\paragraph{Matching conditions.}

As described in \secref{sec:expansion_Master_eq}, we aim for a
prediction of the Higgs pole mass at \NCLO\ with \NCLL\ resummation in
QCD.  This precision requires to relate the SM \MSbar\ top quark mass
$\hat{m}_t$ to the pole mass $M_t$ up to $\ord(\ashat^3)$
\cite{Chetyrkin:1999qi,Melnikov:2000qh}.  We follow the prescription
presented in ref.~\cite{Athron:2017fvs} and express $M_t$ as
\begin{align}
\begin{split}
  M_t &= \hat{m}_t - \re\Sigma_{t,S}^{\SM,\oneL}((M_t^{\text{input}})^2,Q) \\
  &\phantom{={}} -  M_t^{\text{input}}\Big[ \re\Sigma_{t,L}^{\SM,\oneL}((M_t^{\text{input}})^2,Q) +
  \re\Sigma_{t,R}^{\SM,\oneL}((M_t^{\text{input}})^2,Q) \\
  &\phantom{={} + m_t \Big[}
  + \Delta \hat{m}_t^{\QCD,\oneL}(Q) + \Delta \hat{m}_t^{\QCD,\twoL}(Q) + 
  \Delta \hat{m}_t^{\QCD,\thrL}(Q) + \Delta \hat{m}_t^{\QCD,\fourL}(Q) \Big] .
\end{split} \label{eq:mt_MSbar}
\end{align}
Note, that we have included the 4-loop QCD contribution from
ref.~\cite{Martin:2016xsp} for later use in \secref{sec:uncertainty}.
In eq.~\eqref{eq:mt_MSbar} $M_t^{\text{input}}$ denotes the 
observed top quark pole mass, $Q$ is the renormalization scale and
$\Sigma_{t,\{L,R,S\}}^{\SM,\oneL}(p^2,Q)$ are the left-handed,
right-handed and scalar parts of the 1-loop SM top quark self-energy
without QCD contributions.  The separate SM QCD contributions read
\cite{Chetyrkin:1999qi,Melnikov:2000qh,Martin:2016xsp}
\begin{subequations}
\begin{align}
  \Delta \hat{m}_t^{\QCD,\oneL}(Q) &\approx
  \frac{\hat g_3^2}{(4\pi)^2} \left[4 \;\barlog(\hat{t})- 5.333\right],
  \label{eq:top-selfenergy-qcd-1L}
  \\
  \Delta \hat{m}_t^{\QCD,\twoL}(Q) &\approx \frac{\hat g_3^4}{(4 \pi)^4}
  \left[ -6 \;\barlog^2(\hat{t}) +38 \;\barlog(\hat{t}) 
  -103.341\right] ,
  \label{eq:top-selfenergy-qcd-2L}
  \\
  \Delta \hat{m}_t^{\QCD,\thrL}(Q) &\approx
  \frac{\hat g_3^6}{(4\pi)^6}\left[20 \;\barlog^3(\hat{t})
  -86 \;\barlog^2(\hat{t})
   +457.747 \;\barlog(\hat{t})
  -3458.737 \right] ,
  \\
  \begin{split}
  \Delta \hat{m}_t^{\QCD,\fourL}(Q)   &\approx
  \frac{\hat g_3^8}{(4\pi)^8}\, \Big[-85\;\barlog^4(\hat{t}) +323.333\;\barlog^3(\hat{t})
  -1832.501\;\barlog^2(\hat{t}) \\
  &\phantom{\frac{\hat g_3^8}{(4\pi)^8}\, IIIi }+
  45369.45 \;\barlog(\hat{t})
  -154481.798 \Big] ,
  \end{split}
\end{align}
\end{subequations}
with $\hat{t} = \hat{m}_t^2$ and $\barlog(x) \equiv \log(x/Q^2)$.
As referred to before, the construction of eq.~\eqref{eq:mt_MSbar}
contains the observed value of the pole mass $M_t^{\text{input}}$,
which introduces higher-order terms in eq.~\eqref{top-down_relation}
such that the higher-order difference to the low-scale constraint used
in \HSSUSY\ is minimized,  see eq.~(7) from ref.~\cite{Athron:2017fvs}. 
The Fermi constant $G_F$ is calculated similarly to
ref.~\cite{Pierce:1996zz} as
\begin{align}
  \label{eq:GFermi}
  G_F = \frac{\pi\aemhat}{\sqrt{2} M_Z^2 \hat c^2 \hat s^2(1-\Delta \hat r)},
\end{align}
where $\Delta\hat{r}$ contains only SM contributions, including 2-loop
contributions of $\ord(\aemhat\ashat)$ \cite{Fanchiotti:1992tu}.
Because the 2-loop contributions themselves depend on $G_F$, an
iteration is performed in eq.~\eqref{eq:GFermi}.  The remaining
low-energy quantities from \tabref{tab:obs} are calculated at the
scale $Q=M_Z$ as
\begin{subequations}
\begin{align}
  \aemQ &= \frac{\aemhat(Q)}{1 + \Delta\aemhat^{\oneL}(Q)} ,\\
  \asQ &= \ashat(Q) \left[1 - \Delta\ashat^{\oneL}(Q)
              - \Delta\ashat^{\twoL}(Q) - \Delta\ashat^{\thrL}(Q)\right] ,\\
  \label{eq:low_energy_tau}
  M_\tau &= \hat{m}_\tau(Q) - \re\Sigma_\tau^\oneL(\hat{m}_\tau,Q) ,\\
  \label{eq:low_energy_z}
  M_Z^2 &= \hat{m}_Z^2(Q) - \re\Sigma_Z^\oneL(\hat{m}_Z,Q) ,\\
  \mbQ &= \hat{m}_b(Q) \left[ 1 - \re\Sigma_b^{\oneL,\text{heavy}}(\mbQ,Q) \right] ,
  \label{eq:low_energy_mb}
\end{align}
\end{subequations}
where the 1-loop correction $\Delta\aemhat^{\oneL}(Q)$ is parametrized
in terms of the 6 flavor $\aemhat(Q)$ as
\begin{align}
  \Delta\aemhat^{\oneL}(Q) = -\frac{8}{9\pi}
  \aemhat(Q) \log\frac{\hat{m}_t(Q)}{M_Z}.
\end{align}
The loop corrections of $\Delta\ashat^{\nL}(Q)$ are defined in
eqs.~(13)--(15) of ref.~\cite{Athron:2017fvs}.  The 1-loop
\MSbar-renormalized self-energies $\Sigma_\tau^\oneL(p,Q)$ and
$\Sigma_Z^\oneL(p,Q)$ of the $\tau$ lepton and $Z$ boson,
respectively, are evaluated at full 1-loop precision.  In
eq.~\eqref{eq:low_energy_mb} $\Sigma_b^{\oneL,\text{heavy}}(p,Q)$
denotes the 1-loop top quark and electroweak gauge boson contributions
to the bottom quark self-energy as described in
ref.~\cite{Athron:2014yba}.  The predicted $\mbQ$ is evolved to the
scale $Q=\mbfive$ to be compared to the input value $\mbmb$.

In principle, one might treat the Higgs mass similar to all other
low-scale observables and use it as a constraint  
in the sense of eq.~\eqref{eq:pred_observ} to
fix one parameter of the MSSM. 
However, in our application we choose the Higgs mass at the low scale
to be the output of our calculation.
Analogous to eq.~(1) of ref.~\cite{Athron:2017fvs}
we compute the Higgs pole mass in the SM as in eq.~\eqref{eq:SM_higgs_pole}
but instead of using MSSM parameters we express it in terms of
SM parameters at $Q=M_t^{\text{input}}$.
The included corrections are described in \secref{sec:expansion_Master_eq}
and are of $\ord(\oneL+\hat v^2(\twoloopeft) +\hat v^2\thrloopeft)$.

\section{Numerical results}
\label{sec:numerical_results}

In this section we present the numerical results for the light CP-even
Higgs pole mass calculation in the real MSSM, based on the improved
\feft\ approach.  We highlight the numerical impact of the new
included threshold corrections and the parametrization scheme and
describe in particular the $x_t$-resummation.

If not stated otherwise, the dimensionful \DRbarPrime-renormalized
parameters of the MSSM Lagrangian are set to a common SUSY scale
$M_S$,
\begin{subequations}
  \begin{align}
\msfi{f}{i}^2(\MS) &= M_S^2, && (f=q,u,d,l,e; i=1,2,3) \\
M_i(\MS) &= \MS, && (i = 1,2,3) \\
A_f(\MS) &= 0, && (f = b, \tau) \\
\mu(\MS) &= \MS , \\
m_A^2(\MS) &= \frac{B\mu(\MS)}{\sin\beta(\MS) \cos\beta(\MS)} = \MS^2 .\hspace{-12em}&&
\end{align}%
\label{eq:MSSM_MSUSY_scenario}%
\end{subequations}%
For scenarios with non-trivial squark mixing we parametrize our
results in terms of the dimensionless \DRbarPrime\ parameter
$x_t \equiv X_t/M_S$.
In our numerical discussion we choose the input values given in
\tabref{tab:input}.
Effects from \nth{1} and \nth{2} generation (s)fermions are omitted in
our analysis.
\begin{table}[tbh]
  \centering
  \begin{tabular}{ll}
    \toprule
    Quantity & Value \\
    \midrule
    $M_t$ & $173.34\GeV$ \\
    $\mbmb$ & $4.18\GeV$ \\
    $M_\tau$ & $1.777\GeV$ \\
    $M_Z$ & $91.1876\GeV$ \\
    $G_F$ & $1.1663787\cdot 10^{-5} \GeV^{-2}$ \\
    $\asMZ$ & $0.1184$ \\
    $\aemMZ$ & $1/127.944$ \\
    \bottomrule
  \end{tabular}
  \caption{Low-energy input parameters from refs.\
    \cite{Bethke:2009jm,Agashe:2014kda,Beringer:1900zz,ATLAS:2014wva,Tanabashi:2018oca}}
  \label{tab:input}
\end{table}

\subsection{Impact of higher orders, the new parametrization and the 
	$x_t$-resummation}

\begin{figure}[tb]
  \centering
  \includegraphics[width=0.49\textwidth]{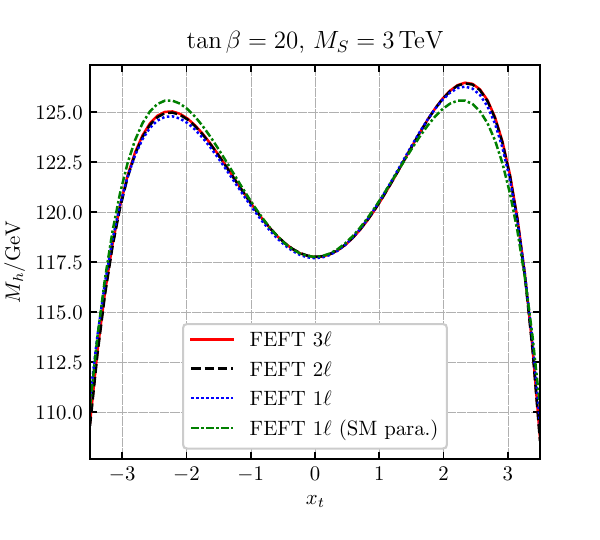} \hfill
  \includegraphics[width=0.49\textwidth]{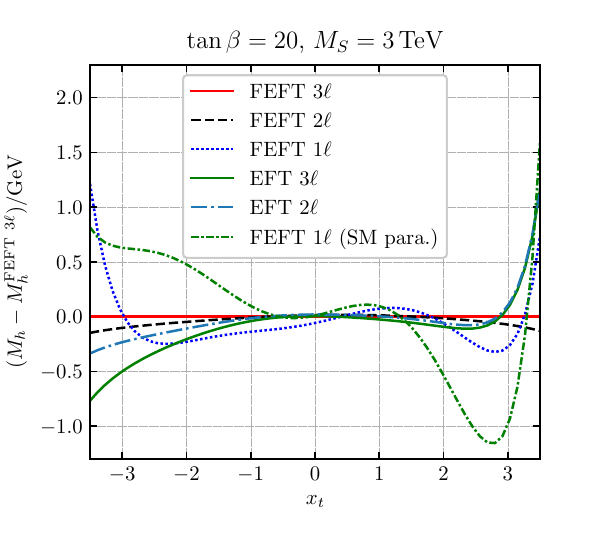}
  \includegraphics[width=0.49\textwidth]{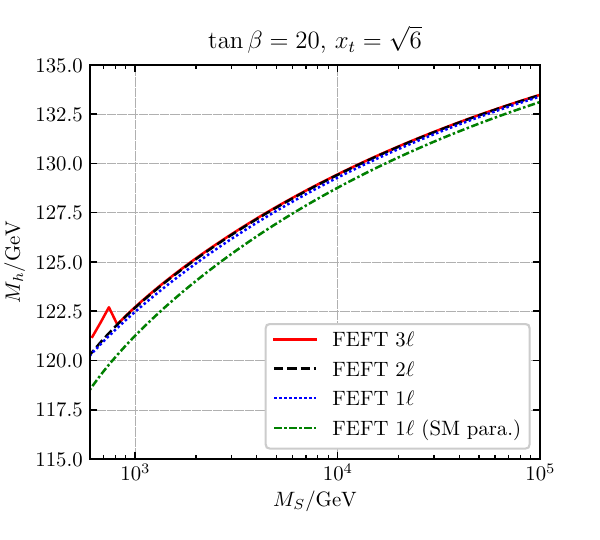} \hfill
  \includegraphics[width=0.49\textwidth]{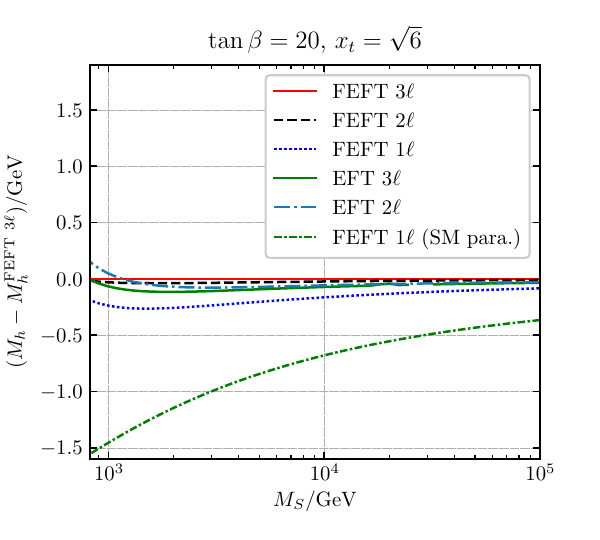}
  \caption{Prediction of the light CP-even Higgs pole mass in the MSSM
    as computed by the \fefts\ \oneL\ (SM para.\@) calculation as
    implemented in \fs 2.4 (green dashed-dotted line) and by the new
    \fefts\ calculations at 1-, 2- and 3-loop precision in the
    $\hat\lambda$ matching for $\tan\beta=20$.  In the left panels we show
    the absolute value of $M_h$ and in the right panels we show the
    difference w.r.t.\ the \fefts\ \thrL\ prediction. On the right panels
    we show additional lines \cite{Harlander:2018yhj}, EFT 2$\ell$ (green solid) and EFT 3$\ell$ (turquoise), which represent the EFT calculation in EFT parametrization 
    at 2-loop and 3-loop respectively (see \secref{sec:sota} for the characterization of EFT $3\ell$).}
  \label{fig:feft_comparison}
\end{figure}

We begin with the discussion of the impact of higher-order corrections
in the matching and the impact of the new full-model parametrization
and the resulting $x_t$-resummation.  \Figref{fig:feft_comparison}
shows the light CP-even Higgs pole mass calculated by different
versions of \feft:
\begin{itemize}
\item \underline{\fefts\ \oneL/\twoL/\thrL}: The new \feft\ hybrid
  calculation developed in the present paper with full-model (MSSM)
  parametrization of the matching calculation and $x_t$-resummation.
  The only difference among these
  calculations stems from the orders taken into account in the
  calculation of $\hat\lambda$, see eq.~\eqref{eq:lambdamatchinggeneral}.
  The \oneL\ version
  (blue dotted line) contains all 1-loop corrections.  The \twoL\
  version (black dashed line) contains in addition the 2-loop
  contributions of
  $\ord(\twoloopfm)$.  
  The \thrL\ version (red solid line) contains in addition the 3-loop
  contributions of $\ord(\thrloopfm)$.
  All three versions match the top Yukawa coupling at full 1-loop
  level and 2-loop $\order{y_tg_3^4}$ at the SUSY scale, whereas the remaining couplings
  are determined at 1-loop level as described in
  \secref{sec:expansion_Master_eq}.  The included corrections to 
  the low-energy input quantities (including the Higgs boson pole
  mass) are the ones described in \secref{sec:matching_low_scale}.
\item \underline{\fefts\ \oneL\ (SM para.)}: The previous \feft\
  calculation (green dashed-dotted line), presented in
  ref.~\cite{Athron:2017fvs} and included in \fs\ since version 2.0.0.
  This calculation is based on the same matching conditions, but
  employs the EFT (SM) parametrization and includes only 1-loop threshold
  corrections.
\end{itemize}
We first discuss the two 1-loop versions (blue dotted and green
dashed-dotted lines).  They differ essentially by the full-model 
versus EFT
parametrization of the high-scale matching.  As discussed in
\secref{sec:resummation}, in the full-model parametrization certain leading
$x_t$ terms are correctly taken into account.
Eq.~\eqref{eq:resummationgeneral} provides the general form of those
terms.  To exemplify the correctly included terms, consider the 1-loop
threshold correction $\Delta \lambda^\oneL$ in the MSSM
parametrization.  It contains terms of the order $\ord(y_t^4x_t^4)$
and mixed electroweak terms of the form
$\ord(y_t^2g_{1,2}^2x_t^2)$. Upon reparametrization of these
contributions in terms of SM parameters, using in particular the
1-loop top Yukawa threshold correction of $\ord(y_tg_3^2 x_t)$, the terms
shown in \tabref{tab:xt-resum} are generated.  As discussed in
\secref{sec:resummation}, these terms are not modified by genuine
$n$-loop contributions.  Hence, already the new \fefts\ \oneL\
calculation correctly takes into account the leading QCD $(n+1)$-loop
contributions of the order $x_t^{(4+n)}$, and the leading mixed
QCD--electroweak $(n+1)$-loop contributions of the order
$x_t^{(2+n)}$.  In contrast, none of the terms in
\tabref{tab:xt-resum} is correctly taken into account in the previous
\fefts\ \oneL\ (SM para.\@) calculation.
\begin{table}[h]
\centering
\begin{tabular}{ccc}
  \toprule
  loop order & $\Delta \lambda^\QCD$ & $\Delta \lambda^{\text{QCD--EW}}$\\
  \midrule
  \twoL & $\hat y_t^4 \hat g_3^2 x_t^5 $ & $\hat y_t^2\hat g_3^2\hat g_{1,2}^2 x_t^3 $\\[0.5em]
  \thrL & $\hat y_t^4 \hat g_3^4 x_t^6 $ & $\hat y_t^2 \hat g_3^4 \hat g_{1,2}^2 x_t^4$\\[0.5em]
  \fourL & $\hat y_t^4 \hat g_3^6 x_t^7 $ & $\hat y_t^2 \hat g_3^6 \hat g_{1,2}^2 x_t^5$\\
  \vdots & \vdots & \vdots \\
  \bottomrule
\end{tabular}
\caption{Contributions to $\Delta \lambda^{\nL}$, which are
  correctly (implicitly) included by the $x_t$-resummation in the new
  \fefts\ \oneL\ calculation, i.e.~terms contained in $\left.\Delta \lambda\right|_{y_t^4,y_t^2g_{1,2}^2}$.
   Note, that the \fefts\ \oneL\
  calculation is based on the full-model parametrization, but the
  terms in this table are provided in terms of SM parameters, to
  compare with other SM-parametrized calculations.  Note further, that
  the terms in this table are already contained in the 1-loop
  calculation; further, higher-order terms resummed by \fefts\
  \twoL/\thrL\ are determined by eq.~\eqref{eq:resummationgeneral}.}
\label{tab:xt-resum}
\end{table}
As a result of the $x_t$-resummation, we see a dramatic shift between
the two \oneL\ versions in \figref{fig:feft_comparison}.  
\begin{itemize}
\item Without $x_t$-resummation, the \fefts\ \oneL\ (SM para.\@)
  result (green dashed-dotted line) and the EFT $3\ell$ calculation
  (green solid line) deviate up to $\sim 1.1\GeV$ for $|x_t|<3$.
\item With $x_t$-resummation, this deviation between the \fefts\
  $1\ell$ and \fefts\ $3\ell$ calculation decreases to less than
  $0.3\GeV$, compare the blue dotted and the red solid lines.
\end{itemize}
For all values of $x_t$ and $\MS$, the new \fefts\ \oneL\ calculation
is far closer to the \fefts\ \thrL\ calculation; hence the convergence
of perturbation theory is significantly improved.

As a side remark we want to stress that in contrast to the correctly included contributions, the 
$1\ell$ calculations in both parametrizations fail to capture the 
pure Yukawa $(n+1)$-loop contributions 
leading in the stop-mixing parameter of $\ord(\hat y_t^{4+2n} x_t^{4+2n})$.  
In order to investigate whether those terms counteract the 
benefits of the resummed QCD-enhanced contributions we show \tabref{tab:xt-incomplete}.
\begin{table}[h]
  \centering
  \begin{tabular}{cccc}
    \toprule
    &terms contained&terms contained&\\
    loop order & in $\left.\Delta\lambda\right|_{y_t^4}$ & in $\left.\Delta\lambda\right|_{\hat y_t^4}$& correct $\Delta\lambda$\\
    \midrule
    \twoL &$-\frac{1}{2}\kappa^2\hat y_t^6 x_t^6 $& 0 & $-\frac{3}{2}
                                                        \kappa^2\hat y_t^6 x_t^6 $\\[0.5em]
    \thrL&$-\frac{9}{16}\kappa^3\hat y_t^8 x_t^8 $ & 0 & unknown\,$\cdot\, \kappa^3 \hat y_t^8 x_t^8 $\\[0.5em]
    \vdots & \vdots & \vdots & \vdots \\
    \bottomrule
  \end{tabular}
  \caption{Comparison of highest-power $x_t$ contributions to
    $\Delta\lambda$ of $\ord(\hat y_t^{2+2n} x_t^{2+2n})$ at $n$-loop
    level, implicitly induced by the 1-loop calculations in EFT and
    full-model parametrization ($\kappa = 1/(4\pi)^2$).  The second
    column shows the terms induced in the full-model-parametrized
    calculation at $\left.\Delta\lambda\right|_{y_t^4}$.  The third
    column shows the analogous terms in the EFT-parametrized
    calculation at $\left.\Delta\lambda\right|_{\hat y_t^4}$.  The
    last column represents the correct result in  EFT parametrization.}
  \label{tab:xt-incomplete}
\end{table}
The second and third rows contain 2- and 3-loop terms of highest power
in $x_t$, respectively, which are implicitly contained by the 1-loop
calculation in full-model parametrization (second column) and in EFT
parametrization (third column).  The terms in these columns have been
obtained by inverting the 1-loop relation between $y_t$ and $\hat y_t$
in full-model parametrization perturbatively up to the 3-loop level.
The terms are compared to the known/unknown correct result in EFT
parametrization in the last column.  The contributions in the EFT
parametrization are vanishing by construction.  At 2-loop we see that
the implicitly included term from the 1-loop correction in full-model
parametrization lies in between the correct result and the analogous
one in the EFT parametrization.  Without further information from an
explicit 3-loop (or higher) calculation there is no indication that
the full-model parametrization worsens the convergence of the
perturbative expansion with respect to these orders.
The 3-loop term is studied
numerically together with the reparametrization terms from the 2-loop
correction $\left.\Delta\lambda^\twoL\right|_{y_t^6}$ in
\figref{fig:xtscanrepara5tev3l} and \figref{fig:uncert_sources}.

Second, we discuss the impact of the 2-loop and 3-loop threshold
corrections in the new \feft\ calculation.  As can be seen in
\figref{fig:feft_comparison}, the impact of the higher-order
corrections is very small and below $0.3\GeV$ for all values of $\MS$
and $|x_t|<3$ in the shown scenarios.  The main reason is again the
$x_t$-resummation: When going from \oneL\ to \twoL, the \oneL\
calculation already contains the leading 2-loop QCD $x_t^5$ term of
\tabref{tab:xt-resum}, and the actual \twoL\ calculation only adds
subleading $x_t^{\le4}$ terms. Similarly, one can show that the \twoL\
calculation already correctly contains the leading 3-loop QCD $x_t^5$
and $x_t^6$ terms, and the actual \thrL\ calculation only adds
subleading $x_t^{\le4}$ terms.\footnote{The kink in the 3-loop line of
  the lower left plot around $\MS \approx 750\GeV$ is due to a
  hierarchy switch in the 3-loop calculation of \Himalaya.}

\subsection{Comparison to state-of-the-art calculations}
\label{sec:sota}
\begin{figure}
  \centering
  \includegraphics[width=0.48\textwidth]{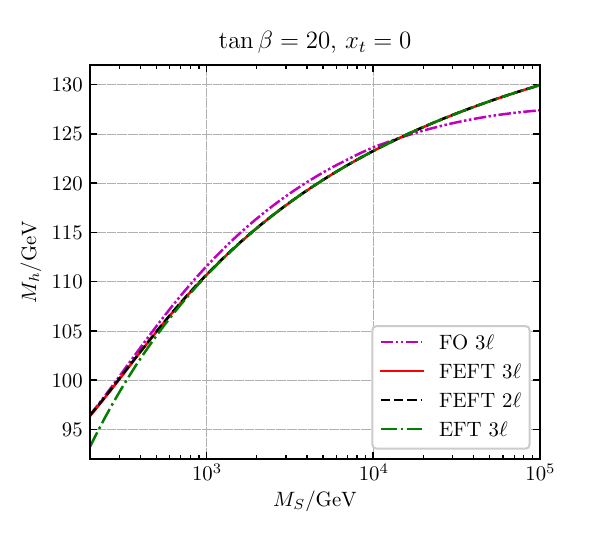} \hfill
  \includegraphics[width=0.48\textwidth]{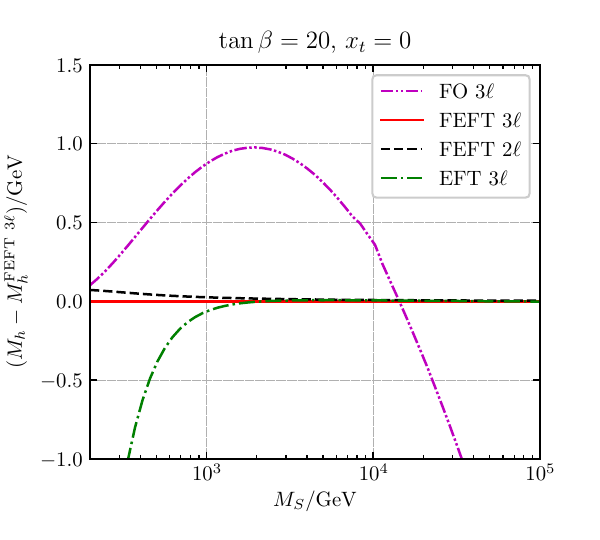}
  \includegraphics[width=0.48\textwidth]{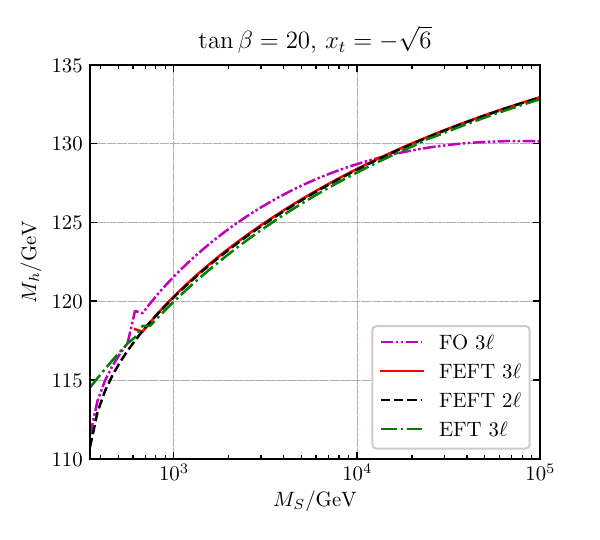} \hfill
  \includegraphics[width=0.48\textwidth]{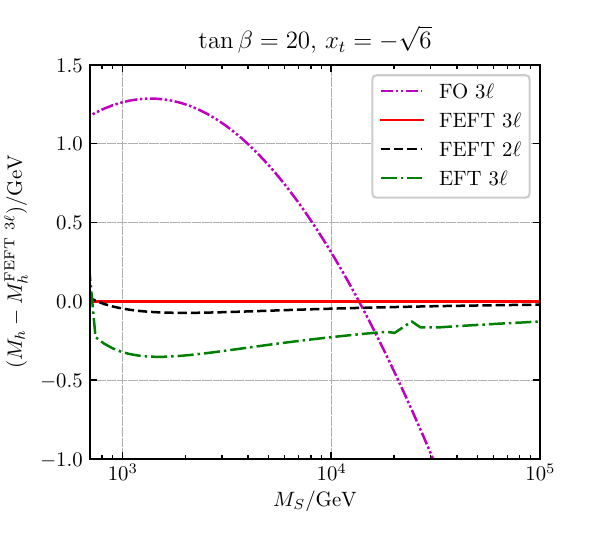}
  \includegraphics[width=0.48\textwidth]{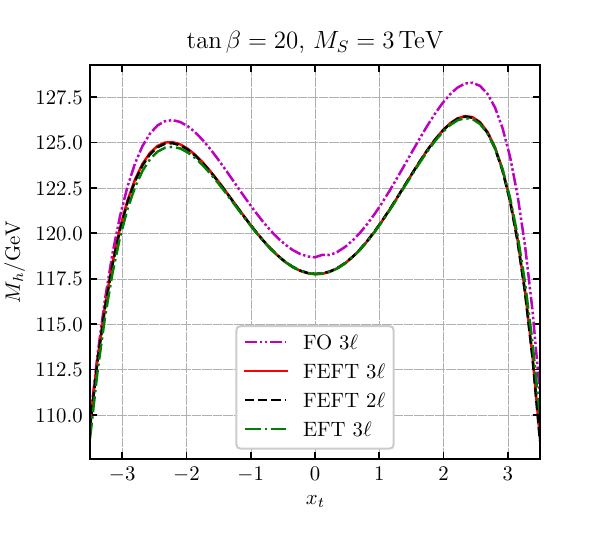} \hfill
  \includegraphics[width=0.48\textwidth]{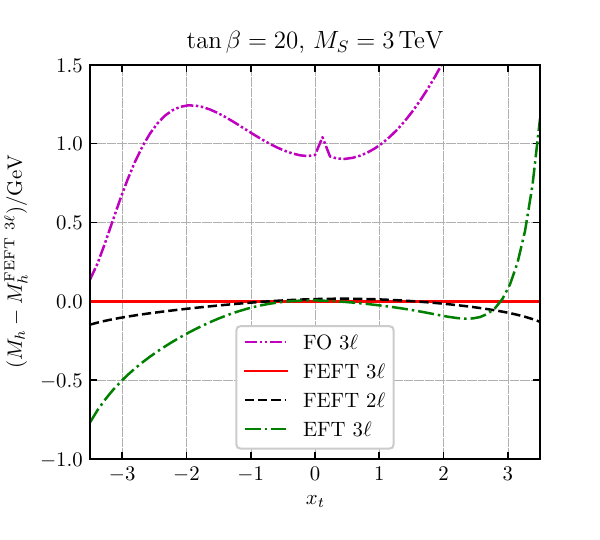}
  \caption{Prediction of the lightest CP-even Higgs pole mass in the
    MSSM as a function of $x_t$ and $M_S$ for $\tan\beta=20$.  In the
    left panels we show the absolute value of $M_h$ and in the right
    panels the difference to the \fefts\ \thrL\ prediction.}
  \label{fig:FO-EFT_compare}
\end{figure}

In this subsection we compare our new improved \feft\ calculation with
the two state-of-the-art 3-loop fixed-order and EFT calculations from
refs.~\cite{Harlander:2017kuc,Harlander:2018yhj}.  Both of these
calculations are also based on the \fs\ framework
\cite{Athron:2014yba,Athron:2017fvs}, which facilitates the
comparison.  In detail, these calculations are:
\begin{itemize}
\item \underline{FO \thrL}: This is the fixed-order calculation, which
  has been presented in ref.~\cite{Harlander:2017kuc}
  (dashed-double-dotted magenta line in \figref{fig:FO-EFT_compare}).
  It includes loop corrections to the Higgs pole mass in the full-model (MSSM)
  parametrization at full 1-loop level and 2- and 3-loop corrections
  in the gaugeless limit at
  $\ord(v^2(\twoloopfm))$ and
  $\ord(v^2(\thrloopfm))$, respectively.

\item \underline{EFT \thrL}: This calculation is the pure EFT
  calculation from ref.~\cite{Harlander:2018yhj}, where a matching at
  the SUSY scale is performed in the EFT (SM) parametrization (dashed-dotted
  green line in \figref{fig:FO-EFT_compare}).
  The threshold correction $\Delta\lambda$ includes the known 1-loop
  contributions from ref.~\cite{Bagnaschi:2014rsa}, 2-loop
  contributions at
  $\ord(\hat g_3^2(\hat y_t^4+\hat y_b^4) + (\hat y_t^2+\hat y_b^2
  +\hat y_\tau^2)^3)$ from
  ref.~\cite{Bagnaschi:2017xid} and 3-loop contributions at
  $\ord(\thrloopeft)$ from ref.~\cite{Harlander:2018yhj},
  all expressed in terms SM parameters.  Note, that this calculation
  neglects all suppressed $v^2/\MS^2$ terms.
\end{itemize}
Note, that these two 3-loop calculations take into account loop
corrections at the same orders as the presented new \fefts\ \thrL\
calculation, except for 2-loop terms suppressed by powers of
$y_b^n y_\tau^m$, which are only included in the pure EFT calculation.

The left column of \figref{fig:FO-EFT_compare} provides a first
overview of the behavior of the three calculations, for large and
small \MS\ and large and small $x_t$. The figure confirms the expected
behavior: For large $\MS$, the pure EFT \thrL\ calculation and the
hybrid \fefts\ \thrL\ calculation agree well, while the FO \thrL\
calculation deviates by several GeV. For small $\MS$, the hybrid
\fefts\ \thrL\ calculation agrees well with the FO \thrL\ calculation,
while the pure EFT \thrL\ calculation deviates by several GeV.  The
kinks in the 3-loop lines in the middle left plot around
$M_S\approx 750\GeV$ are due to a switch of the mass hierarchy in the
3-loop calculation of \Himalaya. 
The kink in the 3-loop FO prediction of the bottom right plot at small
$|x_t|$ originates from the occurrence of tachyonic running
\DRbarPrime\ Higgs boson masses at the electroweak scale, see the
discussion in ref.~\cite{Harlander:2019dge}.

The right column of \figref{fig:FO-EFT_compare} shows the differences
between the calculations in more detail. In the following we discuss
these differences.
We first focus on the differences between the new \feft\ and the pure
EFT calculation at SUSY scales above a few TeV. For such values of
$\MS$, the power-suppressed $v^2/\MS^2$-terms included in the \fefts\
hybrid calculation are numerically insignificant. Further investigations 
revealed that the
numerical impact of the mixed $\ord(y_t^4 y_\tau^2)$ threshold
corrections included in \fefts\ are of the order
$\Delta M_h \approx 10\MeV$ for $x_t=-\sqrt{6}$, $\tan\beta=20$ and
$M_S=2\TeV$ and the additional corrections in the EFT calculation 
$\ord(\hat y_b^4 \hat y_\tau^2 + \hat y_b^2 \hat y_\tau^4)$ have an impact
of $\Delta M_h \approx 10^{-3}\MeV$ and
 are thus negligible.  The essential difference between
the \fefts\ \thrL\ calculation (red line) and EFT \thrL\ (green
dashed-dotted line) is the different parametrization of the matching
in terms of either MSSM or SM parameters, and the resulting leading
higher-order $x_t$ terms included in \fefts\ \thrL.
The precise origin of this difference in the threshold correction
$\Delta\lambda$ can be inferred from \tabref{tab:xt-resum}.  The
\fefts\ \thrL\ calculation correctly includes all terms of the table,
while the EFT \thrL\ calculation only includes the terms of the left
column at most up to the 3-loop level, but neither includes the
4-loop term nor any term of the right column. As will be shown in the
following subsection, the numerically dominant effect comes from the
mixed QCD--EW 2-loop terms of the form
$(\lambda^{\text{\fefts\ \thrL}} - \lambda^{\text{EFT\ \thrL}})
\supset \hat y_t^4 \hat g_{1,2}^2 \hat g_3^2 x_t^3$.  As shown in the
middle row of \figref{fig:FO-EFT_compare}, the numerical difference
originating mainly from these terms remains around $200\MeV$ for
$\MS=100\TeV$ (and $x_t=-\sqrt{6}$).\footnote{%
  Since the couplings $\hat y_t$ and $\hat g_3$ are asymptotically
  free, the difference between the calculations does not approach a
  constant but shrinks slowly for higher $\MS$.}  On the other hand,
the lowest row of \figref{fig:FO-EFT_compare} shows that for fixed
$\MS$ the numerical difference is below $200\unit{MeV}$ for
$|x_t|\le2$ and $\MS=3\TeV$, but for larger $|x_t|$ the difference
rises strongly.

Secondly, we focus on the comparison between the fixed-order and the
\feft\ calculations for SUSY scales below around $1\unit{TeV}$, where
both calculations should be valid. By construction, both calculations
include the same Higgs pole mass contributions of the orders
$\ord(\oneL + v^2 (\twoloopfm) + v^2 (\thrloopfm))$,
 including terms suppressed by $v^2/\MS^2$. However, they differ
at other orders. Numerically, the difference is below $0.5\GeV$ for
small $x_t$ and $\MS \lesssim 500\GeV$ (see top row of
\figref{fig:FO-EFT_compare}), but the difference reaches around
$1\unit{GeV}$ for large $|x_t|$ and small $\MS$ (see middle row of
\figref{fig:FO-EFT_compare}). The origins of these differences are the
following:
\begin{itemize}
\item \underline{Parametrization:} In contrast to our hybrid approach,
  the determination of the \DRbarPrime\ MSSM top quark mass $m_t$ in
  the fixed-order calculation consists of the following expanded
  version of the exact relation
  \begin{align}\label{eq:FO_mt_match}
    m_t = M_t (1 + \Delta m_t^{\oneL} + \Delta m_t^{\twoL} ),	
  \end{align}
  where $\Delta m_t^{\oneL,\twoL}$ represent the 1- and 2-loop
  corrections to the \DRbarPrime\ top quark mass as described in
  refs.~\cite{Athron:2017fvs,Allanach:2018fif}.  Analogously to
  \secref{sec:resummation}, eq.~\eqref{eq:FO_mt_match} does \emph{not}
  represent an all order resummation of terms in the top mass
  parameter of $m_t \supset \hat m_t\times(\hat{g}_3^2 x_t)^n$.
  Consequently, eq.~\eqref{eq:FO_mt_match} does \emph{not} lead to an
  all order resummation of terms in the Higgs pole mass of the form
  \begin{align}
    \left(M_h^{\text{\fefts\ \thrL}}\right)^2 -
    \left(M_h^{\text{FO\ \thrL}}\right)^2
    \supset \hat m_t^2 \left(\hat y_t^2x_t^4 + \hat g_{1,2}^2x_t^2 \right)(\hat g_3^2 x_t)^{n}
  \end{align}
  for $n>2$.
  Besides these non-resummed terms, our new \feft\ hybrid calculation
  includes further incomplete higher-order contributions with high
  powers in $x_t$, which will be discussed in
  \secref{sec:uncertainty}.
\item \underline{Momentum iteration:} The double loop expansion in our
  Higgs pole-mass matching condition \eqref{eq:Higgs_pole_mass_match}
  made it necessary to strictly truncate the momentum iteration in
  order to avoid incomplete contributions, which could potentially
  spoil the resummation of the large logarithms.  The FO \thrL\
  calculation, however, does partially include higher-order effects by
  numerically solving eq.~\eqref{eq:BSM_pole_mass_definition} for
  $M_h^2$.  This includes non-logarithmic contributions, for example
  from the 2-loop electroweak sector and 3-loop top-Yukawa enhanced
  contributions of the form
  \begin{align}
    \left(M_h^{\text{FO\ \thrL}}\right)^2 -
    \left( M_h^{2,\text{\fefts\ \thrL}}\right)^2
    \supset m_t^2 \left( y_t^2 g_{1,2}^2 x_t^4+ y_t^4 g_3^2 x_t^6 + y_t^6 x_t^8\right).
  \end{align}
\item \underline{$\log$-resummation:} For low SUSY scales, the
  smallness of $\log(M_S/m_t)$ leads to a suppression of the resummed
  tower of large logarithms.  However, additional factors of $x_t$
  might counteract this effect, which potentially increase the
  relevance of the resummed logarithms, which are correctly included
  in the EFT-based approaches, such as
  \begin{align}
    \left(M_h^{\text{\fefts\ \thrL}}\right)^2 -
    \left(M_h^{\text{FO\ \thrL}}\right)^2
    \supset \hat m_t^2 \left( \hat y_t^2\hat g_{1,2}^2 x_t^2 + \hat y_t^4 \hat g_3^2 x_t^5 +\hat y_t^6 x_t^{6}\right) \log\frac{M_S}{\hat{m}_t}.
  \end{align}
\end{itemize}

\subsection{Further details on the comparison of hybrid and pure EFT calculations}
\label{sec:numfurtherdetails}
In the lower-right panel of \figref{fig:FO-EFT_compare} one can see a
deviation between the hybrid \fefts\ \thrL\ calculation and EFT \thrL\
for large $|x_t|$.  In the following we elaborate on the large-$x_t$ behavior in
more detail.

\begin{figure}[tbh]
  \centering
  \includegraphics[width=0.6\textwidth]{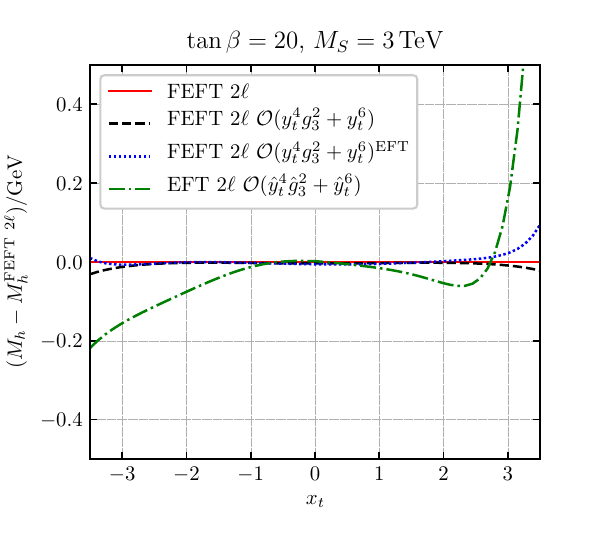}
  \caption{Comparison of our 2-loop hybrid approach (red line and
    black dashed line) to the 1-loop hybrid approach with 2-loop
    threshold corrections at $\ord(g_3^2 y_t^4+y_t^6)$ in the
    EFT-limit (blue dotted line) and to the pure EFT-calculation
    \HSSUSY\ with $\Delta \lambda$ included at
    $\ord(\hat g_3^2 \hat y_t^4+ \hat y_t^6)$ (green dashed-dotted line).}
  \label{fig:xtplothssusytopdown5tev}
\end{figure}

For the discussion it is sufficient to consider the 2-loop
calculations.  \Figref{fig:xtplothssusytopdown5tev} shows the Higgs
pole mass of different 2-loop calculations w.r.t.\ the \fefts\ \twoL\
calculation (red solid line).  The black dashed line corresponds to
the same 2-loop calculation, where 2-loop threshold corrections
proportional to powers of $y_b$ and/or $y_\tau$ have been omitted.
One finds that the difference between these lines is smaller than
$50\MeV$ for the shown parameter scenario.
The blue dotted line represents a modified calculation of the black
dashed line, where the 2-loop threshold correction to $\hat\lambda$ has
been replaced by the analytic 2-loop expressions from
eqs.~\eqref{eq:Delta_as*at^2} and \eqref{eq:Delta_at^3_alt}
of the order $\ord(g_3^2 y_t^4+y_t^6)$, where terms of
$\ord(v^2/\MS^2)$ have been neglected.  Thus, the difference between
the blue dotted and the black dashed lines corresponds to the impact
of some 2-loop higher-dimensional operators.  The effect of these
higher dimensional operators has been discussed in
ref.~\cite{Bagnaschi:2017xid}, where it has been shown that they are
of high relevance for large stop mixing.  For small $|x_t| \lesssim 3$
and the shown value $\MS = 3\TeV$, however, their effect is
negligible.
Note, that in \figref{fig:xtplothssusytopdown5tev} the hybrid 2-loop
result is subtracted from each calculation.  Hence, the blue dotted
line represents the negative correction due to power suppressed terms.
In contrast, figure~4 of ref.~\cite{Bagnaschi:2017xid} shows the
positive influence of higher dimensional operators.  From the figure
we draw the following conclusions:
\begin{itemize}
\item The excellent agreement between the black dashed and the blue
  dotted lines for $|x_t| \lesssim 3$ confirms numerically the
  correctness of our automatized \feft\ pole-mass matching procedure
  for $\hat\lambda$ at $\ord(g_3^2 y_t^4 + y_t^6)$.
\item For $|x_t| \gtrsim 3$ the effect of the higher-dimensional
  2-loop operators is in line with the numerical results of
  ref.~\cite{Bagnaschi:2017xid}.
\end{itemize}
For reference we also show in \figref{fig:xtplothssusytopdown5tev} the
EFT \twoL\ calculation, represented by the green dashed-dotted line.
One finds that EFT \twoL\ deviates numerically from \fefts\ \twoL\ for
$|x_t| \gtrsim 1$.  This discrepancy can be explained by contributions
originating from the different parametrization schemes.  As motivated
above, we categorize the higher-order corrections in two classes of
terms; the ones which are incomplete in both approaches and the ones
which are captured correctly in our full-model parametrization
scheme, but not in the other EFT  parametrization.

\begin{figure}[tbh]
  \centering
  \includegraphics[width=0.6\textwidth]{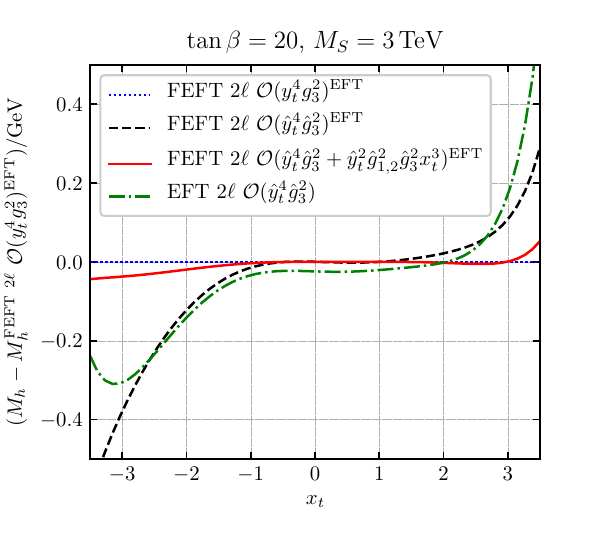}
  \caption{
  	Influence of contributions originating from
    reparametrization.  The plot shows the
    Higgs pole mass as predicted by different 1-loop calculations with
    additional 2-loop threshold contributions to $\Delta\lambda$ at
    $\ord(y_t^4g_3^2)$.  
   	The superscript ``EFT'' in the legend indicates that all 2-loop
    contributions are evaluated in the limit $v^2/\MS^2\to 0$.
    The blue dotted line represents the \feft\
    calculation in full-model parametrization.  The black dashed line
    represents the EFT-reparametrized calculation, truncated at
    $\ord(\hat y_t^4\hat g_3^2)$.  The red solid line corresponds to
    the black dashed line with the additional 2-loop electroweak
    $x_t^3$ contribution $\Delta\lambda^{\text{QCD--EW}}$ from
    \tabref{tab:xt-resum}.
    In the green dashed-dotted line we show the numerical results of the pure EFT calculation
  including 2-loop corrections to $\Delta\lambda$ in EFT parametrization 
  at $\ord(\hat y_t^4 \hat g_3^2)$.}
  \label{fig:xtplothssusytopdown5tevrep}
\end{figure}

Concerning the higher-order terms correctly captured by our new \feft\
hybrid calculation, we find the most dominant contribution to the
numerical difference between the EFT \twoL\ prediction and \fefts\
\twoL\ to be the 2-loop mixed QCD--EW term from \tabref{tab:xt-resum}.
To illustrate this effect we have created a reparametrized version of
the \fefts\ \twoL\ calculation in the EFT parametrization and compare
it with EFT \twoL\ in \figref{fig:xtplothssusytopdown5tevrep}.  The
figure shows different 2-loop calculations w.r.t.~\fefts\ \twoL,
where at 2-loop level only terms of $\ord(y_t^4g_3^2)$ in the
EFT-limit $v^2 \ll \MS^2$ are taken into account (blue dotted line).
The black dashed line represents the reparametrized version of the
blue dotted line, where $\hat\lambda$ is expressed in terms of SM
parameters.  In this calculation only 2-loop contributions of
$\ord(\hat y_t^4\hat g_3^2)$ are taken into account.  One finds that
this reparametrized calculation agrees well with the corresponding EFT
\twoL\ calculation (green dashed-dotted line), which uses the same
parametrization.  The only difference between the blue dotted and the
green dashed-dotted line are power suppressed contributions in the
Higgs mass at 1-loop, which become significant for $|x_t| \gtrsim 3$,
as discussed above.
When adding the 2-loop leading $x_t$ mixed QCD--EW contribution from
eq.~\eqref{eq:xt_pred} to the black dashed line, one obtains the red
solid line.  The so obtained result agrees very well with the
MSSM-parametrized calculation (blue dotted line), which explains the
dominant part of the deviation between the MSSM-parametrized \fefts\
\twoL\ calculation and the EFT-parametrized EFT \twoL\ calculation.
Thus, the numerical effect coming from the correct inclusion of
highest power $x_t$ contributions in our new \feft\ approach improves
the precision for large $|x_t|$ in comparison to the calculation
performed in the EFT parametrization.

\begin{figure}
  \centering
  \includegraphics[width=0.6\textwidth]{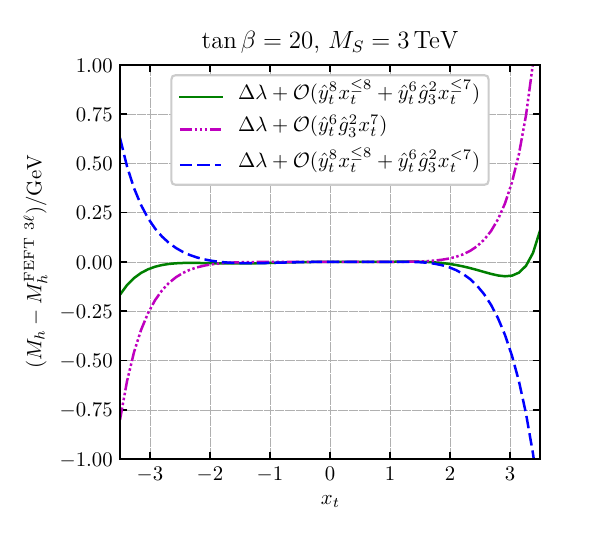}
  \caption{Impact of incomplete higher-order contributions to
    $\Delta\lambda$ from reparametrization in \feft.}
  \label{fig:xtscanrepara5tev3l}
\end{figure}

Besides the higher-order terms correctly taken into account by our new
\feft\ calculation, the threshold corrections $\Delta\lambda$ differ
in both approaches by further terms, which are incomplete both in
the full-model parametrization and in the EFT parametrization.  Such
incomplete higher-order terms are for example top Yukawa 
enhanced 3-loop terms with high $x_t$ powers of the form
$(\hat\lambda^{\text{\fefts\ \thrL}} - \hat\lambda^{\text{EFT\ \thrL}})
\supset \hat y_t^8 x_t^{\leq8} + \hat y_t^6 \hat g_3^2
x_t^{\leq7}$.\footnote{Note that in order to investigate the complete
  reparametrization contributions of this order, the inclusion of
  2-loop threshold corrections to $\Delta y_t$ at
  $\ord(y_t^5 +g_3^2y_t^3)$ is required.} 
The reparametrization of the 1-loop correction alone was discussed in 
\tabref{tab:xt-incomplete}. The discussion here is extended by the 
gauge-less 2-loop contributions to $\Delta \lambda$ in MSSM parametrization.
In
\figref{fig:xtscanrepara5tev3l} we show the numerical influence of
such terms.
When these (incomplete) higher-order terms are added coherently (green
solid line), both contributions almost cancel up to a remaining effect
of $\sim 150\MeV$ in the Higgs pole mass for $|x_t| < 3.5$ and
$M_S= 3\TeV$.  Thus, the numerical effect from the $x_t$-resummation
terms in $\Delta\lambda^{\text{QCD--EW}}$ from
\figref{fig:xtplothssusytopdown5tevrep} remains the dominant
reparametrization effect.  However, when the numerical effect of each
incomplete higher-order term is drawn individually, the contributions
have a higher impact on the Higgs pole mass, see
\figref{fig:xtscanrepara5tev3l}.  The magenta dashed-triple-dotted
line corresponds to the effect of the terms of
$\ord(\hat y_t^6\hat g_3^2 x_t^7)$ and the blue dashed line
corresponds to
$\ord(\hat y_t^8 x_t^{\leq8} + \hat y_t^6 \hat g_3^2 x_t^{<7})$.
There is a cancellation between these incomplete contributions, which
should be kept in mind when using such terms as an uncertainty
estimate of missing higher-order corrections.  Using the maximum
effect of all terms provides a more conservative estimate of the
remaining uncertainty than the coherent sum.

\section{Uncertainty estimation}
\label{sec:uncertainty}

In this section we analyze missing higher-order contributions in our
new \feft\ approach in order to estimate the remaining theory
uncertainty of our calculation.  In accordance with
refs.~\cite{Bagnaschi:2014rsa,Vega:2015fna,Allanach:2018fif} we
distinguish between missing higher-order contributions in the matching
at the SUSY scale, which we denote as \emph{high-scale uncertainty},
and missing loop corrections at the electroweak scale, denoted as
\emph{low-scale uncertainty}.  Note, that since \feft\ is a hybrid
calculation, we do not assign an \emph{EFT uncertainty} to our
calculation from missing terms of $\ord(v^2/\MS^2)$.%
\footnote{Our calculation of the Higgs mass does not include 
	suppressed logarithms beyond the 2-loop gaugeless limit.
	In fact, in ref.~\cite{Bagnaschi:2017xid} it has been demonstrated that 
their impact is very small $\Delta M_h \leq 20\MeV$ for
the studied scenarios.}

\subsection{High-scale uncertainty}
\label{sec:hs_uncert}

We begin our discussion by presenting our methods to estimate the
\emph{high-scale uncertainty}, i.e.\ the numerical impact of the
missing higher-order corrections in the matching of the MSSM
\DRbarPrime\ to the SM \MSbar\ parameters at the SUSY scale.  We
discuss three different approaches: the variation of the matching
scale, implicit higher-order corrections from the double loop
expansion and reparametrization terms.

\paragraph{Variation of the matching scale.}

A commonly applied strategy to estimate higher-order contributions is
to vary the renormalization scale $\Qmatch$ at which the 
threshold corrections are computed.  For reasons of comparability,
we use the conventional range of $\Qmatch\in[\MS/2,2\MS]$
and take the maximum deviation from the value obtained at
$\Qmatch=M_S$ as an estimate
\begin{align}
  \label{eq:match_uncer}
  \DMhQmatch = \max_{Q\in[M_S/2,2M_S]}\left\{|M_h(\Qmatch=\MS)-M_h(\Qmatch=Q)|\right\}.
\end{align}
The numerical variation of $M_h$ results from the fact that the
matching corrections contain explicit dependencies of $\log Q^2$ at
fixed order, while the RGE running cancels those logarithms but also
generates $\log Q^2$ terms at higher orders.  The quantity \DMhQmatch\
thus represents an estimate for these missing logarithmic higher-order
terms.
In particular, in the matching of $\hat\lambda$, the following 2-loop and
3-loop terms are generated:
\begin{align}
  \label{eq:match_uncer_terms}
  \DlQmatch &\supset\ \propto y_t^2 \left(g_3^2 g_{1,2}^2 +
  y_t^2 g_{1,2}^2 + g_{1,2}^4+ y_t^2g_3^4 x_t^4  + y_t^4 g_3^2 + y_t^6\right) +\ord(g_{1,2}^6).
\end{align}
The matching-scale variation thus provides an estimate of the theory
uncertainty related to these terms, at least to their $\log
Q^2$-dependent parts.
We have omitted the specification of the powers of $x_t$ in most
of the terms.  The term of $\ord(y_t^4 g_3^4 x_t^4)$ deserves special
attention: In the degenerate mass case the \Himalaya\ library up to
version 3.0.1 does not provide the correct term in the Higgs mass
correction at this order \cite{Harlander:2017kuc}.  Since this is an
important missing term of higher order in $x_t$, but not of higher
order in the couplings, we have verified that this missing term of
this order has a non-vanishing $\log Q^2$ dependence.  Indeed,
employing 2-loop $\beta$ functions from ref.~\cite{Martin:1993zk} on
the 2-loop Higgs pole mass, derived from the effective potential of
ref.~\cite{Espinosa:1999zm}, the renormalization scale dependence in
the degenerate mass case is given by
\begin{align}
  \label{eq:3L-xt4}
  \frac{\partial}{\partial \log Q^2} \Delta s_h^{\MSSM, y_t^4 g_3^4} 
  = \frac{g_3^4 y_t^4 v^2}{(4 \pi)^6} \frac{224}{9} \left(x_t^4\ + \propto x_t^{\leq 3}\right).
\end{align}
Thus, the
matching scale variation in our calculation provides an estimate of
the uncertainty originating from missing logarithmic terms at
$\ord(y_t^4 g_3^4 x_t^4)$ in particular.

We'd like to point out a technical difficulty in this matching scale
variation.  The evolution of RGEs in the MSSM requires the numerical
input values of MSSM \DRbarPrime\ parameters as a boundary condition.
However, in the MSSM two parameters cannot be fixed by the input;
rather they have to be eliminated by imposing the two electroweak
symmetry breaking conditions.  Solving these so-called tadpole
equations at the loop level will introduce logarithms which contain
light masses.  Hence, it is a legitimate question to ask whether such
contributions spoil the automatized cancellation of large logarithms
in the matching correction.
In our calculation, the tadpole
equations at the SUSY scale are solved for the dimensionful
soft-breaking Higgs-doublet mass parameters
$m_{H_u}^2$ and $m_{H_d}^2$. 
An explicit calculation up to leading 2-loop QCD order
shows that large logarithms enter into $\hat\lambda$ 
with a suppression of $v^2/M_S^2$ beyond the considered order.
These contributions would be absent in a pure EFT calculation 
and they can be regarded as a power-suppressed contribution in a 
hybrid calculation.

\paragraph{Implicit corrections at higher order.}

In \secref{sec:expansion_Master_eq} we discussed the expansion of the
master formula \eqref{eq:lambda_from_Mh} and explained how ``explicit''
contributions from genuine multi-loop diagrams are accompanied by ``implicit''
corrections in the double loop expansion, i.e.\ from the
reparametrization of the SM self-energy in terms of MSSM
parameters. These implicit corrections have the form of products of
derivatives of the SM Higgs pole mass shift $\Delta s_h^{\SM}$  times parameter shifts.

Hence, as another estimate of
missing higher-order corrections, we compute further  terms with such
a structure at orders beyond
the precision of the included threshold corrections and discard terms
which contain logarithms of the form $\log(m_t/Q)$.  The resulting
contributions take the form
\begin{align}
  \label{eq:3l_implicit}
  \frac{1}{\hat v^2}\left[\sum_{P} \left(\frac{\partial}{\partial P} \Delta s_h^{\SM}\right) \Delta P \right]_{\log(m_t/Q)=0}
   = \DltwoEW + \DlthrGL,
\end{align}
where $\DltwoEW $ denotes terms which would arise in an actual 2-loop calculation beyond the gaugeless limit, and
$\DlthrGL$ contains terms which would arise in an actual 3-loop
calculation in the gaugeless limit. The corresponding orders in
couplings are
\begin{align}
  \DltwoEW &\supset\ \propto g_{1,2}^2\left[g_3^2y_t^2 x_t^{\leq 1}+ y_t^4 x_t^{\leq 4} + y_t^2g_{1,2}^2 x_t^{\leq 2} +g_{1,2}^4 \right],
  \label{eq:2l_gauge_uncer}
  \\
  \DlthrGL &\equiv \Dlthr_{ g_3^4 y_t^4 x_t^{\leq 2} } + \Dlthr_{ g_3^2 y_t^6 x_t^{\leq 5}, y_t^8 x_t^{\leq 8}} \,.
  \label{eq:3l_gaugeless}
\end{align}
The 3-loop gaugeless contributions contained in the generated terms on
the r.h.s.\ of eq.~\eqref{eq:3l_gaugeless} are of the order as
indicated in the subscript.

We can thus first define an estimate of the size of the missing 2-loop electroweak SUSY
corrections as
\begin{align}
  \label{eq:DMhtwEW}
  \DMhEW = \left| M_h(\Delta\lambda^{\thrL}) - M_h(\Delta\lambda^{\thrL} + \DltwoEW) \right|,
\end{align}
where $M_h(\Delta\lambda^{\thrL})$ denotes the \fefts\ \thrL\
calculation.
Next, we can define an estimate of the size of missing higher-order
SUSY-QCD contributions as 
\begin{subequations}
	  \label{eq:DMhtwoQCD}
\begin{align}
  \DMhtwoQCD &= \left| M_h(\Delta\lambda^{\twoL}) - M_h(\Delta\lambda^{\twoL} + \Dlthr_{g_3^4 y_t^4 x_t^{\leq 2} } + \Dlthr_{g_3^2 y_t^6 x_t^{\leq 5}, y_t^8 x_t^{\leq 8}}) \right|, \\
  \DMhthrQCD &= \left| M_h(\Delta\lambda^{\thrL}) - M_h(\Delta\lambda^{\thrL} + \Dlthr_{g_3^2 y_t^6 x_t^{\leq 5}, y_t^8 x_t^{\leq 8}}) \right|,
  \label{eq:DMhthrQCD}
\end{align}
\end{subequations}
of the \fefts\ \twoL\ and \thrL\ calculations, respectively.  Note,
that for the uncertainty estimate of the \fefts\ \thrL\ calculation
\eqref{eq:DMhthrQCD}, we do not use the 3-loop terms
$\Dlthr_{g_3^4 y_t^4 x_t^{\leq 2}}$, since they are already included
in the known 3-loop threshold corrections $\Delta\lambda^{\thrL}$ at
$\mathcal O(y_t^4g_3^4)$.

Note, that since the derivatives of the SM self-energy do not depend
on the MSSM parameters, the $x_t$ dependence of the terms contained in
eq.~\eqref{eq:3l_implicit} is only introduced by the shift $\Delta P$.
This is the reason for the particular maximum powers of $x_t$ which 
appear in eqs.~\eqref{eq:2l_gauge_uncer} and \eqref{eq:3l_gaugeless}. 
In particular, at the order $g_3^2y_t^6$, these uncertainty estimates
only contain terms up to $x_t^5$, while the true threshold correction 
at this order is allowed to contain $x_t^6$.  Hence the method of implicit 
corrections cannot reliably estimate the influence of the highest-power
$x_t$ contributions.

\paragraph{Reparametrization terms.}

For the reasons discussed in the previous sections, we chose to
express the threshold corrections in terms of MSSM parameters.  When
computed at all orders in perturbation theory, both the full-model and the
EFT parametrization do not differ by definition.  Hence,
it is possible to estimate the uncertainty of missing higher-order 
contributions by the numerical difference of the Higgs mass prediction
in both parametrizations. The full-model parametrization
is preferred because at some finite order in MSSM parameters
it already resums highest power 
$x_t$ corrections of QCD-enhanced orders in SM parameters.
However, we can use reparametrization to generate terms of orders
which are missing or incomplete in our calculation. Specifically,
already in
\secref{sec:numfurtherdetails}, in the context of
\figref{fig:xtscanrepara5tev3l}, such reparametrization terms 
of the orders
\begin{align}
  \Dlrep \supset \hat y_t^8 x_t^{\leq 8} + \hat y_t^6 \hat g_3^2 x_t^{\leq 7} 
  \label{eq:3l_reparam}
\end{align}
were discussed.
In contrast to the implicit corrections, reparametrization generates
terms of highest order in $x_t$ which can appear in the true threshold
correction, and the reparametrization terms in
eq.~\eqref{eq:3l_reparam} can thus more reliably estimate the 
influence of missing highest-power $x_t$ contributions.
Because of the nature of reparametrization, this method
also estimates missing higher-order terms in the threshold corrections
$\Delta y_t$ and $\Delta g_3$.
For later discussion of the size of the reparametrization terms of
eq.~\eqref{eq:3l_reparam}, we define the following uncertainty
estimates,
\begin{subequations}
\begin{align}
  \DMhrepQCD &= \left| M_h(\Delta\lambda^{\thrL}) - M_h(\Delta\lambda^{\thrL} - \ord(\hat y_t^6 \hat g_3^2 x_t^7)) \right|, \\
  \DMhrepHiggs &= \left| M_h(\Delta\lambda^{\thrL}) - M_h(\Delta\lambda^{\thrL} - \ord(\hat y_t^6 \hat g_3^2 x_t^{<7} + \hat y_t^8 x_t^{\leq 8})) \right|,
  \label{eq:DMh_rep}
\end{align}
\end{subequations}
where we subtract the reparametrization terms from the \fefts\ \thrL\
calculation in order to reproduce the truncation of the EFT parametrization of $\hat \lambda$ at $\order{\oneL + \twoloopeft +\thrloopeft}$.  
Up to a sign,
the dashed-triple-dotted magenta line and the dashed blue line in
\figref{fig:xtscanrepara5tev3l}  shows equivalently
the numerical influence of the terms estimated by $\DMhrepQCD$ 
and $\DMhrepHiggs$.

At this point it is worthwhile to discuss the difference in the
estimation of the uncertainty of an EFT-parametrized calculation at
similar order, i.e.\ with a matching of $\hat \lambda$ at
$\ord(1\ell + \hat g_3^2(\hat y_t^4+\hat y_b^4) + (\hat y_t^2+\hat
y_b^2 +y_\tau^2)^3 +\thrloopeft)$.  The reparametrization provides a
way to estimate higher-order terms in this calculation, which are
sensitive to high powers of $x_t$.  Furthermore, the uncertainty
estimation should also cover terms of
$\ord(\hat y_t^8, \hat y_t^6 \hat g_3^2)$, which are incomplete in
both parametrizations, c.f.\ \tabref{tab:xt-incomplete}.
Consequently, if the discussed techniques are applied to construct
higher-order terms for the uncertainty estimation of the
EFT-parametrized calculation, we expect that they lead to very similar
expressions for
$\Delta \lambda \supset \hat y_t^8 x_t^{\leq 8}+\hat y_t^6 \hat g_3^2
x_t^{\leq 7}$.  Note, that in contrast to the full-model-parametrized
calculation, the EFT-parametrized one would in addition have to
estimate the size of the terms of
$\Delta \lambda \supset \hat y_t^2 \hat g_1^2 \hat g_3^2 x_t^3
+\hat y_t^2  \hat g_2^2 \hat g_3^2 x_t^3 + \hat y_t^4
\hat g_3^6 x_t^7$, which are implicitly captured in full-model
parametrization.  Thus, in EFT parametrization more higher order
contributions would be needed to estimate the uncertainty for large
$|x_t|$.

\subsection{Low-scale uncertainty}
\label{sec:LS_uncer}

In this section we describe our method to estimate the \emph{low-scale
  uncertainty}, i.e.\ the theory uncertainty from missing higher-order
loop corrections in the matching to the SM input parameters at the
electroweak scale.  We consider two different approaches: the
variation of the renormalization scale of the Higgs pole mass
calculation and the variation of loop orders in the determination of
the top Yukawa coupling.

\paragraph{Variation of the pole mass scale.}

First we discuss the variation of the renormalization scale at which
the pole mass $M_h$ is computed in the SM.  By default the scale
$\Qpole=M_t$ is chosen, which we vary by factor of two,
\begin{align}
  \DMhQpole = \max_{Q\in[M_t/2,2M_t]}\left\{|M_h(\Qpole=M_t)-M_h(\Qpole=Q)|\right\}.
  \label{eq:DMhQpole}
\end{align}
This procedure estimates the impact of missing logarithmic
higher-order corrections to the Higgs pole mass shift in the SM.

\paragraph{Variation of the loop order of threshold corrections at the low scale.}

As described in \secref{sec:matching_low_scale}, the relation between
low-energy observables and \MSbar-renormalized SM couplings contains
corrections that can be switched off in the calculation without
reducing the precision of the result for $M_h$.  As was shown in
refs.~\cite{Bagnaschi:2014rsa,Vega:2015fna,Allanach:2018fif,Bahl:2019hmm},
the dominant uncertainty obtained from
this procedure is driven by the higher-order threshold correction in
the relation between the top quark pole mass and the top Yukawa
coupling.  We define our estimation of missing threshold corrections
at the electroweak scale in accordance with that reference as
\begin{subequations}
\label{eq:uncer_ls_Mt}
\begin{align}
  \DMhtwoYt &= \left|M_h^{y_t,\twoL} - M_h^{y_t,\thrL} \right|,\\
  \DMhthrYt &= \left|M_h^{y_t,\thrL} - M_h^{y_t,\fourL} \right|,
\end{align}
\end{subequations}
where the superscript of the symbols $M_h^{y_t,\nL}$ indicates that
eq.~\eqref{eq:mt_MSbar} is evaluated at $n$-loop level.  Since the
consistent resummation of NNLL/\NCLL logarithms requires an evaluation
of eq.~\eqref{eq:mt_MSbar} at 2-/3-loop level, we estimate the
uncertainty of the \fefts\ \twoL/\thrL\ calculation by
$\DMhtwoYt$ and $\DMhthrYt$, respectively.

\subsection{Numerical size of individual uncertainties}

\begin{figure}
  \centering
  \includegraphics[width=0.49\linewidth]{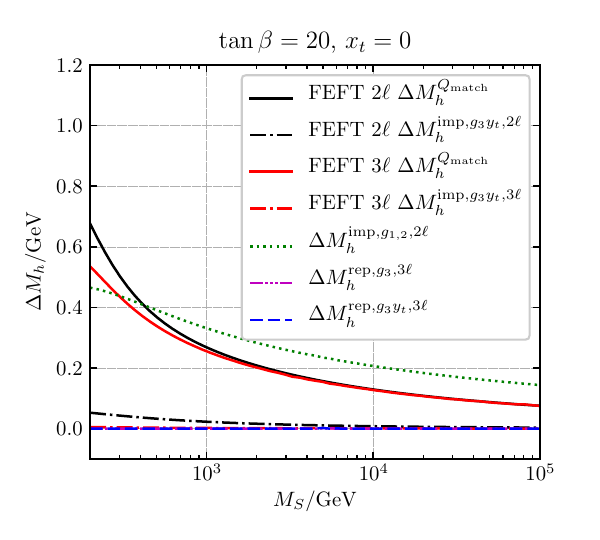} \hfill
  \includegraphics[width=0.49\linewidth]{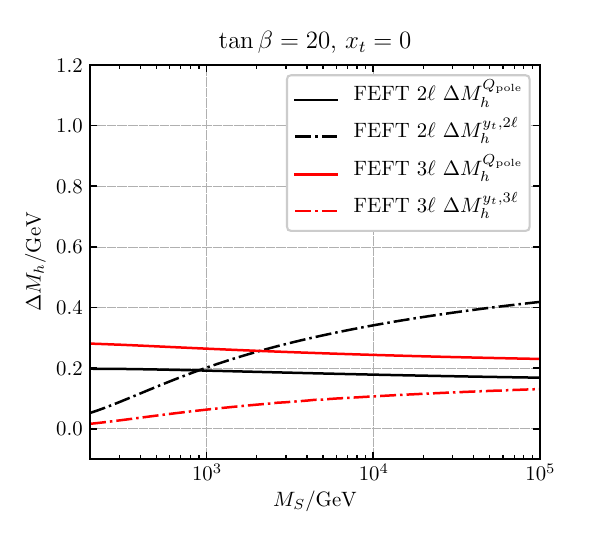}
  \includegraphics[width=0.49\linewidth]{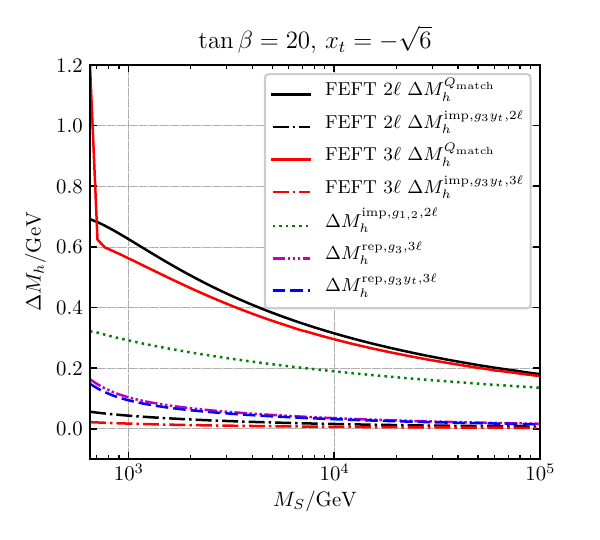}\hfill
  \includegraphics[width=0.49\linewidth]{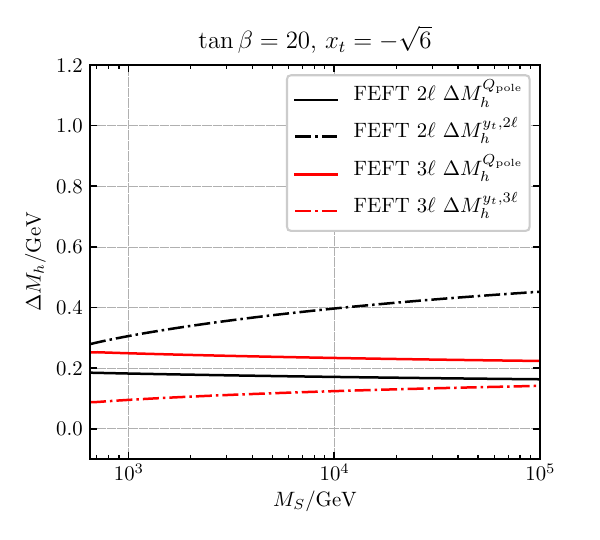}
  \includegraphics[width=0.49\linewidth]{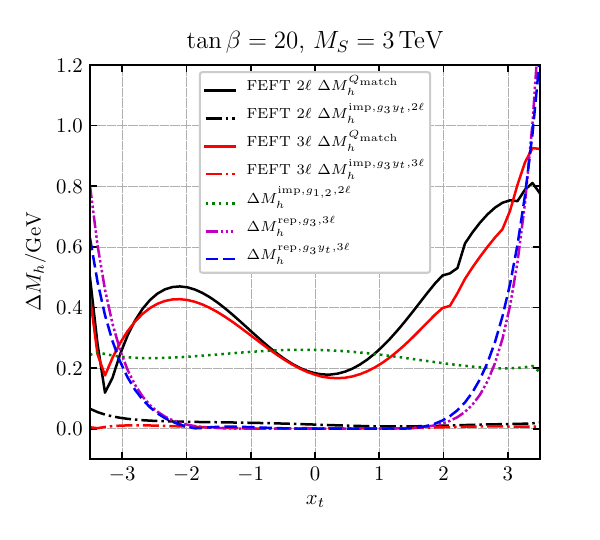}\hfill
  \includegraphics[width=0.49\linewidth]{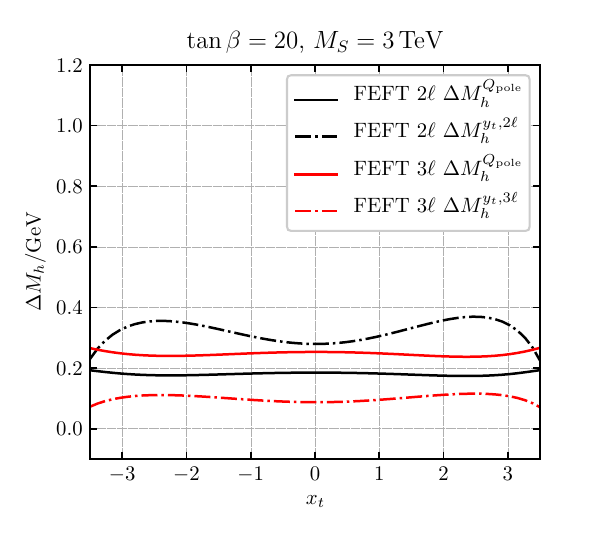}
  \caption{Individual contributions to the high-scale uncertainty
    (left column) and to the low-scale uncertainty (right column).}
  \label{fig:uncert_sources}
\end{figure}
In \figref{fig:uncert_sources} we show the individual sizes of the
uncertainty estimates discussed above for the parameter scenarios from
\figref{fig:FO-EFT_compare}.  The two black lines correspond to
uncertainties for \fefts\ \twoL\ and the other lines correspond to
\fefts\ \thrL.

\subsubsection{High-scale uncertainty}

We start with a discussion of the high-scale uncertainty, shown in the
left column of \figref{fig:uncert_sources}.

\paragraph{Estimate of missing 3-loop QCD and $y_t$-enhanced
  contributions beyond $\ord(y_t^4g_3^4x_t^{\leq 4})$.}
\label{sec:est_HS_uncer}
In \figref{fig:uncert_sources}, the black and red solid lines
represent the matching-scale uncertainties $\DMhQmatch$ of the 2- and
3-loop \fefts\ calculation, respectively.  The matching-scale
uncertainty provides a global estimate of many kinds of terms, see
eq.~\eqref{eq:match_uncer_terms}. The \emph{difference} between the black
and red solid lines corresponds to the inclusion
of the known leading-QCD 3-loop contributions of $\ord(y_t^4g_3^4x_t^{\leq 3})$ to
$\Delta\lambda$. We find that this inclusion reduces the uncertainty very little,
less than $0.2\GeV$ for all studied scenarios.
In particular, since terms of the order $\ord(y_t^4g_3^4x_t^4)$ are
not known for all parameter scenarios, we expect a remaining
uncertainty of significant size for large $|x_t|$ (see lower left
panel of \figref{fig:uncert_sources}).  Note, that $\DMhQmatch$ is
sensitive to terms of $\ord(y_t^4g_3^4x_t^4)$ (c.f.\
eq.~\eqref{eq:3L-xt4}) and thus includes an estimate of these missing
terms.

To provide a direct estimate the size of the missing non-logarithmic 3-loop QCD and
$y_t$-enhanced contributions, we show as black and red dashed-dotted
lines the uncertainties $\DMhtwoQCD$ and $\DMhthrQCD$ for the 2- and
3-loop \fefts\ calculations, respectively.  We find that these
QCD uncertainties are very small already for the 2-loop calculation,
$\DMhtwoQCD \lesssim 0.1\GeV$.  This is fully in line with the small
\emph{difference} between the 2-loop and 3-loop matching-scale
uncertainty described above.
The 3-loop QCD and $y_t$-enhanced corrections missing in \fefts\
\thrL, $\DMhthrQCD$, including terms with fewer powers of $g_3$, are
found to be negligible (red dashed-dotted line).

Taken together, all these results provide strong evidence
that the contributions of leading QCD-type are already very well under
control and inclusion of higher-order
leading-QCD threshold corrections of $\ord(y_t^4g_3^6)$ will not improve the
precision of the calculation significantly.
	
\paragraph{Importance of 2-loop electroweak contributions.}
The size of the missing 2-loop electroweak SUSY contributions to
$\Delta\lambda$ are estimated by the matching-scale variation,
$\DMhQmatch$, and more directly by the generated implicit contributions, $\DMhEW$,
defined in \secref{sec:hs_uncert}.
The implicit contributions, shown as green dotted line in
\figref{fig:uncert_sources}, have a sizable numerical effect of
$\DMhEW \lesssim 0.4\GeV$.  Further investigations of the induced
terms from eq.~\eqref{eq:2l_gauge_uncer}
indicate that the terms of $\ord(y_t^4 g_{1,2}^2)$ are typically
dominant for the parameter scenarios studied here.  For vanishing stop
mixing, $\DMhEW$ is of the same order as $\DMhQmatch$.  More
precisely, the offset of the solid lines, i.e.\ the smallest value of
$\DMhQmatch$ at $x_t\approx -3$, in the lower left panel of
\figref{fig:uncert_sources} is of the same magnitude as the almost
constant green dotted line.  For $x_t=0$ and $\MS>1\TeV$, both
$\DMhQmatch$ and $\DMhEW$ predict that the uncertainty decreases at
the same rate when going to higher $M_S$.\footnote{The middle-left
  plot in \figref{fig:uncert_sources} shows a numerical instability at
  $M_S< 750\GeV$ in the \fefts\ \thrL\ calculation due to a hierarchy
  switch in \Himalaya.  The kink in the curves for $\DMhQmatch$ in the
  lower left panel at $x_t \approx -3.2$ is due to a numerical
  artifact of our definition of the uncertainty.  The irregularities
  at $x_t\approx 2$ in the middle-left plot are due to a numerical
  instability in our code, which is absent for lower values of
  $\tan\beta$.}

This indicates that missing electroweak 2-loop terms contribute a
theory uncertainty which is typically around $0.2$--$0.3\GeV$, has a weak $x_t$-dependence, and which is the
dominant theory uncertainty for small $|x_t|$.

\paragraph{Relevant higher-order contributions for large $|x_t|$.}
For large $|x_t|\gtrsim2$  the matching-scale uncertainty $\DMhQmatch$
is larger than for small $x_t$. This cannot be attributed exclusively to the
missing leading-QCD and 2-loop electroweak terms discussed so far.
As discussed in \secref{sec:hs_uncert}
and ref.~\cite{Athron:2016fuq}, this is not unexpected because of the low
powers of $x_t$ appearing in $\DMhthrQCD$. On the other hand, the
increased uncertainty for large $x_t$ is in line with the discussion of
the impact of non-resummed large-$x_t$ contributions in
\secref{sec:numfurtherdetails}.
In order to estimate missing terms
with high $x_t$-dependence, we employ the uncertainty estimates
based on
reparametrization terms. Indeed, reparametrization terms $\DMhrepQCD$
(magenta dashed-triple-dotted 
line) and $\DMhrepHiggs$ (blue dashed line) in
\figref{fig:uncert_sources} do contain the maximal powers of $x_t$ at
their respective loop order.  In fact, the combinations
$\DMhEW + \DMhrepQCD$ and $\DMhEW + \DMhrepHiggs$ are of the order of
$\DMhQmatch$, see the middle panel in the left column of
\figref{fig:uncert_sources}.  This suggests that both electroweak and
QCD and $y_t$-enhanced terms with high powers of $x_t$ are the
dominant source of uncertainty for large stop mixing,
which must be brought under control to reduce the high-scale
uncertainty further.

For $|x_t|\approx 3$, the uncertainty estimate obtained from
reparametrization becomes dominant.  However, this is not specific to
performing the calculation in full-model parametrization and it cannot
be interpreted as an indication that the EFT parametrization would
perform better with regard to missing contributions at
$\order{\hat y_t^8 +\hat y_t^6 \hat g_3^2}$.  In fact, as discussed at
the end of \secref{sec:hs_uncert}, the estimation of the uncertainty
for a calculation performed in SM parametrization would lead to a
similar result at these orders.

\subsubsection{Low-scale uncertainty}
\label{sec:low-scale_uncertainty}

Now we discuss the size of the low-scale uncertainty as defined by the
measures in \secref{sec:LS_uncer}.  The individual sources of the
low-scale uncertainty are shown in the right column of
\figref{fig:uncert_sources}.  The variation of the pole mass scale,
$\DMhQpole$, is shown by the solid lines for \fefts\ \twoL\ (black
solid line) and \fefts\ \thrL\ (red solid line).  We find excellent
agreement of the pole mass uncertainty of \fefts\ \twoL\ with the
corresponding result shown in figure~3 of ref.~\cite{Allanach:2018fif}.
Concerning the \fefts\ \thrL\ calculation we find a larger uncertainty
of $\DMhQpole$ than the corresponding \fefts\ \twoL\ calculation,
which is surprising at first sight.  The reason for this is the
inclusion of the 3-loop Higgs pole mass shift in the SM of
$\ord(\hat v^2 \hat y_t^4 \hat g_3^4)$ from
ref.~\cite{Martin:2014cxa}, which has the particular property that it
increases the sensitivity of the Higgs pole mass on renormalization
scale, if the scale is varied within $\Qpole\in[M_t/2, 2M_t]$.
However, if the scale $\Qpole$ is varied within a larger range, the
inclusion of this 3-loop correction leads to a significantly reduced
dependence of the Higgs pole mass on $\Qpole$.  In order to keep our
results comparable with the literature, we stick to the convention of
using the $\Qpole\in[M_t/2, 2M_t]$.  As a result, we find
$\DMhQpole \lesssim 0.3\GeV$ for \fefts\ \thrL\ in the shown parameter
scenarios.

Our second measure to estimate part of the low-scale uncertainty is
given by the influence of higher-order correction in the relation
between the Yukawa coupling and the pole mass of the top quark,
$\DMhnYt$, defined in eqs.~\eqref{eq:uncer_ls_Mt}.  The uncertainties
$\DMhtwoYt$ and $\DMhthrYt$ of the \fefts\ \twoL\ and \thrL\
calculations are shown as black and red dashed-dotted lines in the
right column of \figref{fig:uncert_sources}, respectively.  Again, by
comparing the uncertainty $\DMhtwoYt$ with the corresponding result
from figure~3 of ref.~\cite{Allanach:2018fif}, we find excellent
agreement.  Compared to the \fefts\ \twoL\ calculation, the \fefts\
\thrL\ calculation has a strong reduction of the uncertainty with
$\DMhthrYt \lesssim 0.2\GeV$.  This is the main source of the improved
precision of our 3-loop calculation of $M_h$ in the studied scenarios.

\subsection{Combined Uncertainty}

In this subsection we combine the individual uncertainty estimates
presented in the previous subsections to obtain a total uncertainty
estimate of our new 2-loop and 3-loop \feft\ calculations.  Since the
individual uncertainty estimates at the high- and low-energy scales
are sensitive to an overlap of higher-order terms, we define the
following combined high-scale uncertainty, $\DMhHS$, and low-scale
uncertainty, $\DMhLS$, for the \fefts\ $\nL$ calculation:
\begin{subequations}
\begin{align}
  \DMhHS &= \max\left\{\DMhQmatch, \DMhlam\right\}, \label{eq:DMhHS}\\
  \DMhLS &= \max\left\{\DMhQpole , \DMhnYt\right\}. \label{eq:DMhLS}
\end{align}
\end{subequations}
In eq.~\eqref{eq:DMhHS}, $\DMhlam$ refers to the following combination
of our different approaches of generating higher-order terms in
$\hat\lambda$ as described in \secref{sec:hs_uncert},
\begin{align}
  \DMhlam = \DMhEW + \max\left\{
  \DMhnQCD, \DMhrepHiggs, \DMhrepQCD \right\} .
  \label{eq:DMhlam}
\end{align}%
Since the uncertainty estimates $\DMhnQCD$, $\DMhrepHiggs$ and
$\DMhrepQCD$ are sensitive to an overlap of higher-order contributions
to $\hat\lambda$ that involve terms of $\ord(y_t^ng_3^m)$, we take
their maximum in eq.~\eqref{eq:DMhlam}.  On the other hand, the
electroweak contributions $\DMhEW$ are an independent subset of
higher-order terms that involve electroweak gauge couplings, so we add
it linearly to the other terms in eq.~\eqref{eq:DMhlam}.  To obtain
the total combined uncertainty, $\Delta M_h$, of our calculations, we
add the high-scale and low-scale uncertainties linearly,
\begin{align}
  \DMh &= \DMhHS + \DMhLS.
\end{align}
\begin{figure}[tb]
  \centering
  \includegraphics[width=.49\linewidth]{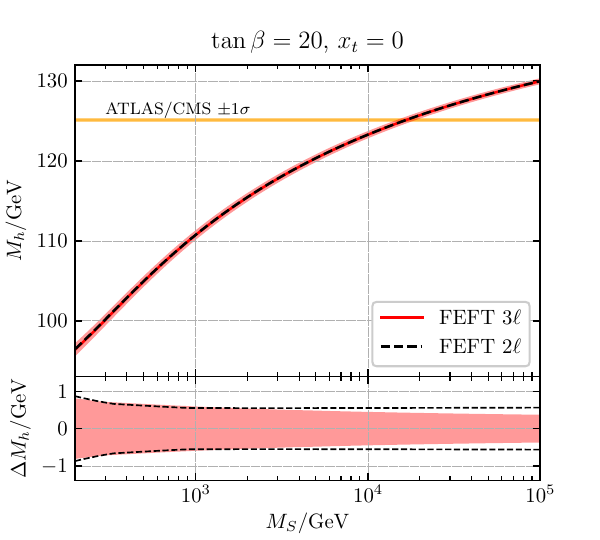}
  \hfill
  \includegraphics[width=0.49\linewidth]{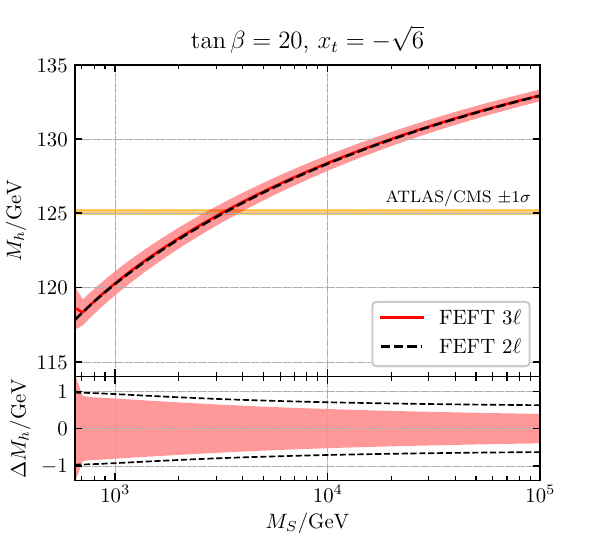}
  \includegraphics[width=0.49\linewidth]{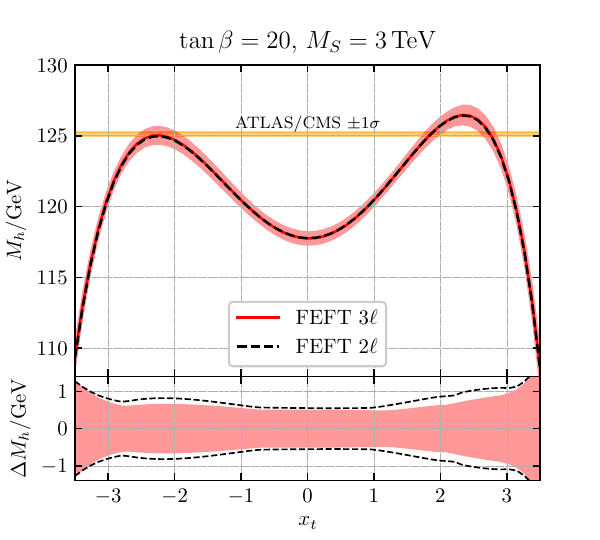}\hfill
  \caption{Light CP-even Higgs pole mass predictions with \fefts\
    \twoL/\thrL\ including the combined uncertainty estimates.  The
    orange band represents the experimentally measured value of the
    Higgs mass, $M_h = (125.10 \pm 0.14)\GeV$,
    including the experimental uncertainty.}
  \label{fig:uncert_all}
\end{figure}%
For the degenerate SUSY mass scenarios defined in
\secref{sec:numerical_results}, the results of our combined
uncertainty estimates are shown in \figref{fig:uncert_all}.  The red
solid line represents the Higgs pole mass $M_h$ obtained with the
\fefts\ \thrL\ calculation and the red band in the lower sub-plots
denotes the corresponding combined uncertainty $\DMh$.  The
black dashed lines correspond to the \fefts\ \twoL\ calculation
accordingly.
The difference between the \fefts\ \thrL\ and \twoL\ calculations is
of the order $|M_h^{\thrL} - M_h^{\twoL}| \lesssim 0.3\GeV$.
Compared to the 2-loop calculation, we find a more pronounced decrease
of the uncertainty of the 3-loop calculation for large stop mixing
$|x_t| \sim 2$ and $M_S \gtrsim 5\TeV$.  The dominant reduction of the
total uncertainty of the 3-loop calculation is achieved in the
low-scale uncertainty $\DMhLS$, where $\DMhtwoYt$ is the dominant
uncertainty of the 2-loop calculation.

In general we find for the studied degenerate \DRbarPrime\ SUSY mass
parameter scenarios a combined uncertainty of the \fefts\ \thrL\
calculation of $\DMh \lesssim 1\GeV$ for $\MS \gtrsim 1\TeV$ and
$|x_t| \lesssim 3$.%
\footnote{Note, that $\DMh$ is a measure of missing higher-order
  corrections in the relation between the predicted light CP-even
  Higgs pole mass and the \DRbarPrime\ input parameters.  As was
  stressed in ref.~\cite{Bahl:2019hmm}, there are additional
  uncertainties when the \DRbarPrime\ input parameters are related to
  other physical observables.}
This combined uncertainty becomes smaller for $|x_t| \to 0$ and larger
$\MS$, where it can reach $\DMh \sim 0.5\GeV$.  These findings are
compatible with the uncertainty estimates of
refs.~\cite{Bahl:2019hmm,Harlander:2019dge}, where hybrid calculations
with a comparable precision were studied.
For large SUSY scales of $\MS\gtrsim 5\TeV$ we find that the remaining
uncertainty of the \fefts\ \thrL\ calculation is dominated by the
low-scale uncertainty induced by the determination of the top Yukawa
coupling and the electroweak part of the high-scale uncertainty, which
can be of similar size.

\section{Conclusions}

We have presented an extension of the \feft\ method to calculate the
SM-like Higgs pole mass in the MSSM.  The
method combines the virtues of an 
EFT and fixed-order calculation, resulting in a prediction that
includes power-suppressed corrections and a resummation of large
logarithms.
We have applied our method to perform a state-of-the-art calculation
of the light CP-even Higgs pole mass in the MSSM, including
corrections up to the 3-loop level and resummation of large
logarithmic corrections up to \NCLL.

The key of our extension is the generation of a consistent automatized
pole mass matching procedure beyond the 1-loop level.  The consistency
of the \feft\ method in this regard refers the cancellation of large
logarithmic loop corrections and the inclusion of power-suppressed
contributions in the matching to the EFT (assumed to be the Standard
Model), thereby avoiding problems of double counting.
Conceptually, this was achieved by a paradigm shift where the usually
applied EFT-parametrized formulation of the high-scale matching was
replaced by a parametrization in terms of full-model (MSSM)
parameters. Technically, it required the inclusion
of derivatives of the SM self energies and tadpoles in the multi-loop
matching relations as described in \secref{sec:expansion_Master_eq}.

A thorough study of the new full-model parametrization shows that the
new approach automatically resums leading contributions in the
stop-mixing parameter $x_t$, analogously to the well known
$\tan\beta$-resummation.  
This $x_t$-resummation leads to  significantly stabilized convergence
of the perturbation series.
For instance, in standard parameter scenarios such as in
\figref{fig:feft_comparison} and \ref{fig:FO-EFT_compare}, the
numerical impact of the known 2-loop 
(gaugeless) and 3-loop (leading QCD) threshold corrections is reduced
to less
than $\sim 0.3\GeV$, compared to an impact of order $0.5$--$1.5\GeV$
in EFT-parametrized calculations.

Next, we have performed a detailed analysis of missing higher-order
contributions of our 3-loop \feft\ calculation. We have employed 
several different methods of uncertainty estimates, which have a
complementary sensitivity to different types of missing higher-order
contributions. Our analysis 
indicates that the remaining theory uncertainty of our calculation is
dominated by $(i)$ missing loop corrections to the top Yukawa coupling at
the electroweak scale and $(ii)$ missing electroweak 2-loop
corrections to the quartic Higgs coupling at the SUSY
scale, as shown in \figref{fig:uncert_sources}.  Numerically, we find
that the remaining theory uncertainty of our 3-loop \feft\ calculation
amounts to $\DMh \lesssim 1\GeV$ for SUSY scales above $1\TeV$ and a
stop-mixing of $x_t \lesssim 3$.  This uncertainty is reduced to
$\DMh\sim 0.5\GeV$ for vanishing stop-mixing and/or SUSY scales of
$\MS\gtrsim 10\TeV$.

Finally, we note that the resummation effects might be of high
relevance for non-minimal supersymmetric extensions of the Standard
Model, where the loop corrections to the Higgs mass are not known to
the same order as in the MSSM.  There, the matching correction in the
full-model parametrization at NLO, for example, would result in a
resummation of highest stop-mixing contributions of
$\ord(\hat y_t^2(\hat y_t^2 +\hat g_{1,2}^2 ) \hat g_3^{2n})$ with $n>1$, making resummation
effects more advisable.

\acknowledgments

We are grateful to Jonas Klappert for helpful communication on the
\Himalaya\ library and to Henning Bahl for discussions about details
of hybrid calculations.
We thank Ulrich Nierste for detailed discussions of 
$\tan\beta$-resummation and of refs.~\cite{Carena:1999py,Hofer:2009xb}.
This research was supported by the German Research Foundation (DFG)
under grant number STO 876/2-2 and by the high-performance computing
cluster Taurus at ZIH, TU Dresden.

\clearpage

\appendix

\section{Higgs pole mass matching for $\Delta\lambda$ at $\ord(y_t^6)$}
\label{App:lambda_atat}

In this appendix we show how logarithmic contributions cancel against each other
in the (implemented) master formula \eqref{eq:lambda_from_Mh} at $\ord(y_t^6)$,
 yielding a
threshold correction in eq.~\eqref{eq:Delta_at^3_alt} which is equivalent to the results presented in the literature.
For the sake of brevity we consider the single scale scenario, where
all \DRbarPrime\ SUSY mass parameters and the matching scale are set
equal to $\MS$, i.e.\
$\msf{f}^2 = M_i^2 = \mu^2 = m_A^2 = Q^2 = M_S^2$ ($f=q,u,d,l,e$).
Furthermore, to keep the expressions short, we consider a scenario
with a sufficiently large value of
$\tan\beta$, such that a power expansion in
$\cot\beta \equiv ct_\beta$ up to terms $\propto ct_\beta^2$ is
reasonable.
According to eq.~\eqref{eq:double-loop-exp-2L} the 2-loop threshold
correction is obtained as
\begin{align}
\left.\Delta \lambda^\twoL\right|_{y_t^6}&=\frac{1}{v^2}\left(\Delta s_h^{\MSSM,y_t^6}-\Delta s_h^{\SM,y_t^6} \right) 
-2\frac{\Delta v^{\at}}{v^3} \left(\Delta s_h^{\MSSM,y_t^4}-\Delta s_h^{\SM,y_t^4} \right)\,.
\label{eq:app:Dl2}
\end{align}
At the considered order, the Higgs pole mass correction in the MSSM in
the gaugeless limit is given by
\begin{align}
\Delta s_h^{\MSSM,y_t^6} = (\Delta m^{2,\MSSM}_{h, \EP})^{y_t^6}
+ (\Delta m^{2,\MSSM}_{h, \text{p}})^{y_t^6} ,
\label{eq:split_at^2_contr}
\end{align}
where the first term on the r.h.s.\ of eq.~\eqref{eq:split_at^2_contr}
represents the MSSM effective potential contribution from
ref.~\cite{Espinosa:2000df},
\begin{align}
\begin{split}
(\Delta m^{2,\MSSM}_{h, \EP})^{y_t^6}=\frac{y_t^6 v^2}{4(4\pi)^4}   \Big\{
&-4 \left[48 K+9 L_{St}^2 + 21 L_{St}+\pi ^2-12\right] \\&  +
(59-96 K) x_t^4+8 (36 K-17) x_t^2-6 x_t^6
\\& -ct_{\beta}\, 2 x_t \left[(96 K+19) x_t^4-16 (24 K+5) x_t^2+36 (16 K+3)\right]\\
&+ ct_{\beta}^2\, \big[-(96 K+1) x_t^6+(576 K-105) x_t^4+ 4 (73-384 K)
x_t^2
\\& ~~~~~~~~+24 \left(38 K+\pi ^2-7\right)+216 L_{St}+108 L_{St}^2\big]\Big\},
\end{split}
\end{align}
where $L_{St} \equiv \log(\MS^2/m_t^2)$.
The second term on the r.h.s.\ of eq.~\eqref{eq:split_at^2_contr}
originates from the momentum-dependence of the 1-loop Higgs
self-energy.  It can be regarded as the difference between the pole
mass and the mass shift induced by the MSSM effective potential.  The
SM 2-loop contributions at the considered order are given by
\begin{align}
\begin{split}
  \label{eq:app:DSM}
  \Delta s_h^{\SM,y_t^6} ={}& (\Delta m^{2,\SM}_{h, \EP})^{y_t^6}  
 + \left(\frac{\partial}{\partial p^2 }\Delta s^{\SM,\hat y_t^4}(0)\right) \Delta (p^2)^{\at}    \\
 & +
  \left(\frac{\partial}{\partial \hat y_t }\Delta s^{\SM, \hat y_t^4 }(0)\right) \Delta y_t^{\at}+
   \left(\frac{\partial}{\partial \hat v }\Delta s^{\SM,\hat y_t^4} (0)\right) \Delta v^{\at} \,,
   \end{split}\\
      (\Delta m^{2,\SM}_{h, \EP})^{y_t^6} ={}& \frac{3  y_t^6 v^2 (3\,ct_\beta^2 -1 )}{(4\pi)^4}  
   \left[3L_{St}^2 + 7 L_{St} +2+\frac{\pi^2}{3}\right]\,,
\end{align}
where the first term on the r.h.s.\ of eq.~\eqref{eq:app:DSM}
represents the contribution from the SM effective potential
\cite{Ford:1992pn} and the other terms are the implicit contributions.
The combination of the MSSM and SM momentum contributions reads,
\begin{multline}
\label{eq:mom_iter_atat}
(\Delta m^{2,\MSSM}_{h, \text{p}})^{y_t^6}
-\left(\frac{\partial}{\partial p^2 }\Delta s^{\SM,\hat y_t^4}(0)\right) \Delta (p^2)^{\at}\\
=\frac{y_t^6 v^2 (3\,ct_\beta^2 -1 )}{4(4\pi)^4}x_t^2
\left[ 12L_{St} +12x_t^2 - x_t^4 \right] .
\end{multline}
This contribution were for example
presented in eq.~(A.9) of ref.~\cite{Vega:2015fna} and have been denoted as
WFR contributions.
The combination of the remaining (implicit) terms reads
\begin{multline}
  \frac{2\Delta v^{\at}}{v}
  \left(\Delta s_h^{\SM, y_t^4}-\Delta s_h^{\MSSM,y_t^4} \right)
  -\left(\frac{\partial}{\partial \hat v }\Delta s^{\SM, \hat y_t^4}(0)\right) \Delta v^{\at}
  -  \left(\frac{\partial}{\partial \hat y_t }\Delta s^{\SM,\hat y_t^4}(0)\right) \Delta y_t^{\at}\\
   = \frac{y_t^6 v^2}{4(4\pi)^4} \left[18\, ct_\beta^2\,(2 L_{St} -1)
   +(1-3\,ct_\beta^2)x_t^2 (12 L_{St} -12x_t^2 +x_t^4)\right].
\end{multline}
Note, that these (implicit) contributions arise in our calculation due
to our choice of the full-model parametrization of $\hat\lambda$.
Inserting all contributions from above into eq.~\eqref{eq:app:Dl2},
all large logarithms cancel and one obtains
\begin{align}
\label{eq:Delta_at^3}
\begin{split}
\left.\Delta\lambda^\twoL\right|_{y_t^6} 
= \frac{y_t^6}{4(4\pi)^4} \Big\{
& - 4 x_t^6 + (35-96 K) x_t^4 + 8 (36 K-17) x_t^2 - 192 K + 72 \\
& + ct_\beta\, \big[-2 (96 K+19) x_t^5+32 (24K+5) x_t^3-72 (16 K+3) x_t\big]\\
& + ct_{\beta}^2\,\big[-(96 K+7) x_t^6+(576 K-33) x_t^4+4 (73-384 K) x_t^2\\
& \phantom {+ct_{\beta}^2  \big[\,}+6 (152K+2 \pi ^2-43)\big]
\Big\}.
\end{split}
\end{align}
Inserting the numerical value for the constant $K \simeq -0.1953256$
\cite{Espinosa:2000df}, one arrives at the expression in
eq.~\eqref{eq:Delta_at^3_alt}.

\newpage
\addcontentsline{toc}{section}{References}

\bibliographystyle{JHEP}
\bibliography{flexibleefthiggs}


\newpage

\section{Addendum}

The following text in this section has been published by the authors
as a separate addendum \cite{Kwasnitza:2023} to this publication. We
include it here for convenience.

\subsection{Introduction}

We present the C++ program \prog, which implements the 3-loop \feft\
state-of-the art calculation of $M_h$ in the real MSSM at \NCLL\ and
\NCLO\ with $x_q$ resummation.  The program is based on the \fs\ model
\texttt{NUHMSSMNoFVHimalaya} and implements the matching and running
described in our original publication, thus reproducing the results
presented there. The program provides an easy-to-use SLHA interface
for the MSSM input parameters and prints the value of $M_h$ as a
single number to \texttt{stdout}.

We have structured the addendum as follows. In \secref{sec:building}
we describe the technical details relevant for building the
program. In \secref{sec:interface} we discuss the user interface
and relevant configuration options.  Finally, we comment on the
upcoming integration of the refined FlexibleEFTHiggs approach with
full-model parametrization into the general \fs\ package.

\subsection{Installation and usage of the stand-alone code}
\label{sec:building}

The \prog\ program can be downloaded as compressed package from
\begin{center}
  \footnotesize\url{https://flexiblesusy.hepforge.org/downloads/FlexibleEFTHiggs/MSSMEFTHiggs3L.tar.gz}
\end{center}
To build \prog, the boost C++ library, the Eigen3 library, the GNU
Scientific Library and the Himalaya library~\cite{Harlander:2017kuc} (version 4.0.0 or
higher) are required. For installation instructions of the Himalaya
library see e.\,g.\ ref.~\cite{Harlander:2017kuc}.

After the package has been extracted, it can be configured and
compiled by running the following commands:
\begin{lstlisting}
$ ./configure --enable-himalaya --enable-fflite \
     --with-himalaya-incdir=${HIMALAYA_DIR}/include \
     --with-himalaya-libdir=${HIMALAYA_DIR}/build \
     --with-models=NUHMSSMNoFVHimalaya
$ make
\end{lstlisting}
The variable \texttt{HIMALAYA\_DIR} contains the path to
Himalaya root directory, required for the 3-loop pole-mass matching.
Due to an improved numerical robustness, we recommend the
configuration with the shipped 1-loop integral library
\texttt{FFLite}. For more options see \texttt{./configure -h}.  After
the compilation has finished, the program can be run with the shipped SLHA
input file as follows:
\begin{lstlisting}
$ SLHA_INPUT=models/NUHMSSMNoFVHimalaya/LesHouches.in.NUHMSSMNoFVHimalaya
$ models/NUHMSSMNoFVHimalaya/run_NUHMSSMNoFV_fefthiggs.x \
     --slha-input-file=$SLHA_INPUT
\end{lstlisting}
Running the program with the shipped SLHA input file yields the
following output for the lightest $CP$-even Higgs pole mass $M_h$ on
command line:
\begin{lstlisting}
123.522878
\end{lstlisting}

\subsection{Interface and configuration options}
\label{sec:interface}

The \prog\ program expects the MSSM input parameters in SLHA-1 format,
see \figref{fig:interface}. It calculates the lightest $CP$-even Higgs
boson pole mass $M_h$ in the real MSSM with fermion and sfermion
flavour conservation and with the non-universal Higgs mass parameters
$m_{H_u}^2$ and $m_{H_d}^2$ fixed by the electroweak symmetry breaking
conditions, as described in the original publication.  When the
calculation has finished successfully, the program writes the decimal
floating-point value of the $M_h$ to \texttt{stdout}.

\begin{figure}[tb]
	\centering
	\begin{tikzpicture}
	\path [line,line width=3pt, draw=black!40] (-5.5,2.5) --node[above] {\textbf{Input}} node[below] {SLHA file}  (-3,2.5);
	\path [line,line width=3pt, draw=black!40 ]
	(6,2.5)-- node[above] {\textbf{Output}} node[below] {$M_h$}  (8.5,2.5);
	\begin{scope}
	\path[arrow] (0,0) -- (0,5) node[left]{$Q$};
	\draw[thick] (0,4) node[left]{$\Qmatch$} -- node[above = 0.5cm]{MSSM} (3,4) node[right]{matching};
	\draw[thick] (0,1) node[left]{$\Qlow$} -- (3,1) node[right]{calculate $M_h$};
	\draw[thick] (1.5,2.5) node[]{SM};
	\path[arrow,latex-] (-1.5,1) -- node[above,rotate=90]{RG running} (-1.5,4);
	\draw[line width=1, draw=gray,rounded corners] (-2.5,6) rectangle (5.5,-.5);
	\node[color=black!85, anchor=west] at (-2.5,5.7)  {\prog};
	\end{scope}
	\end{tikzpicture}
	\caption{Interface of the stand-alone code}
	\label{fig:interface}
\end{figure}
The multi-loop contributions entering the Higgs mass calculation are
controlled by the configuration options in the \fs\ block of the SLHA
input.  A detailed documentation of the flags is given in ref.~\cite{Athron:2017fvs}.
Here, we discuss the relevant options in the \fs\ block of the SLHA
input, which controls the individual corrections of the Higgs pole
mass calculation. Depending on the desired precision of the Higgs pole
mass calculation, we present two configurations.

\paragraph{Default 3-loop precision (i.\,e.\ $M_h$ at \NCLO, \NCLL and with $x_q$-resummation):}
For a consistent \feft~calculation at this order, the following relevant flags have to be set in the SLHA input:
\begin{lstlisting}
Block FlexibleSUSY
 4   3           # pole mass loop order
 5   3           # EWSB loop order
 6   4           # beta-functions loop order
 7   3           # threshold corrections loop order
 8   1           # Higgs 2-loop corrections O(alpha_t alpha_s)
 9   1           # Higgs 2-loop corrections O(alpha_b alpha_s)
10   1           # Higgs 2-loop  O((alpha_t + alpha_b)^2)
11   1           # Higgs 2-loop corrections O(alpha_tau^2)
13   2           # Top pole mass QCD corrections (1 = 2L, 2 = 3L)
18   0           # pole mass scale in the EFT (0 = Mt)) 
19   0           # EFT matching scale (0 = SUSY scale)
20   2           # EFT loop order for yt matching
21   3           # EFT loop order for lambda matching
24   124111321   # individual threshold correction loop orders
26   1           # Higgs 3-loop corrections O(alpha_t alpha_s^2)
27   0           # Higgs 3-loop corrections O(alpha_b alpha_s^2)
28   0           # Higgs 3-loop corrections O(alpha_t^2 alpha_s)
29   0           # Higgs 3-loop corrections O(alpha_t^3)
30   0           # Higgs 4-loop corrections O(alpha_t alpha_s^3)
\end{lstlisting}
The meaning of each flag is described in the associated comment. The
user should be aware that deviations from the displayed flag
configuration usually result in a reduced precision of the
calculation.  In the following we briefly describe a selection of
adjustments:
\begin{itemize}
\item \textbf{Flag 18} This flag can be used to set the
  renormalization scale $\Qpole$ (in GeV), at which the Higgs pole
  mass $M_h$ is calculated in the SM. Possible values are $\Qpole=0$,
  which corresponds to $\Qpole=M_t$, or any positive value $\Qpole>0$.
  This flag can be used to vary the renormalization scale in order to
  estimate the low-scale uncertainty as described in
  \secref{sec:low-scale_uncertainty}.

\item \textbf{Flag 19} This flag can be used to set the matching scale
  $\Qmatch$ (in GeV) at which $\lambda$ is determined. Possible values
  are $\Qmatch=0$, which corresponds to $\Qmatch=\MS$, or any positive
  value $\Qmatch>0$.
  This flag can be used to vary the matching scale in order to
  estimate the high-scale uncertainty as described in the vicinity of
  eq.~\eqref{eq:DMhQpole}.

\item \textbf{Flag 20} This flag has a different meaning than
  described in the documentation in ref.~\cite{Athron:2017fvs}, where
  it controls the loop order of the upwards matching from the SM to
  the full model.  Our calculation does not require any upwards
  matching and we use it to control the downwards matching of SM-like
  gauge and Yukawa couplings.  Possible values are 0 (tree-level), 1
  (1-loop) and 2 (2-loop).  For a calculation of $M_h$ at \NCLO\ and
  \NCLL, the flag must be set to $2$. Reducing the value to 1 or 0
  reduces the large-log resummation to NNLL or LL, respectively.

\item \textbf{Flag 21} This flag controls the loop order for the
  calculation of $\lambda$ and specifies the contributions in
  eq.~\eqref{eq:hat-lambda}. Possible values are 0 (tree-level), 1
  (1-loop), 2 (2-loop) and 3 (3-loop).  For a calculation of $M_h$ at
  \NCLO\ and \NCLL, the flag must be set to $3$.  If numerical
  instabilities occur, it may be beneficial to reduce the loop order
  of the calculation of $\lambda$ to 2-loop (gauge-less limit) and
  therefore restrict the precision to NNLL and NNLO.
\end{itemize}

\paragraph{Minimal 2-loop precision (i.\,e.\ $M_h$ at NNLO, NNLL and with $x_q$-resummation):}
The minimal flag configuration to achieve a \feft~calculation at this
precision requires the following configuration settings in the SLHA
input:
\begin{lstlisting}
Block FlexibleSUSY
 4   2           # pole mass loop order
 5   2           # EWSB loop order
 6   3           # beta-functions loop order
 7   2           # threshold corrections loop order
 8   1           # Higgs 2-loop corrections O(alpha_t alpha_s)
 9   1           # Higgs 2-loop corrections O(alpha_b alpha_s)
10   1           # Higgs 2-loop  O((alpha_t + alpha_b)^2)
11   1           # Higgs 2-loop corrections O(alpha_tau^2)
13   1           # Top pole mass QCD corrections (1 = 2L, 2 = 3L)
18   0           # pole mass scale in the EFT (0 = Mt)) 
19   0           # EFT matching scale (0 = SUSY scale)
20   1           # EFT loop order for yt matching
21   2           # EFT loop order for lambda matching
24   112111111   # individual threshold correction loop orders
26   0           # Higgs 3-loop corrections O(alpha_t alpha_s^2)
27   0           # Higgs 3-loop corrections O(alpha_b alpha_s^2)
28   0           # Higgs 3-loop corrections O(alpha_t^2 alpha_s)
29   0           # Higgs 3-loop corrections O(alpha_t^3)
30   0           # Higgs 4-loop corrections O(alpha_t alpha_s^3)
\end{lstlisting}

\subsection{Outlook}

In this addendum, we have presented the stand-alone program \prog,
which has been developed for the Higgs mass calculation presented in
the original publication.

We plan to implement the refined FlexibleEFTHiggs approach with
full-model parametrization into the general \fs\ package. This allows
to apply the calculation to models beyond the real MSSM, such as the
NMSSM etc.  The planned integrated version will also allow access to
the full pole-mass spectrum of the model as well as the computation of
other observables.

\end{document}